\documentclass[11pt, reqno]{amsart}

\usepackage{amsmath}
\usepackage{amsfonts}
\usepackage{amssymb}
\usepackage{amsthm}
\usepackage{mathtools}
\usepackage[scr=dutchcal, bb = stixtwo]{mathalpha}
\usepackage{tikz-cd}
\usepackage{tikz}
\usetikzlibrary{arrows.meta,decorations.pathreplacing, positioning,calc}
\usepackage{bm}
\usepackage{bbm}
\usepackage[all]{xy}
\usepackage{pdfpages}
\usepackage{enumerate}
\usepackage{float}
\usepackage{todonotes}
\usepackage{graphicx}
\usepackage{color}
\usepackage{zref-clever}
\usepackage[
colorlinks,
linkcolor=blue,
urlcolor=blue,
citecolor=blue,
destlabel=true,
bookmarksdepth=3
]{hyperref}
\usepackage{enumitem}
\setlist[itemize]{leftmargin=*}
\setlist[enumerate]{leftmargin=*}

\numberwithin{equation}{section}

\usepackage{quiver}

\newtheorem{theorem}{Theorem}[section]
\newtheorem{corollary}[theorem]{Corollary}
\newtheorem{proposition}[theorem]{Proposition}
\newtheorem{lemma}[theorem]{Lemma}

\theoremstyle{definition}
\newtheorem{definition}[theorem]{Definition}
\newtheorem{remark}[theorem]{Remark}

\zcsetup{nameinlink = false, cap}

\AddToHook{env/theorem/begin}{\zcsetup{reftype = theorem}}
\AddToHook{env/corollary/begin}{\zcsetup{reftype = corollary}}
\AddToHook{env/proposition/begin}{\zcsetup{reftype = proposition}}
\AddToHook{env/lemma/begin}{\zcsetup{reftype = lemma}}
\AddToHook{env/definition/begin}{\zcsetup{reftype = definition}}
\AddToHook{env/remark/begin}{\zcsetup{reftype = remark}}

\AddToHook{env/figure/begin}{\zcsetup{reftype = figure}}

\newcommand{\sT}{\mathscr{T}}

\newcommand{\id}{\mathrm{id}}

\DeclareMathOperator{\Hom}{Hom}

\let\Re\relax
\DeclareMathOperator{\Re}{Re}

\newcommand{\Z}{\mathbb{Z}}

\newcommand{\R}{\mathbb{R}}

\newcommand{\cL}{\mathcal{L}}

\newcommand{\fA}{\mathfrak{A}}

\newcommand{\cE}{\mathcal{E}}

\newcommand{\sH}{\mathscr{H}}

\newcommand{\cB}{\mathcal{B}}
\newcommand{\cI}{\mathcal{I}}
\newcommand{\sL}{\mathscr{L}}

\newcommand{\sK}{\mathscr{K}}

\newcommand{\cA}{\mathcal{A}}

\newcommand{\Hslab}[1]{H^{\mathrm{slab},a}}

\DeclareMathOperator{\Tr}{\mathrm{Tr}}

\newcommand{\sQ}{\mathscr{Q}}

\newcommand{\cO}{\mathcal{O}}

\newcommand{\dd}{\,d}

\newcommand{\tn}[1]{\textnormal{#1}}

\newcommand{\defeq}{\vcentcolon=}
\renewcommand{\1}{\mathbb{1}}
\DeclareMathOperator{\Ad}{\mathrm{Ad}}
\newcommand{\bbA}{\mathbb{A}}

\newcommand{\bbC}{\mathbb{C}}
\newcommand{\bbS}{\mathbb{S}}
\newcommand{\bbV}{\mathbb{V}}

\newcommand{\cJ}{\mathcal{J}}

\newcommand{\cV}{\mathcal{V}}

\newcommand{\bbP}{\mathbb{P}}

\newcommand{\bbN}{\mathbb{N}}
\newcommand{\bbZ}{\mathbb{Z}}

\newcommand{\bbR}{\mathbb{R}}
\newcommand{\Unitary}{\mathrm{U}}

\DeclareMathOperator{\diag}{\mathrm{diag}}

\newcommand{\cW}{\mathcal{W}}
\newcommand{\cM}{\mathcal{M}}

\newcommand{\cU}{\mathcal{U}}

\newcommand{\E}{\mathcal{E}}

\DeclareMathOperator{\vecspan}{\mathrm{span}}

\newcommand{\SU}{\mathrm{S}\Unitary}
\newcommand{\PU}{\mathbb{P}\Unitary}
\newcommand{\PM}[1]{\mathbb{P}M_{#1}}
\newcommand{\SM}[1]{\mathbb{S}M_{#1}}

\newcommand{\RP}{\mathbb{RP}}
\newcommand{\act}{\triangleright}

\newcommand{\norm}[1]{\left\| #1 \right\|}
\newcommand{\mqty}[1]{\begin{pmatrix} #1 \end{pmatrix}}
\newcommand{\smallmat}[1]{\begin{smallmatrix} #1 \end{smallmatrix}}
\newcommand{\ketbra}[2]{\lvert #1 \rangle\! \langle #2 \rvert}
\newcommand{\ket}[1]{\lvert#1\rangle}
\newcommand{\bra}[1]{\langle #1 \rvert}
\newcommand{\qqtext}[1]{\quad \tn{#1} \quad}
\newcommand{\abs}[1]{\left| #1 \right|}
\newcommand{\ev}[1]{\left\langle #1 \right\rangle}

\newcommand{\del}[1]{\left(#1\right)}
\newcommand{\sbr}[1]{\left[#1\right]}

\newcommand{\mat}[1]{M_{#1}}

\newcommand{\func}[2]{#2^{#1}}
\newcommand{\cont}[2]{C(#1, #2)}

\newcommand{\vecspace}[1]{{\mathscr{#1}}}

\newcommand{\MPSk}{k}
\newcommand{\MPSl}{\ell}
\newcommand{\targ}[1]{t(#1)}
\newcommand{\smallt}{t}
\newcommand{\Targ}[1]{T(#1)}
\newcommand{\bigT}{T}
\newcommand{\MPSA}{A}

\newcommand{\MPSm}{m}
\newcommand{\MPSn}{n}
\newcommand{\smallv}{v}

\newcommand{\varg}[2][]{v_{#1}(#2)}
\newcommand{\Varg}[2][]{V_{#1}(#2)}
\newcommand{\bbv}{\mathbb{v}}
\newcommand{\bbw}{\mathbb{w}}
\newcommand{\bbvarg}[1]{\mathbb{v}(#1)}
\newcommand{\bbVarg}[1]{\mathbb{V}(#1)}
\newcommand{\phasearg}[1]{\lambda(#1)}
\newcommand{\altphasearg}[1]{\Lambda(#1)}
\newcommand{\altphase}{\Lambda}
\newcommand{\smallz}{z}
\newcommand{\smallw}{w}

\newcommand{\pstate}{\mathcal{P}}
\newcommand{\qstate}{\mathcal{Q}}

\newcommand{\Cgrp}{C_{\mathrm{grp}}}
\newcommand{\Zgrp}{Z_{\mathrm{grp}}}
\newcommand{\Hgrp}{H_{\mathrm{grp}}}

\pgfmathsetmacro\mathaxis{-height("$\vcenter{}$")}

\title[Parametric and Universal Thouless Invariants]{Parametric and Universal Thouless Invariants in Matrix Product States}

\author[Goller]{Leonardo Goller\textsuperscript{1,*}}
\author[Singh]{Amartya H.\ Singh\textsuperscript{2,3}}
\author[Spiegel]{Daniel D.\ Spiegel\textsuperscript{4}}
\author[Venuti]{Marco Venuti\textsuperscript{2,3}}

\thanks{\textsuperscript{1}Mathematics Area, SISSA, Via Bonomea 265, 34136 Trieste, Italy}
\thanks{\textsuperscript{2}Physics Area, SISSA, Via Bonomea 265, 34136 Trieste, Italy}
\thanks{\textsuperscript{3}INFN, Sezione di Trieste, Via Valerio 2, 34127 Trieste, Italy}
\thanks{\textsuperscript{4}Department of Mathematics, Harvard University}
\thanks{\textsuperscript{*}Present address: IHES, 35 Route de Chartres, 91440 Bures-sur-Yvette, France}
\thanks{%
  Email addresses: \texttt{goller@ihes.fr}, \texttt{amsingh@sissa.it}, \texttt{dspiegel@math.harvard.edu}, \indent \texttt{marco.venuti.phys@gmail.com}\spacefactor 1001
}

\begin{document}

\begin{abstract}
    We investigate the classification of continuous families of translation-invariant injective matrix product states on an infinite chain, parametrized by a paracompact Hausdorff space $X$ and subject to a compact Hausdorff symmetry $G$. We construct an invariant for phases of such families, valued in $\check{H}^1(X;\widehat{G})$, that generalizes the known  $\widehat{G}$  phase invariant for $G$-symmetric Thouless pumps. The phase invariant is constructed as the pullback to $X$ of a single universal \v{C}ech cohomology class on a classifying space of $G$-symmetric translation-invariant injective matrix product states. Several examples are given over various parameter spaces where this invariant is nontrivial.
\end{abstract}

\maketitle

\tableofcontents

\section{Introduction}

Thouless pumps are a well-known class of periodic processes in the study of topological phases of quantum matter. In a typical Thouless pump, a many-body quantum system in one spatial dimension with a $\Unitary(1)$ symmetry undergoes a periodic and adiabatic evolution in which charge is transported along the spatial dimension. In the original work of Thouless \cite{OriginalThouless}, it is shown that for certain infinitely extended systems, the amount of charge transported in one period is quantized and has a topological interpretation.

Numerous works in the physics literature have extended the original Thouless charge pump in many directions, including to interacting systems \cite{NiuThouless}, to systems in higher spatial dimensions \cite{PhysRevB.82.115120,KS_HigherThouless}, to the pumping of Chern number \cite{qpump}, and to group-cohomological classifications for systems with many-body localization and  more general symmetry group $G$ \cite{ElseNayak_MBL,RoyFenner,PotterMorimotoVishwanath}. A handful of works in the physics literature have focused on classifying pumps of matrix product states  \cite{Shiozaki_AdiabaticCycles,PhysRevB.106.165115,Shiozaki_EquivariantMPS}; such states are the focus of this work as well. In the recent mathematical physics literature, loops of invertible states on spin chains with a compact symmetry group $G$ have been fully classified by an invariant taking values in the Pontryagin dual $\widehat{G} \cong \Hgrp^1(G;\Unitary(1))$ \cite{Bachmann2024}. That work proves that two loops have the same value of the invariant if and only if they are related to each other by a specialized notion of homotopy.

This article generalizes the $\widehat{G}$ invariant that classifies Thouless pumps of spin chains to an invariant for families of translation-invariant injective matrix product states parametrized by an arbitrary paracompact Hausdorff topological space $X$. Loosely speaking, a parametrized quantum system is a family of Hamiltonians or states (in this article we focus almost exclusively on states) that vary continuously with parameters in $X$. Thus, a Thouless pump can be thought of as a parametrized system with $X = S^1$. We will show how to assign an invariant in the degree one \v{C}ech cohomology group $\check{H}^1(X;\widehat{G})$ to an $X$-parametrized family of translation-invariant injective matrix product states invariant under a compact symmetry $G$. When $X = S^1$, one has $\check{H}^1(S^1; \widehat{G}) \cong \widehat{G}$, recovering the previously known invariant for Thouless pumps.

Our interest in studying parametrized systems largely stems from Kitaev's conjecture \cite{kitaevSimonsCenter,KitaevHomotopy,kitaev}. Roughly, his conjecture states that in each spatial dimension $d$ there should be a space $\mathcal{Q}_d$ of gapped bosonic invertible quantum states, and that there should be weak homotopy equivalences $\mathcal{Q}_d \to \Omega \mathcal{Q}_{d+1}$, where $\Omega \mathcal{Q}_{d+1}$ is the loop space of $\mathcal{Q}_{d+1}$. Thus, the spaces $\mathcal{Q}_d$ form a loop spectrum. A parametrized family of such states in $d$ spatial dimensions could then be realizable as a continuous function $X \to \mathcal{Q}_d$, whereas a phase could be identified as a homotopy class of maps $X \to \mathcal{Q}_d$. The topology of $\mathcal{Q}_d$ determines what kinds of invariants distinguish the different phases of parametrized systems. We note that Kitaev also made the analogous conjecture for fermionic systems, but we will restrict our attention to bosonic systems here.

Only in recent years have the first rigorous definitions of spaces $\mathcal{Q}_d$ for $d > 0$ been given in accordance with Kitaev's description \cite{HomotopicalFoundations,KZ3,kubota,JonesSopenkoThorngren_QCA}. In this paper we will use the topological space $\mathcal{B}$ defined in \cite{KZ3}; it is a model for $\mathcal{Q}_1$ that is concrete, relatively easy to define, and specifically designed for matrix product states. Every point of $\cB$ corresponds to a unique translation-invariant injective matrix product state on an infinite 1d lattice and vice versa. In \cite{KZ3} it was proven that $\cB$ has the weak homotopy type $K(\bbZ, 2) \times K(\bbZ, 3)$, leading to an $H^2(X;\bbZ) \times H^3(X;\bbZ)$ classification of families of matrix product states parametrized by $X$. The $H^3(X;\bbZ)$ phase invariant is sometimes called the Kapustin--Spodyneiko invariant or higher Berry class \cite{KapustinSpodyneiko, ArtymowiczKapustinSopenko}.

Points of $\cB$ have no symmetry imposed on them, so even if a parametrized family $X \to \cB$ possesses a certain symmetry, its homotopy class will contain maps that violate the symmetry.
Thus, it is not surprising that $\cB$ is connected and simply connected: this is the statement that there are no nontrivial unparametrized phases or Thouless pumps in 1d spin systems without symmetry.
To impose a symmetry, one must restrict to the subspace of $\cB$ consisting of those states that respect the symmetry.
This is in accordance with \cite{qpump} wherein it is proposed that the symmetry be built into the space of allowable states.
Our symmetry will take the form of a representation $U: G \to \Unitary(\bbC^\infty)$, and the subspace of symmetric states will be called $\cB_U$; we refer the reader to \zcref{sec:regarding_symmetries} for detailed definitions.
In contrast with $\cB$, the space $\cB_U$ need not be connected or simply connected.
Indeed, the SPT index in $\Hgrp^2(G;\Unitary(1))$ \cite{PollmannEntanglementSpectrum,PollmannBergTurnerOshikawa12,ChenGuWen2011_1,ChenGuWen2011_2,SchuchClassifyingQuantumPhasesMPS,Ogata20,Ogata21} is constant on the connected components of $\cB_U$ (\zcref{cor:SPT_locally_constant}), so states with different SPT indices lie in different components. For example, for $G = \Z_2 \times \Z_2$, this is the case for the product state and the cluster state of \zcref{sec:clsta}.

With $\cB_U$ in hand, this paper will consider continuous families $X \to \cB_U$ as the objects of study and will assign a \v{C}ech cohomology class in $\check{H}^1(X;\widehat{G})$ to such a family, generalizing the Thouless pumping invariant that describes the flow of charge.
The $\check{H}^1(X;\widehat{G})$ invariant, which we call a \emph{parametric Thouless invariant}, is obtained as the pullback along the parametrized family $X \to \cB_U$ of a single \v{C}ech class in $\check{H}^1(\cB_U; \widehat{G})$---it follows that the parametric Thouless class is an invariant of the homotopy class of the map $X \to \cB_U$. In particular, when $X = S^1$ this cohomology class detects nontrivial loops in $\cB_U$. Since every parametric Thouless invariant comes from a single \v{C}ech class on $\cB_U$, we call this element of $\check{H}^1(\cB_U;\widehat{G})$ a \emph{universal Thouless invariant}. The universal Thouless invariant can also be thought of as the parametric Thouless invariant coming from the parametrized family with $X = \cB_U$ and $X \to \cB_U$ as the identity map.

Aside from the novel generalization of the Thouless invariant, this work is somewhat unique in the methods used. Namely, we will work almost exclusively in the category of topological spaces and continuous maps (with a brief excursion into the category of topological spaces and Borel measurable maps). While this is certainly a very familiar category, most other works on Thouless pumps and topological phases in the mathematical physics literature use much more specialized notions of path, parametrized family, and homotopy (e.g., \cite{Bachmann2024,KapustinSopenko_LocalNoether}) that are designed to admit quasi-adiabatic evolutions along paths. Working in the more familiar category of topological spaces and continuous maps allows us to bring to bear tools from algebraic topology and reflects more closely the intuitive definition of topological phases as path components, or more generally homotopy classes of maps, in a phase diagram.

The contents of the paper are as follows.
In \zcref{sec:Definitions_Notation} we lay out definitions and notation needed for the remainder of the paper, briefly review the construction of the space $\cB$ from \cite{KZ3}, and provide an exposition of some abstract results about representations on inner product spaces of countably infinite dimension.
In \zcref{sec:phase_def}, we clarify the definition of a parametrized family and justify our use of the classifying space $\cB_U$ by showing how it can be used to rephrase a more familiar looking notion of phase---one defined in terms of stacking.
We proceed to the general construction of the universal Thouless invariant in \zcref{sec:general_construction} and we provide examples over a variety of topological spaces where the parametric Thouless invariant is nontrivial in \zcref{sec:examples}.

Finally, let us take a moment to constrast in more detail the present work with prior works on Thouless pumps and parametrized families of matrix product states. The work \cite{KS_HigherThouless} generalizes Thouless pumps to parametrized families beyond the circle, but requires the dimension of the parameter space to match the spatial dimension of the system; here we will discuss only one-dimensional systems but will consider arbitrary parameter spaces. Thouless pumps of matrix product states parametrized by \(S^1\) and protected by a finite symmetry group \(G\) were studied in \cite{Shiozaki_AdiabaticCycles}. In that article, the construction of the pumping invariant relies on the existence of a globally continuous and periodic choice of MPS tensors and associated virtual symmetry operators, which need not be available for general parametrized families.
The recent work \cite{Shiozaki_EquivariantMPS} develops a framework for equivariant families of matrix product states, allowing $G$ to act nontrivially on the parameter space. The approach there is based on discretizing the parameter space. For the subgroup $G_0$ of unitary symmetries acting trivially on the parameters, \cite{Shiozaki_EquivariantMPS} associates a an element of $\widehat{G}$ to each discretized loop, yielding an invariant in $\operatorname{Hom}(\pi_1(X),\widehat{G_0})$ and recovering the usual Thouless pump for $G_0=U(1)$. For sufficiently regular parameter spaces, such as connected CW complexes, this group is naturally isomorphic to $\check H^1(X;\widehat{G_0})$, so that the two approaches yield invariants taking values in the same cohomology group.
Finally, \cite{Bachmann2024} provides a complete classification of \(G\)-symmetric loops of one-dimensional invertible states for compact symmetry groups \(G\), proving that two loops with stably \(G\)-equivalent basepoints are stably \(G\)-homotopic if and only if their pumping invariants coincide. This result relies on a specialized notion of loops and stable homotopy, defined in terms of almost-local evolutions, rather than the ordinary topological notions. Likewise, \cite{KapustinSopenko_LocalNoether} constructs equivariant higher Berry invariants for a specialized notion of smooth parametrized families of gapped states, encompassing a broader class of states than those considered in our work and recovering Thouless pumping as a special case. The invariants in \cite{KapustinSopenko_LocalNoether} take values in real equivariant cohomology, whereas our construction yields classes in \(\check H^1(X,\widehat G)\), allowing us, in particular, to capture torsion pumping invariants that are invisible to real cohomology.

Our contribution is a general and mathematically rigorous approach that does not require discretizing the parameter space or finding a globally continuous representation of the state in terms of MPS tensors. The intrinsic construction of the parametric Thouless invariant directly on the space $\cB_U$ of physical states is also novel, and the use of ordinary continuous maps and homotopies is natural but uncommon.

%!TEX root = MPS_Symmetry-master.tex

\section{Preliminaries}\label{sec:Definitions_Notation}

\subsection{General Notation and Terminology}

\leavevmode
\smallskip

\textit{Function Spaces and their Topology.}
If $X$ and $Y$ are sets, we let $\func{X}{Y}$ denote the set of functions $X \to Y$.
If $(Y,d)$ is a metric space and $X$ is any nonempty set, then a function $f:X \to Y$ is said to be \emph{bounded} if the image $f(X)$ has finite diameter. The set of bounded functions $X \to Y$ can be equipped with the \emph{uniform metric}:
\[
	D(f_1, f_2) \defeq \sup_{x \in X} d(f_1(x), f_2(x)).
\]
The topology induced by this metric will be called the \emph{uniform topology}. In \zcref{sec:general_construction} we will commonly be in a situation where $Y$ itself has finite diameter, in which case the set of bounded functions coincides with $Y^X$ and thus $Y^X$ may be equipped with the uniform metric and uniform topology.

If $X$ and $Y$ are topological spaces, then we denote by $\cont{X}{Y}$ the space of continuous functions $X \to Y$, equipped with the compact-open topology, namely the topology generated by sets of the form:
\begin{equation*}
O_{K, U} \defeq \{f \in \cont{X}{Y}: f(K) \subset U\}, \quad K \subset X \, \, \text{compact and} \,\, U \subset Y \,\, \text{open}.
\end{equation*}

If $(Y,d)$ is a metric space, then the compact-open topology on
$\cont{X}{Y}$ coincides with the topology of uniform convergence on
compact subsets of $X$; see, e.g., \cite[Chapter XII, Sec.~7]{DugundjiTopology}.
In particular,
\begin{equation*}
    f_n \to f
    \quad\Longleftrightarrow\quad
    \forall\, K \subseteq X \text{ compact},\qquad D(f_n|_K, f|_K) \to 0.
\end{equation*}
If $X$ is compact, the compact-open topology on $\cont{X}{Y}$
therefore coincides with the uniform topology. Moreover, if $X$ is any Hausdorff space, $Y$ is a locally compact Hausdorff space, and $Z$ is any topological space, the
compact-open topology satisfies the exponential law
\begin{equation*}
    \cont{X \times Y}{Z}
    \cong
    \cont{X}{\cont{Y}{Z}},
\end{equation*}
where all spaces of continuous maps are endowed with the compact-open
topology; see \cite[Ch.~5, \S 3]{Hu_Topology}. The homeomorphism
is given by \emph{currying},
\begin{equation*}
    f \longmapsto
    \bigl(x \longmapsto (y \longmapsto f(x,y))\bigr).
\end{equation*}

Finally, a function between topological spaces $X$ and $Y$ will be called \emph{Borel measurable} if it is measurable with respect to the Borel $\sigma$-algebras on domain and codomain.

\smallskip

\textit{Projective Vector Spaces and their Topology.}
If ${\vecspace{V}}$ is a nonzero topological vector space, then $\bbP {\vecspace{V}}$ will denote the projectivization of ${\vecspace{V}}$, i.e., the space of one-dimensional subspaces of ${\vecspace{V}}$, equipped with the quotient topology obtained from the canonical projection ${\vecspace{V}} \setminus \left\{0\right\} \to \bbP {\vecspace{V}}$. If ${\vecspace{V}}$ is equipped with a norm, then we notate
\[
\bbS {\vecspace{V}} = \left\{v \in {\vecspace{V}} : \norm{v} = 1\right\}.
\]
If the norm is continuous (not necessarily assuming it induces the topology of ${\vecspace{V}}$), then we note that the above topology on $\bbP{\vecspace{V}}$ coincides with the quotient topology obtained from the restriction $\bbS {\vecspace{V}} \to \bbP {\vecspace{V}}$ of the canonical projection. Both the restricted and unrestricted projections are fiber bundles.

\smallskip
\textit{Colimits and the Colimit Topology.}
Recall that a \emph{directed set} $(I, \leq)$ is a nonempty set $I$ endowed with a reflexive and transitive relation $\leq$ such that, for every $i,j \in I$, there exists $k \in I$ satisfying $i \leq k$ and $j \leq k$. Further recall that a subset $J \subset I$ is \emph{cofinal} if for every $i \in I$ there exists $j \in J$ such that $i \leq j$.

Suppose $X$ is a nonempty set with a collection $\cA$ of subsets of $X$ such that $\cA$ is a directed set when partially ordered by inclusion of subsets. Further assume that $X = \bigcup \cA$ and that each $A \in \cA$ is equipped with a topology such that whenever $A, B \in \cA$ and $A \subset B$, the inclusion map $\iota:A \to B$ is continuous. We may endow $X$ with the final topology induced by the inclusion maps $A \hookrightarrow X$ across all $A \in \cA$. With this topology, $X$ together with the inclusions $A \hookrightarrow X$ is a colimit of the collection $\cA$ with the inclusion maps $A \hookrightarrow B$, in the category of topological spaces and continuous maps. We will say that $X$ is \emph{topologized as a colimit of $\cA$} when we wish to give $X$ this final topology. Note that any cofinal subset of $\cA$ induces the same topology as $\cA$.

If each inclusion $A \hookrightarrow B$ is a closed embedding, then each inclusion $A \hookrightarrow X$ is also a closed embedding. If there exists an increasing cofinal sequence in $\cA$ consisting of $T_1$ spaces, then any compact subset of $X$ is contained in some $A \in \cA$. These hypotheses will always be satisfied in cases of interest to us.

For example, if $\vecspace{V}$ is a vector space with a countable Hamel basis, then we may set $\cA$ to be the set of finite-dimensional subspaces of $\vecspace{V}$ and topologize $\vecspace{V}$ as a colimit of its finite-dimensional subspaces. All hypotheses stated above are satisfied for this example.

In \zcref{sec:general_construction}, colimits will also appear in the category of abelian groups and homomorphisms in the context of \v{C}ech cohomology. Let us review the general construction of a directed colimit in this category.

\begin{definition}
A \emph{directed system of sets} is a collection of sets $(X_i)_{i \in I}$ labeled by a directed set $I$ endowed with so-called \textit{transition maps} $\phi_{ji}: X_i \to X_j$ whenever $i \leq j$ satisfying the following consistency relations:
\begin{enumerate}
    \item $\phi_{ii} = \id_{X_i}$ for all $i \in I$,
    \item $\phi_{kj} \circ \phi_{ji} = \phi_{ki}$ whenever $i \leq j \leq k$.
\end{enumerate}
These maps are illustrated in the commutative diagram below.
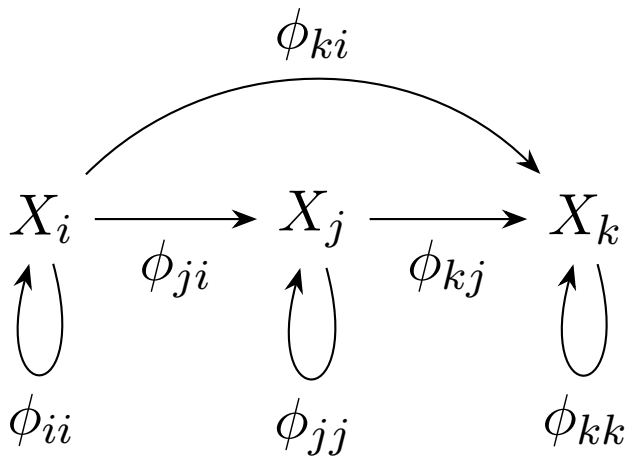
\begin{center}
\begin{tikzpicture}[>=Stealth]
    \node (X_i) at (0,0)   {$X_i$};
    \node (X_j) at (1.8,0) {$X_j$};
    \node (X_k) at (3.6,0) {$X_k$};

    \draw[->] (X_i) -- node[below] {$\phi_{ji}$} (X_j);
    \draw[->] (X_j) -- node[below] {$\phi_{kj}$} (X_k);
    \draw[->] (X_i) to[out=45,in=135]
        node[above] {$\phi_{ki}$} (X_k);

    \draw[->] (X_i) to[loop below, min distance=10mm]
        node[below] {$\phi_{ii}$} (X_i);
    \draw[->] (X_j) to[loop below, min distance=10mm]
        node[below] {$\phi_{jj}$} (X_j);
    \draw[->] (X_k) to[loop below, min distance=10mm]
        node[below] {$\phi_{kk}$} (X_k);
\end{tikzpicture}
\end{center}
\end{definition}

Given a directed system of sets, we can define a set $X$ by \cite[Section 4.15]{stacks-project}:
\[
    X \defeq
    \left(\bigsqcup_{i\in I} X_i\right)\big/\sim,
\]
where two elements $x_i \in X_i$ and $x_j \in X_j$ are identified if for some $k \geq i,j$ it holds that $\phi_{ki}(x_i)=\phi_{kj}(x_j)$ in $X_k$ \cite[Section 4.19]{stacks-project} (once this equality holds for a single $k \geq i,j$, it holds for every $\ell \geq k$ due to the compatibility condition between transition maps). The colimit $X$ is naturally equipped with canonical maps $\phi_i: X_i \to X$ given by the composition of the natural inclusion $\iota_i:X_i \to \sqcup_{j \in I} X_j$ and the quotient $\pi: \sqcup_{i \in I} X_i \to X$.

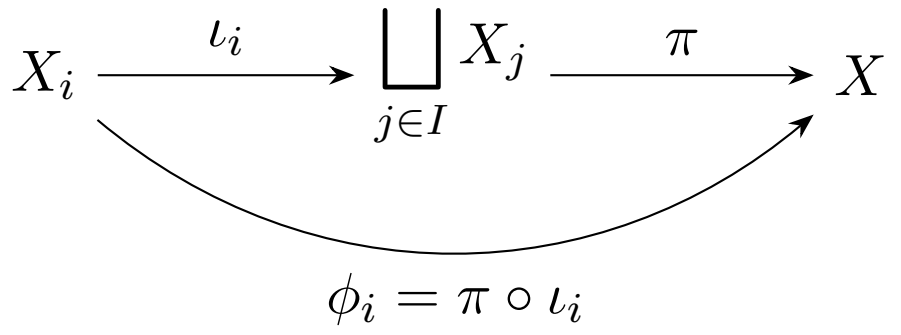
\begin{center}
\begin{tikzpicture}[>=Stealth]
    \node (Xi) at (0,0) {$X_i$};
    \node (U)  at (2.7,0) {$\displaystyle\bigsqcup_{j\in I}X_j$};
    \node (X)  at (5.4,0) {$X$};

    \draw[->] (Xi) -- node[above] {$\iota_i$} (U);
    \draw[->] (U) -- node[above] {$\pi$} (X);
    \draw[->] (Xi) to[out=-40,in=-140]
        node[below] {$\phi_i=\pi\circ\iota_i$} (X);
\end{tikzpicture}
\end{center}

The canonical maps are compatible with the transition maps:
for every $i\leq j$, $\phi_j\circ\phi_{ji}=\phi_i$. Equivalently, the following diagram commutes:

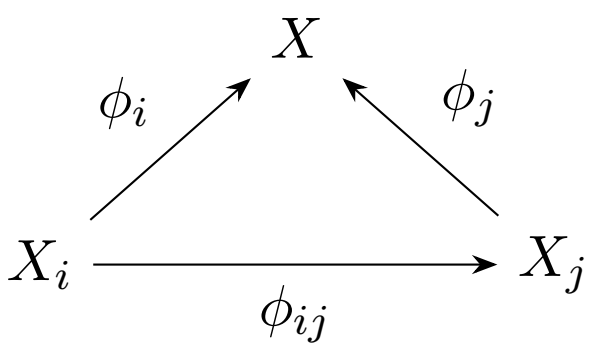
\begin{center}
\begin{tikzpicture}[>=Stealth]
    \node (Xi) at (-1.7,0) {$X_i$};
    \node (Xj) at ( 1.7,0) {$X_j$};
    \node (X)  at (0,1.5)  {$X$};

    \draw[->] (Xi) -- node[below] {$\phi_{ij}$} (Xj);
    \draw[->] (Xi) -- node[above left] {$\phi_i$} (X);
    \draw[->] (Xj) -- node[above right] {$\phi_j$} (X);
\end{tikzpicture}
\end{center}

The set $X$, together with the canonical maps $(\phi_i)_{i\in I}$, is a colimit for the directed system in the category of sets and functions. If each $X_i$ is a group and each $\phi_{ji}$ is a homomorphism, then there exists a unique binary operation on $X$ such that $X$ is a group and each $\phi_i$ is a homomorphism. With respect to this group structure, $X$ is a colimit of the directed system in the category of groups and homomorphisms. If, in addition, each group $X_i$ is abelian (as it will be in our use of \v{C}ech cohomology), then $X$ is abelian and is the colimit of the directed system in the category of abelian groups and homomorphisms.

\bigskip

\textit{Spaces of matrices.}
We use the following notation.

\begin{definition}\label{def:infinite_M_spaces}
Given $\chi \in \bbN = \{1,2,\ldots\}$, we let $\mat{\chi}$ denote the set of $\chi \times \chi$ complex matrices.
We let $\mat{\infty}$ denote the set of $\infty \times \infty$ complex matrices with only finitely many nonzero entries, topologized as a colimit of its finite-dimensional subspaces, or equivalently as the colimit of the embeddings
$\mat{\chi} \hookrightarrow \mat{\infty}$ defined by embedding $M_\chi$ in the top-left corner and adding countably infinitely many rows and columns of zeros. We write
\[
	\cM = \left\{\MPSA \in \prod_{i=1}^\infty \mat{\infty} : \MPSA^i \neq 0 \tn{ for only finitely many $i \in \bbN$}\right\},
\]
where $\MPSA^i$ is the $i$th matrix in the sequence $\MPSA$.
Given $d, D \in \bbN$, we define
\[
	\cM(d, D) = \left\{\MPSA \in \cM: \MPSA^i_{\alpha \beta} = 0 \tn{ if $i > d$ or $\alpha > D$ or $\beta > D$}\right\}.
\]
We let $\Unitary(\infty)$ denote the set of $\infty \times \infty$ matrices $Y$ of the form
\[
	Y = \mqty{ Y_0 & 0 \\0 & \1_\infty}
\]
where $Y_0 \in \Unitary(n)$ for some $n \in \bbN$ and $\1_\infty$ is the $\infty \times \infty$ identity matrix. We can regard $\Unitary(n)$ as a subgroup of $\Unitary(\infty)$ by embedding $Y_0 \in \Unitary(n)$ in the upper left corner of an infinite matrix, as above. We topologize $\Unitary(\infty)$ as the colimit of the inclusions $\Unitary(n) \hookrightarrow \Unitary(\infty)$ across all $n \in \bbN$.
\end{definition}

\subsection{Regarding MPS Tensors}\label{sec:mps-tensors}

We adhere to the definitions and notation laid out in \cite{KZ3}, which we recall here for the reader's convenience. The notation in \zcref{def:MPS_tensor} through \zcref{def:B} is essentially the same as that used in \cite[\S2]{KZ3} and the reader may find more detail there.

\begin{definition}\label{def:MPS_tensor}
We define an \emph{MPS tensor} to be a finite or infinite sequence $\MPSk = (\MPSk^1, \MPSk^2, \ldots) $ of complex square matrices $\MPSk^i$, each of the same finite or infinite dimension, such that only finitely many of the matrices $\MPSk^i$ are nonzero and, for each $i$, only finitely many of the matrix entries $\MPSk^i_{\alpha \beta}$ are nonzero. When referring to the array of numbers $\MPSk^i_{\alpha \beta}$, we call $i$ the \emph{physical index} and $\alpha$ and $\beta$ the \emph{bond indices} of $\MPSk$. The length $d$ of the sequence $(\MPSk^1,\MPSk^2, \ldots)$ is called the \emph{physical dimension} of $\MPSk$, while the dimension $\chi$ of the matrices $\MPSk^i$ is called the \emph{bond dimension}.

An MPS tensor $\MPSk$ of finite bond dimension $\chi$ is called \emph{injective} if
    \[
	\vecspan\{\MPSk^1, \MPSk^2, \ldots \} = \mat{\chi},
    \]
and \emph{right-normalized} if
    \[
	\sum_{i} \MPSk^i\MPSk^{i*} = \1_\chi
    \]
    where we use a star to denote Hermitian conjugation and $\1_\chi$ denotes the $\chi \times \chi$ identity matrix.

The set of injective right-normalized MPS tensors of finite physical dimension $d$ and finite bond dimension $\chi$ is denoted $\cI(d, \chi)$ and is topologized as a subspace of $\mat{\chi}^d$. Note that $\cI(d,\chi)$ is locally compact since it is the intersection of an open set with a closed set (namely, the set of injective tensors and the set of right-normalized tensors). The set of injective right-normalized MPS tensors of infinite physical dimension and finite bond dimension $\chi$ is denoted $\cI(\chi)$. We topologize $\cI(\chi)$ as the colimit of the maps $\cI(d, \chi) \hookrightarrow \cI(\chi)$ defined by adding infinitely many zero matrices to a sequence in $\cI(d, \chi)$.
\end{definition}

Since an MPS tensor $\MPSk$ of infinite bond and/or physical dimension must satisfy $\MPSk^i_{\alpha \beta} = 0$ for all but finitely many indices, it may be regarded as an MPS tensor of finite bond and physical dimension by truncating zeros from the array. The purpose of allowing infinite bond and physical dimension is to be able to continuously move between MPS tensors of all finite values of physical and bond dimensions.

The reader may have noticed that we used capital letters for matrices in \zcref{def:infinite_M_spaces} and lowercase letters for matrices in \zcref{def:MPS_tensor}. Our rough convention henceforth is to use lowercase letters for matrices that appear as blocks in larger matrices, with the larger matrices denoted with capital letters. We will now see this convention in use in the definition below.

\begin{definition}\label{def:E}
Given $d, D, \chi \in \bbN$ with $\chi \leq D$, we define $\cE(d, D, \chi)$ to be the set of $\MPSA \in \cM(d, D)$ of the form
\begin{equation}\label{eq:A_form_simplified}
\MPSA^i = \mu Y\mqty{ \MPSk^i & 0 \\ \MPSm^i & 0 }Y^*
\end{equation}
where $\mu \in \Unitary(1)$, $Y \in \Unitary(D)$, $\MPSk \in \cI(d, \chi)$, and each $\MPSm^i$ is an arbitrary $(D - \chi) \times \chi$ matrix. The phase $\mu$ may be absorbed into $\MPSk$ and $\MPSm$ when convenient.
The size $\chi$ of the matrix $\MPSk$ is called the \emph{essential rank} of $A$.

Taking unions over $\chi$, we define
\[
	\cE(d, D) \defeq \bigcup_{\chi \leq D} \cE(d, D, \chi)
\]
We topologize $\cE(d, D)$ as a subspace of the finite-dimensional vector space $\cM(d, D)$. We then take unions over $d$ and $D$ and define
\[
	\cE \defeq \bigcup_{d, D \in \bbN} \cE(d, D).
\]
We give $\cE$ the final topology induced by the inclusion maps $\cE(d, D) \hookrightarrow \cE$.
We note that a tensor $A \in \cE$ has a well-defined and unique essential rank.
\end{definition}

\begin{definition}\label{def:L_and_Q}
Given an MPS tensor $\MPSA$, we define $Q(\MPSA)$ to be the projection onto the image of the matrix $\sum_i \MPSA^{i*}\MPSA^i$. Note that if $\MPSA$ is of the form \eqref{eq:A_form_simplified}, then
\[
	Q(\MPSA) = Y\mqty{ \1_{\chi} & 0 \\ 0 & 0 } Y^*.
\]
We refer the reader to \cite[Def.~2.6]{KZ3} for the justification of this fact.
\end{definition}

\begin{definition}\label{def:B}
We define an equivalence relation on $\cE$ by writing $\MPSA \sim B$ if $\MPSA$ and $B$ can be written in the form \eqref{eq:A_form_simplified} with the same $\MPSk$ but possibly different $\mu$, $Y$, and $\MPSm$. This can be described independent of the decomposition \eqref{eq:A_form_simplified}: we write $\MPSA \sim B$ if there exists $(\mu, Z) \in \Unitary(1) \times \Unitary(\infty)$ and $C \in \cM$ such that $Q(\MPSA)C^i = 0$, $C^i Q(\MPSA) = C^i$, and
\[
	B^i = \mu Z(\MPSA^i + C^i)Z^*
\]
We define $\cB = \cE/{\sim}$, denote the projection map by $p:\cE \to \cB$, and give $\cB$ the resulting quotient topology. We define $\cB(d, D) = p(\cE(d, D))$ and we give this space the quotient topology it inherits from $\cE(d,D)$. This topology coincides with the subspace topology obtained from $\cB$ \cite[Rem.~2.14 \& Rem.~2.16]{KZ3}. An element of $\cB$ will be referred to as a \emph{state}. This terminology is justified by the existence of a canonical injection $\cB(d, D) \to \pstate(d)$, where  $\pstate(d)$ is the set of pure states of the quasi-local algebra $\mathfrak{A}(d)$ with lattice $\bbZ$ and on-site algebra $\mat{d}$, commonly written as:
\[
	\mathfrak{A}(d) = \overline{\bigcup_{\Lambda \subset \bbZ} \bigotimes_{i \in \Lambda} \mat{d}}^{\norm{\cdot}},
\]
where the union is taken over all nonempty finite subsets of $\bbZ$. These states on the quasi-local algebra were originally defined and studied in \cite{FannesNachtergaeleWernerFCS,FannesNachtergaeleWernerPureStates} and the reader can see find the details of the aforementioned canonical injection in this context in \cite[Prop.~2.17]{KZ3}, the statement of which is reproduced in \zcref{sec:phase_def}.
\end{definition}

\begin{definition}\label{def:tilde_T}
    Given $\chi \in \bbN$, we define a function $\smallt:\cI(\chi) \to \mat{\chi}$ by setting $\targ{\MPSk}$ to be the unique positive invertible matrix such that
    \[
        \Tr \targ{\MPSk}  = 1 \qquad \mathrm{and} \qquad \sum_i \MPSk^{i*} \targ{\MPSk} \MPSk^i = \targ{\MPSk}.
    \]
    That $\smallt$ is well-defined follows from the noncommutative Frobenius-Perron theorem of Evans and H{\o}eghn-Krohn \cite{MR0482240} (see \cite[\S2.1]{KZ3} for further discussion).

    Given $d, D \in \bbN$, we define $\bigT:\cE(d, D) \to \mat{D}$ by
    \begin{equation}\label{eq:def-tilde-T}
        \Targ{\MPSA} = Y\begin{pmatrix} \targ{\MPSk} & 0 \\ 0 & 0 \end{pmatrix}Y^*
    \end{equation}
    where $\MPSA$ is expressed in terms of $Y$ and $\MPSk$ as in \eqref{eq:A_form_simplified}. In the proof of \cite[Prop.~2.17]{KZ3} it is shown that $\bigT$ is a well-defined (i.e., independent of the representation of $\MPSA$ as in \eqref{eq:A_form_simplified}) and continuous function. We observe that $\bigT$ commutes with the inclusions $\cE(d, D) \hookrightarrow \cE(d', D')$ and $\mat{D} \hookrightarrow \mat{D'}$, where $d \leq d'$ and $D \leq D'$:
    \[
        \begin{tikzcd}
            {\mathcal{E}(d,D)} & {\mathcal{E}(d',D')} \\
            {\mat{D}} & {\mat{D'}}
            \arrow[hook, from=1-1, to=1-2]
            \arrow["T"', from=1-1, to=2-1]
            \arrow["T", from=1-2, to=2-2]
            \arrow[hook, from=2-1, to=2-2]
        \end{tikzcd}\]
    Thus, it extends to a well-defined continuous function $\bigT:\cE \to \mat{\infty}$.
\end{definition}

\subsection{Regarding Symmetries}\label{sec:regarding_symmetries}

Note that the usual inner product on $\bbC^d$ can be extended to an inner product on $\bbC^\infty$:
\[
	\ev{a, b} \defeq \sum_i \overline{a_i}b_i, \quad a, b \in \bbC^\infty.
\]
In general, let ${\vecspace{V}}$ be an inner product space of countable (finite or infinite) algebraic dimension.
A linear map $U:{\vecspace{V}} \to {\vecspace{V}}$ is \emph{unitary} if it is bijective and $\ev{Ua, Ub} = \ev{a,b}$ for all $a,b \in {\vecspace{V}}$. Any inner product space of countable dimension admits an orthonormal Hamel basis $(e_i)$. In terms of the matrix elements $U_{ij} = \ev{e_i, Ue_j}$, one can check that a linear map $U:{\vecspace{V}} \to {\vecspace{V}}$ is unitary if and only if
\begin{enumerate}
\item for each $i$, only finitely many $U_{ij}$ are nonzero,
\item for each $i_1$ and $i_2$,
    \[
	\sum_{j} \overline{U_{ji_1}} U_{ji_2} = \sum_j U_{i_1j}\overline{U_{i_2j}} = \delta_{i_1i_2}.
    \]
\end{enumerate}
We write $\Unitary({\vecspace{V}})$ for the set of unitaries $U:{\vecspace{V}} \to {\vecspace{V}}$. This is not to be confused with $\Unitary(\infty)$ or $\Unitary(\sH)$ for a separable infinite-dimensional Hilbert space $\sH$. For example, an infinite direct sum of finite-size unitaries can be regarded as an element of $\Unitary(\bbC^\infty)$, but is not in general an element of $\Unitary(\infty)$, and a separable infinite-dimensional Hilbert space $\sH$ is not of countable algebraic dimension.

\begin{definition}\label{def:representation}
Let $G$ be a compact Hausdorff topological group and ${\vecspace{V}}$ an inner product space of countable dimension. By a \emph{representation} of $G$ on ${\vecspace{V}}$ we mean a homomorphism $U:G \to \Unitary({\vecspace{V}})$ such that for every $a \in {\vecspace{V}}$, the map $G \ni g \mapsto U(g)a \in {\vecspace{V}}$ is continuous when ${\vecspace{V}}$ is endowed with the final topology induced by the inclusions of its finite-dimensional subspaces.
Since ${\vecspace{V}}$ is a topological vector space with this topology, this is equivalent to assuming that each map $G \to {\vecspace{V}}$, $g \mapsto U(g)e_i$ is continuous, where $(e_i)$ is an orthonormal Hamel basis of ${\vecspace{V}}$. Since $G$ is compact, this is also equivalent to requiring that for each $i \in \bbN$, there exists a finite-dimensional subspace ${\vecspace{V}}_i \subset {\vecspace{V}}$ such that $\{U(g)e_i: g \in G\} \subset {\vecspace{V}}_i$ and $g \mapsto U(g)e_i$ is continuous as a map in ${\vecspace{V}}_i$.
\end{definition}

Let us fix a representation $U$ of $G$ on $\bbC^\infty$.
Given an infinite sequence of objects $\MPSA = (\MPSA^1, \MPSA^2, \ldots)$ in a complex vector space, we define $g \act \MPSA$ to be the infinite sequence whose $i$th element is given by
\begin{equation}\label{eq:action_def}
	(g \act \MPSA)^i = \sum_j U(g)_{ij}\MPSA^j.
\end{equation}
Since $U$ is a homomorphism, we have
\[
1 \act \MPSA = \MPSA \qqtext{and} gh \act \MPSA = (g \act (h \act \MPSA))
\]
for all $g,h \in G$. Moreover, \eqref{eq:action_def} defines a continuous group action on $\cI(\chi)$, as we show in the following proposition.

\begin{proposition}\label{prop:g_act_K}
If $\MPSk \in \cI(\chi)$, then:
\begin{enumerate}
\item $g \act \MPSk \in \cI(\chi)$ for every $g \in G$
\item the group action:
    \begin{equation*}
        \begin{aligned}
            G \times \cI(\chi) &\to \cI(\chi)\\
            (g, k) &\mapsto g \act k
        \end{aligned}
    \end{equation*}
    is continuous.
\end{enumerate}

\end{proposition}

\begin{proof}
Given $b \in \mat{\chi}$, injectivity of $\MPSk$ implies that we may write $b$ as a finite linear combination $b = \sum_i \lambda_i \MPSk^i$. Then
\[
	b = \sum_i \lambda_i (g^{-1} \act (g \act \MPSk))^i = \sum_{j} \left(\sum_i \lambda_i U(g^{-1})_{ij}\right) (g \act \MPSk)^j,
\]
which exhibits $B$ as a finite linear combination of the $(g \act \MPSk)^j$ matrices. Thus, $g \act \MPSk$ is injective. To check the right-normalization condition, we compute:
\[
	\sum_{i} (g \act \MPSk)^i (g \act \MPSk)^{i*} = \sum_{i,j_1,j_2} \overline{U(g)_{ij_2}}U(g)_{ij_1} \MPSk^{j_1}  \MPSk^{j_2*} = \sum_{j_1,j_2} \delta_{j_1j_2} \MPSk^{j_1} \MPSk^{j_2*} = \1_{\chi},
\]
where we have used unitarity of $U(g)$ in the second step and the right-normalization property of $\MPSk$ in the last step.

To check continuity of $G \times \cI(\chi) \to \cI(\chi)$, $(g,\MPSk) \mapsto g \act \MPSk$, we first note that compactness of $G$ implies that the product topology on $G \times \cI(\chi)$ coincides with the colimit topology induced by the inclusions $G \times \cI(d, \chi) \hookrightarrow G \times \cI(\chi)$. Thus, it suffices to check that the restriction of the action to $G \times \cI(d, \chi) \to \cI(\chi)$ is continuous for all $d \in \bbN$. The continuity assumption on $U$ implies that there exists $d' \in \bbN$ such that the action maps $G \times \cI(d, \chi)$ into $\cI(d', \chi)$. Continuity of the restricted map $G \times \cI(d, \chi) \to \cI(d', \chi)$ is manifest from the definition of the action \eqref{eq:action_def}, so we're done.
\end{proof}

As mentioned earlier, an injective right-normalized MPS tensor $\MPSk$ of physical dimension $d$ gives rise to a pure state on the quasi-local algebra $\fA(d)$. It is a well-known fact, called the Fundamental Theorem of Injective MPS, that two injective right-normalized MPS tensors $\MPSk$ and $\MPSl$ of physical dimension $d$ define the same state if and only if they have the same bond dimension $\chi$ and there exists $\lambda \in \Unitary(1)$ and $\smallv \in \Unitary(\chi)$ such that
\[
	\MPSk^i = \lambda \smallv^* \MPSl^i \smallv
\]
for all $i$ \cite{FannesNachtergaeleWernerPureStates,PGVWC_MPS}. Using the injectivity of $\MPSl$, it is easy to see that the $\lambda$ above is unique and the $\smallv$ is unique up to multiplication by a $\Unitary(1)$ phase. In other words, $\smallv$ is unique as an element of $\PU(\chi)$. We will use the ``blackboard bold'' font to distinguish elements of $\Unitary(\chi)$ from elements of $\PU(\chi)$. Thus, $\bbv \in \PU(\chi)$ denotes the image of $v \in \Unitary(\chi)$ under the projection $\Unitary(\chi) \to \PU(\chi)$. The above equation may then be written
\[
    \MPSk^i = \lambda \bbv^* \MPSl^i \bbv
\]
where $\lambda$ and $\bbv$ are unique.

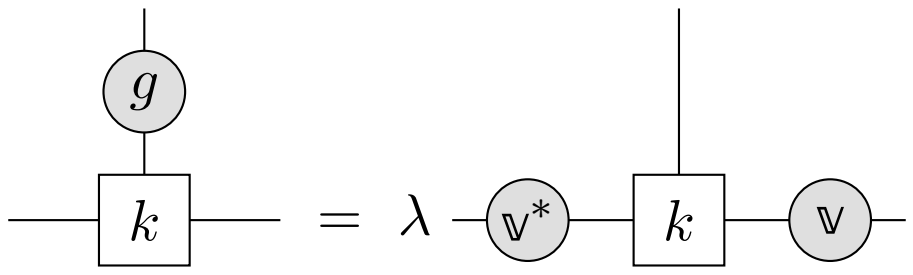
\begin{figure}[t]
    \begin{equation*}
        \begin{tikzpicture}[baseline=\mathaxis]
            \draw (-0.9,0) -- (0.9,0);
            \draw (0,0) -- (0,1.4);
            \filldraw[fill=white] (-0.3,-0.3) rectangle (0.3,0.3)
            node [pos=0.5] {$\MPSk$};
            \filldraw[fill=gray!25] (0,0.85) circle (0.27) node {$g$};
        \end{tikzpicture}
        \ = \ \lambda \
        \begin{tikzpicture}[baseline=\mathaxis]
            \draw (-1.5,0) -- (1.5,0);
            \draw (0,0) -- (0,1.4);
            \filldraw[fill=white] (-0.3,-0.3) rectangle (0.3,0.3)
            node[pos=0.5] {$\MPSk$};
            \filldraw[fill=gray!25] (-1,0) circle (0.27) node {$\bbv^*$};
            \filldraw[fill=gray!25] (1,0) circle (0.27) node {$\bbv$};
        \end{tikzpicture}
    \end{equation*}
    \caption{\label{fig:push_through}%
      The push-through property~\eqref{eq:g_act_K_invariance}: the on-site symmetry acting on the physical leg of $\MPSk$ equals, up to the phase $\lambda = \phasearg{g,\MPSk}$, the conjugation of the virtual legs by a projective unitary $\bbv=\bbv(g,\MPSk)$.
    }
\end{figure}

\begin{definition}\label{def:IU}
    We define $\cI_U(d, \chi)$ and $\cI_U(\chi)$ to be the subsets of $\cI(d, \chi)$ and $\cI(\chi)$, respectively, such that for every $g \in G$, there exists $\lambda \in \Unitary(1)$ and $\bbv \in \PU(\chi)$ such that
    \begin{equation}\label{eq:g_act_K_invariance}
        (g \act \MPSk)^i = \lambda \bbv^* \MPSk^i \bbv
    \end{equation}
for all $i$ (see \zcref{fig:push_through}).

Observe that if $\MPSA \in \cE$ with decomposition \eqref{eq:A_form_simplified}, then
\[
	(g \act \MPSA)^i = \mu Y\mqty{ (g \act \MPSk)^i & 0 \\ (g \act \MPSm)^i & 0 } Y^*
\]
Since $g \act \MPSk \in \cI(\chi)$ by \zcref{prop:g_act_K}, we see that $g \act \MPSA \in \cE$. Thus, $G$ acts on $\cE$, and continuity of this action is proved in exactly the same way as in the proof of \zcref{prop:g_act_K}. We let $\cE_U(d,D)$ and $\cE_U$ denote the subspaces of $\cE(d,D)$ and $\cE$, respectively, consisting of those $\MPSA \in \cE(d,D)$ or $\MPSA \in \cE$ such that $p(g \act \MPSA) = p(\MPSA)$ for every $g \in G$.  Likewise we use $\cB_U(d,D)$ and $\cB_U$ to denote the image of $\cE_U(d,D)$ and $\cE_U$ under $p$. Note that $\MPSA \in \cE_U$ if and only if $\MPSA$ can be written in the form \eqref{eq:A_form_simplified} with $\MPSk \in \cI_U(\chi)$ for some $\chi$.
\end{definition}

\begin{remark}\label{rem:topology_with_U}
    The sets $\cI_U(d, \chi)$ and $\cI_U(\chi)$ are closed in $\cI(d, \chi)$ and $\cI(\chi)$, respectively. To see this, first observe that these sets are the intersection across all $g \in G$ of the sets of $\MPSk$ such that \eqref{eq:g_act_K_invariance} holds for a single given $g$. Closedness of each of these sets follows from continuity of $\MPSk \mapsto g \act \MPSk$, from continuity of the action of $\Unitary(1) \times \PU(\chi)$ on the set of injective MPS tensors,  and from compactness of $\Unitary(1) \times \PU(\chi)$. It follows from \cite[Prop.~A.1]{KZ3} that the subspace topology on $\cI_U(\chi)$ coincides with the final topology induced by the inclusions $\cI_U(d, \chi) \hookrightarrow \cI_U(\chi)$. Furthermore, these inclusions are closed embeddings.

    Since the action of $G$ on $\cE$ is continuous, similar arguments apply to $\cE_U$. Namely, $\cE_U$ is closed in $\cE$ and therefore the subspace topology on $\cE_U$ coincides with the final topology induced by the inclusions $\cE_U \cap \cE(d, D) \hookrightarrow \cE_U$. Note that $\cE_U(d, D) = \cE_U \cap \cE(d, D)$. It is also worth pointing out that since $\cE_U(d, D)$ is closed in $\cE(d, D)$ and is saturated with respect to the projection $p:\cE(d, D) \to \cB(d, D)$, we know $\cB_U(d, D)$ is closed in $\cB(d, D)$ and the quotient topology on $\cB_U(d, D)$ obtained from $\cE_U(d, D)$ coincides with the subspace topology $\cB_U(d, D)$ obtains from $\cB(d, D)$. We will always equip $\cB_U(d, D)$ with this topology. Since $\cB(d, D) \to \cB(d', D')$ is an embedding whenever $d \leq d'$ and $D \leq D'$ \cite[Prop.~2.15]{KZ3}, we know the inclusion $\cB_U(d,D) \to \cB_U(d', D')$ is a closed embedding as well.
\end{remark}

\subsection{Preliminaries on Representations}

It will be useful to collect here some facts about representations on inner product spaces of countable dimension, as defined in \zcref{def:representation}. Let $G$ be a compact Hausdorff group.
If $\vecspace{V}$ and $\vecspace{V}'$ are inner product spaces of countable algebraic dimension that are equipped with representations $U$ and $U'$ of $G$, then a linear isometry $\iota:\cV \to \cV'$ such that $U'(g)\iota = \iota U(g)$ for every $g \in G$ will be called an \emph{isometric intertwiner}.

\begin{proposition}\label{prop:direct_sum_decomposition}
If $U$ is a representation of $G$ on an inner product space ${\vecspace{V}}$ of countably infinite dimension, then there exists a sequence $({\vecspace{V}}_n)_{n \in \bbN}$ of pairwise orthogonal irreducible invariant subspaces such that ${\vecspace{V}} = \vecspan \bigcup_n {\vecspace{V}}_n$.
\end{proposition}

\begin{proof}
Let $(e_n)_{n \in \bbN}$ be a basis of ${\vecspace{V}}$. Let ${\vecspace{V}}_1 = \vecspan\{U(g)e_1 : g \in G\}$, which is a finite-dimensional invariant subspace by the compactness of $G$ and the continuity $g \mapsto U(g)e_1$ with respect to the colimit topology on ${\vecspace{V}}$. Then let $e_2'$ be the orthogonal projection of $e_1$ onto the orthogonal complement of ${\vecspace{V}}_1$. Define ${\vecspace{V}}_2 = \vecspan\{U(g)e_2':g \in G\}$. Let $e_3'$ be the projection of $e_3$ onto the orthogonal complement of ${\vecspace{V}}_1 + {\vecspace{V}}_2$. Then define ${\vecspace{V}}_3 = \vecspan\{U(g)e_3':g \in G\}$. Proceeding in this fashion yields the desired subspaces ${\vecspace{V}}_n$ by recursion, except that they might not be irreducible. However, each ${\vecspace{V}}_n$ is a direct sum of its irreducible invariant subspaces, so the result follows.
\end{proof}

The next preliminary result applies to a special kind of representation $U$ on a countably infinite-dimensional inner product space. The terminology below comes from the homotopy theory literature \cite{LewisMaySteinberger}.

\begin{definition}
    We say a representation $U$ of $G$ on an inner product space ${\vecspace{V}}$ of countably infinite dimension is a \emph{$G$-universe} if for every finite-dimensional invariant subspace ${\vecspace{V}}_0 \subset {\vecspace{V}}$, there exists a sequence $({\vecspace{V}}_n)_{n \in \bbN}$ of pairwise orthogonal invariant subspaces such that the subrepresentations of $U$ on each ${\vecspace{V}}_n$ are all unitarily equivalent to the subrepresentation of $U$ on ${\vecspace{V}}_0$.\footnote{The definition in \cite[Ch.~1, \S2]{LewisMaySteinberger} requires a $G$-universe to have a trivial subrepresentation. We will not require this.} A \emph{complete $G$-universe} is a representation $U$ on ${\vecspace{V}}$ such that for every representation $U_0:G \to \Unitary(\sH)$ of $G$ on an arbitrary finite-dimensional Hilbert space $\sH$, there exists a sequence $({\vecspace{V}}_n)_{n \in \bbN}$ of pairwise orthogonal invariant subspaces such that the subrepresentations of $U$ on each ${\vecspace{V}}_n$ are all unitarily equivalent to $U_0$.
\end{definition}

We note that a complete $G$-universe exists on $\bbC^\infty$ if and only if $G$ is second-countable, because $G$ has countably many inequivalent irreducible representations if and only if $G$ is second-countable. If $\cU$ is a countable collection of representations of $G$, then we may build a $G$-universe $U$ on $\bbC^\infty$ by setting
\[
U(g) = \mqty{U_1(g) && \\ &U_2(g) & \\ && \ddots}
\]
where $U_1, U_2, \ldots$ is a sequence of irreducible representations of $G$ such that for each $U_0 \in \cU$, there exist countably infinitely many $U_n$ such that $U_0$ is unitarily equivalent to $U_n$.

\begin{proposition}\label{prop:G-universe_decomposition}
If $U$ is a $G$-universe on an inner product space ${\vecspace{V}}$ of countably infinite dimension and $({\vecspace{V}}_n)_{n \in \bbN}$ is a sequence of pairwise orthogonal irreducible invariant subspaces such that $X = \vecspan \bigcup_n {\vecspace{V}}_n$, then for every irreducible invariant subspace ${\vecspace{V}}_0 \subset {\vecspace{V}}$, infinitely many ${\vecspace{V}}_n$ are unitarily equivalent to ${\vecspace{V}}_0$.
\end{proposition}

\begin{proof}
By definition, there exists a sequence $(\tilde {{\vecspace{V}}}_m)_{m \in \bbN}$ of orthogonal invariant subspaces such that the subrepresentations of $U$ on $\tilde{{\vecspace{V}}}_m$ are unitarily equivalent to the subrepresentation of $U$ on ${{\vecspace{V}}}_0$. Given $m \in \bbN$, there exists $n_m \in \bbN$ such that we may express any $w \in \tilde {{\vecspace{V}}}_m$ uniquely as $w = v_1(w) + \cdots + v_{n_m}(w)$, where $v_i(w) \in {{\vecspace{V}}}_i$ for all $i$. Observe that
\[
U(g)w = U(g)v_1(w) + \cdots + U(g)v_n(w) = v_1(U(g)w) + \cdots + v_n(U(g)w).
\]
Thus, the functions $v_i$ are intertwiners $\tilde {\vecspace{V}}_m \to {\vecspace{V}}_i$. Since $\tilde {\vecspace{V}}_m$ and ${\vecspace{V}}_i$ are irreducible, it follows that $\tilde {\vecspace{V}}_m$ is unitarily equivalent to any ${\vecspace{V}}_i$ such that $v_i$ is not the zero operator.

Suppose there are only finitely many ${\vecspace{V}}_i$ unitarily equivalent to ${\vecspace{V}}_0$. Then there exists $n$ such that every $\tilde {\vecspace{V}}_m$ is contained in ${\vecspace{V}}_1 + \cdots + {\vecspace{V}}_n$. Since ${\vecspace{V}}_1 + \cdots + {\vecspace{V}}_n$ is finite-dimensional and there are infinitely many pairwise orthogonal $\tilde {\vecspace{V}}_m$, this is impossible. Thus, infinitely many ${\vecspace{V}}_n$ are unitarily equivalent to ${\vecspace{V}}_0$.
\end{proof}

Consider representations $U:G \to \Unitary({\vecspace{V}})$ and $U':G \to \Unitary({\vecspace{V}}')$ on inner product spaces ${\vecspace{V}}$ and ${\vecspace{V}}'$ of countable dimension and topologize ${\vecspace{V}}$ and ${\vecspace{V}}'$ as colimits of their finite-dimensional subspaces. Let $\cJ(U, U')$ denote the set of isometric intertwiners $\iota:{\vecspace{V}} \to {\vecspace{V}}'$. Equip $\cJ(U, U')$ with the subspace topology it obtains from the compact-open topology on $\cont{{\vecspace{V}}}{{\vecspace{V}}'}$. It is a well-known fact in homotopy theory that if $U'$ is a $G$-universe, then $\cJ(U, U')$ is either empty or contractible. Unfortunately, the only published proofs the authors are aware of, cited below, contain an error. Essentially, they claim that the normalization of the straight-line path from the isometry $\bbC^\infty \ni e_i \mapsto e_{2i} \in \bbC^\infty$ to the identity is a path of isometries, but it is not, as can be seen by evaluating this path of operators on $e_1 + e_2$. We will therefore provide a corrected proof, that gives a valid path of isometries from the above isometry to the identity. We will use the following proposition.

\begin{proposition}[{\cite[Prop.~3.1]{KZ3}}]\label{prop:isometry_construction}
Let $\phi:\bbN \to \bbN$ be a strictly increasing function. There exist continuous maps $\Gamma_{ab}:[0,1] \to \bbR$ defined for all $a, b \in \bbN$ with the following properties:
\begin{enumerate}
	\item $\Gamma_{ab}(0) = \delta_{ab}$,
	\item $\Gamma_{ab}(1) = \delta_{a\phi(b)}$,
	\item $\Gamma_{ab}(t) = 0$ for all $t \in [0,1]$ whenever $a > \phi(b)$,
	\item $\Gamma(t)^*\Gamma(t) = \1$ for all $t \in [0,1]$, where $\Gamma(t)$ is the $\bbN \times \bbN$ matrix whose entries are $\Gamma_{ab}(t)$.
\end{enumerate}
Here, $\delta_{ab}$ is the Kronecker delta function.
\end{proposition}

\begin{theorem}[{cf.~\cite[Ch.~2, Lem.~1.5]{LewisMaySteinberger} \& \cite[Ch.~1, Lem.~1.3]{May_RingSpacesRingSpectra}}]\label{thm:isometry_contractible}
    If $U'$ is a $G$-universe, then $\cJ(U, U')$ is either empty or contractible.
\end{theorem}

\begin{proof}
    Let us assume that $\cJ(U,U')$ is nonempty and show that it is contractible.

    By \zcref{prop:direct_sum_decomposition} there exists a sequence $({\vecspace{V}}_n')_{n \in \bbN}$ of pairwise orthogonal irreducible invariant subspaces such that ${\vecspace{V}}' = \vecspan \bigcup_n {\vecspace{V}}_n'$. For each unitary equivalence class $r$ of irreducible representation obtained from the subspaces $({\vecspace{V}}_n')_{n \in \bbN}$, \zcref{prop:G-universe_decomposition} states that there exist infinitely many ${\vecspace{V}}_n'$ defining a representation in the class $r$. Thus, we may relabel the subspaces in the sequence $({\vecspace{V}}_n')_n$ as $({\vecspace{V}}_{rk}')_{r,k}$, where $k \in \bbN$ and $r$ indexes the unitary equivalence classes among $({\vecspace{V}}_n')_{n \in \bbN}$, such that for every $r$ and $k$ the irrep on ${\vecspace{V}}_{rk}'$ is in the unitary equivalence class $r$.

    For each $r$ and $k$, let $T_{rk}:{\vecspace{V}}_{r1}' \to {\vecspace{V}}_{rk}'$ be a unitary intertwiner. Then for each $r$, let $(e_{ri})_i$ be an orthonormal basis for ${\vecspace{V}}_{r1}'$, so that $(T_{rk}e_{ri})_i$ is an orthonormal basis for ${\vecspace{V}}_{rk}'$ and $(T_{rk}e_{ri})_{r,k,i}$ is an orthonormal basis for ${\vecspace{V}}'$.

    Let $\Gamma_{ab}(t)$ be the real numbers obtained from \zcref{prop:isometry_construction} for the function $\phi(n) = 2n$. Then for each $t$ define $S(t):{\vecspace{V}}' \to {\vecspace{V}}'$ to be the unique linear map satisfying
    \[
    S(t)T_{rk}e_{ri} = \sum_m \Gamma_{mk}(t)T_{rm}e_{ri}.
    \]
    Observe that
    \begin{align*}
        \ev{S(t)T_{rk}e_{ri}, S(t)T_{r'k'}e_{r'i'}} &= \sum_{m,m'}\Gamma_{mk}(t) \Gamma_{m'k'}(t) \ev{T_{rm}e_{ri}, T_{r'm'}e_{r'i'}}\\
        &=  \sum_{m,m'} \Gamma_{mk}(t)\Gamma_{m'k'}(t) \delta_{rr'}\delta_{mm'}\delta_{ii'}\\
        &= \delta_{rr'}\delta_{kk'}\delta_{ii'}.
    \end{align*}
    This proves that $S(t)$ is an isometry. To see that it is an intertwiner, observe that invariance of ${\vecspace{V}}_{r1}'$ implies that we can write $U'(g)e_{ri} = \sum_j \lambda_{rij}(g)e_{rj}$ for some $\lambda_{rij}(g) \in \bbC$. Then the fact that $T_{rk}$ is an intertwiner implies
    \begin{align*}
        U'(g)S(t)T_{rk}e_{ri} &= \sum_m \Gamma_{mk}(t) T_{rm}U'(g)e_{ri}\\
        &= \sum_{m,j}\lambda_{rij}(g) \Gamma_{mk}(t) T_{rm}e_{rj}\\
        &= \sum_j \lambda_{rij}(g) S(t)T_{rk}e_{rj}\\
        &= S(t)T_{rk}U'(g)e_{ri}\\
        &= S(t)U'(g)T_{rk}e_{ri}.
    \end{align*}
    We see that $U'(g)S(t) = S(t)U'(g)$ since $(T_{rk}e_{ri})_{r,k,i}$ is a basis for ${\vecspace{V}}'$.

    The map $\cont{{\vecspace{V}}}{{\vecspace{V}}'} \times I \to \cont{{\vecspace{V}}}{{\vecspace{V}}'}$, $(R, t) \mapsto S(t)R$ is a well-defined function and it restricts to a well-defined function $\cJ(U, U') \times I \to \cJ(U, U')$. We must argue that it is continuous. Since ${\vecspace{V}}$ may be topologized as the sequential colimit of an increasing family of finite-dimensional subspaces ${\vecspace{V}}_1 \subset {\vecspace{V}}_2 \subset \cdots \subset {\vecspace{V}}$, the compact-open topology on $\cont{{\vecspace{V}}}{{\vecspace{V}}'}$ is the categorical limit of the restriction maps $\cont{{\vecspace{V}}}{{\vecspace{V}}'} \to \cont{{\vecspace{V}}_n}{{\vecspace{V}}'}$, $R \mapsto R|_{{\vecspace{V}}_n}$ across all $n$. Thus, it suffices to prove that the composition $\cont{{\vecspace{V}}}{{\vecspace{V}}'} \times I \to \cont{{\vecspace{V}}}{{\vecspace{V}}'} \to \cont{{\vecspace{V}}_n}{{\vecspace{V}}'}$ is continuous. We have a commutative diagram:
    \[
    \begin{tikzcd}
        \cont{{\vecspace{V}}}{{\vecspace{V}}'} \times I \rar \dar & \cont{{\vecspace{V}}}{{\vecspace{V}}'}\dar \\
        \cont{{\vecspace{V}}_n}{{\vecspace{V}}'} \times I \rar & \cont{{\vecspace{V}}_n}{{\vecspace{V}}'}
    \end{tikzcd}
    \]
    where the left vertical arrow is the restriction map on the first component and the identity on the second component. Thus, it suffices to show that the lower map $\cont{{\vecspace{V}}_n}{{\vecspace{V}}'} \times I \to \cont{{\vecspace{V}}_n}{{\vecspace{V}}'}$ is continuous. Since ${\vecspace{V}}_n$ is finite-dimensional, hence locally compact Hausdorff, this is equivalent to the associated map $\cont{{\vecspace{V}}_n}{{\vecspace{V}}'} \times {\vecspace{V}}_n \times I \to {\vecspace{V}}'$, $(R, v, t) \mapsto S(t)Rv$ being continuous. Again since ${\vecspace{V}}_n$ is locally compact Hausdorff, the evaluation map $\cont{{\vecspace{V}}_n}{{\vecspace{V}}'} \times {\vecspace{V}}_n \to {\vecspace{V}}'$ is continuous, so it suffices to show that the map ${\vecspace{V}}' \times I \to {\vecspace{V}}'$, $(w, t) \mapsto S(t)w$ is continuous. Since ${\vecspace{V}}'$ is the colimit of its finite-dimensional subspaces, it suffices to show that the restricted map ${\vecspace{V}}'_0 \times I \to {\vecspace{V}}'$, $(w, t) \mapsto S(t)w$ is continuous for every finite-dimensional subspace ${\vecspace{V}}_0' \subset {\vecspace{V}}'$. The image of this map is contained in a finite-dimensional subspace ${\vecspace{V}}_{00}'$ of ${\vecspace{V}}'$, so it suffices to check continuity of the map ${\vecspace{V}}_0' \times I \to {\vecspace{V}}_{00}'$, $(w, t) \mapsto S(t)w$. This follows from the fact that every matrix element of $S(t)$ with respect to the basis $(T_{rk}e_{ri})_{r,k,i}$ is continuous.

    Thus, the map $\cJ(U, U') \times I \to \cJ(U, U')$, $(R, t) \mapsto S(t)R$ is a homotopy of the identity on $\cJ(U, U')$ with the map $R \mapsto S(1)R$. Note that for every $R \in \cJ(U, U')$ the image of $S(1)R$ is contained in $\bigoplus_{r,k} {\vecspace{V}}'_{r,2k}$. We now define an isometry $\tilde S:{\vecspace{V}}' \to {\vecspace{V}}'$ as the unique linear map such that
    \[
    \tilde S T_{rk}e_{ri} = T_{r,2k-1}e_{ri}.
    \]
    It is easily checked that $\tilde S \in \cJ(U', U')$. Finally, we let $R_0 \in \cJ(U, U')$ be arbitrary and we define a homotopy $H:\cJ(U, U') \times I \to \cJ(U, U')$ by
    \[
    H(R, t) = \sin\left( \frac{\pi t}{2}\right)\tilde SR_0 + \cos\left( \frac{\pi t}{2} \right)S(1)R.
    \]
    That this is an isometry follows from the fact that $S(1)$ and $\tilde S$ have orthogonal images and it is an intertwiner since all maps involved are intertwiners. Continuity of $H$ follows from a similar argument to the previous paragraph. Thus, the map $R \mapsto S(1)R$ is nullhomotopic, so $\cJ(U, U')$ is contractible.
\end{proof}

\section{The Notion of Phase}\label{sec:phase_def}

Our goal in this section is to make precise an intuitive notion of a \emph{system} and \emph{phase} of $X$-parametrized $G$-symmetric translation-invariant injective matrix product states, where $X$ is a topological space and $G$ is a compact Hausdorff group. A system will be a continuous map from $X$ into a suitable topological space of states on a quasi-local algebra. Thus, our first task is to clearly articulate this space. We will then show how one can use $\cB_U$ to provide a convenient rephrasal of the notion of phase.

As detailed in \cite[\S 2.1]{KZ3}, an injective MPS tensor $\MPSk^1,\ldots, \MPSk^d$ defines a translation-invariant pure state $\omega_\MPSk$ on the quasi-local algebra $\fA(d)$ whose value on $C = C_1 \otimes \cdots \otimes C_n \in \fA(d)$ is
\begin{equation}\label{eq:state_def}
    \omega_\MPSk(C) = \underset{j_1,\ldots, j_n}{\sum_{i_1,\ldots, i_n}} C_{1,i_1j_1}\cdots C_{n,i_nj_n} \Tr[\MPSk^{i_n*}\cdots \MPSk^{i_1*} \targ{\MPSk} \MPSk^{j_1} \cdots \MPSk^{j_n}].
\end{equation}
One can shown that from a tensor $A \in \cE(d, D)$ with a representation as in \eqref{eq:A_form_simplified}, the state associated with $\MPSk$ can be extracted. It is worth reproducing the statement given in \cite{KZ3}. Below, $\pstate(d)$ is the set of pure states on $\fA(d)$.

\begin{proposition}[{\cite[Prop.~2.17]{KZ3}}]
    Given $d, D \in \bbN$, the map
    \[
    \cE(d, D) \to \pstate(d), \quad A = Y\mqty{\MPSk & 0 \\ \MPSm & 0} Y^* \mapsto \omega_A \defeq \omega_\MPSk
    \]
    is well-defined and continuous with respect to the weak* topology on $\pstate(d)$. There exists a unique function $\cB(d, D) \to \pstate(d)$ such that the diagram below commutes.
    \[
    \begin{tikzcd}
        \cE(d, D) \arrow[d,"p"] \arrow[r]& \pstate(d)\\
        \cB(d, D) \arrow[ur] &
    \end{tikzcd}
    \]
    Moreover, $\cB(d, D) \to \pstate(d)$ is a weak*-continuous injection.
\end{proposition}

Let $U$ be a representation of $G$ on $\bbC^\infty$. Note that an injective MPS tensor of physical dimension $d$ must have a bond dimension less than or equal to $\sqrt{d}$. Thus, for $D, D' \geq \sqrt{d}$, we have $\cB_U(d, D) = \cB_U(d, D')$ as sets and, by \zcref{rem:topology_with_U}, as topological spaces as well. We will write $\cB_U(d) \defeq \cB_U(d, D)$ for any $D \geq \sqrt{d}$. Then, let us write $\qstate_U(d)$ for the image of the map $\cB_U(d) \to \pstate(d)$. We equip $\qstate_U(d)$ with the unique topology making the map $\cB_U(d) \to \qstate_U(d)$ a homeomorphism. We will identify $\cB_U(d)$ with $\qstate_U(d)$ and leave the map $\cB_U(d) \to \qstate_U(d)$ implicit when convenient.

A general spin system will have a finite-dimensional on-site Hilbert space but need not have a canonical basis. Given a finite-dimensional Hilbert space $\sH$, let $\fA(\sH)$ be the corresponding quasi-local algebra
\[
\fA(\sH) \defeq \overline{\bigcup_{\Lambda \subset \bbZ} \bigotimes_{i \in \Lambda} \cB(\sH)}^{\norm{\cdot}}.
\]
We will also assume that our system comes with a continuous representation $W:G \to \Unitary(\sH)$. We need to identify and topologize an appropriate subset of the set $\pstate(\sH)$ of pure states of $\fA(\sH)$ that will correspond to the symmetric, translation-invariant, injective matrix product states with on-site Hilbert space $\sH$.  To do this, we embed $\sH$ into $\bbC^d$, which will induce a map between $\pstate(\sH)$ and $\pstate(d)$ as follows.

In general, if $\sH$ and $\sK$ are finite-dimensional Hilbert spaces and $\iota : \sH \to \sK$ is a linear isometry, then there exists a unique unital completely positive map $\alpha_\iota :\fA(\sK) \to \fA(\sH)$ such that
\begin{equation}\label{eq:isometry_automorphism}
\alpha_\iota(C_{-m} \otimes \cdots \otimes C_n) = \iota^*C_{-m}\iota \otimes \cdots \otimes \iota^* C_{n}\iota
\end{equation}
for all $C_{-m} \otimes \cdots \otimes C_n \in \fA(\sK)$ supported in the arbitrary interval $[-m,n]$ \cite[Lem.~2.3]{HomotopicalFoundations}. Furthermore, the assignment $\iota \mapsto \alpha_\iota$ is contravariant; that is, if $\iota_1 : \sH_1 \to \sH_2$ and $\iota_2 :\sH_2 \to \sH_3$ are linear isometries between finite-dimensional Hilbert spaces, then $\alpha_{\iota_2 \circ \iota_1} = \alpha_{\iota_1} \circ \alpha_{\iota_2}$ \cite[Prop.~2.4]{HomotopicalFoundations}. On pure states, precomposition with $\alpha_\iota$ defines an injective function $\iota_*:\pstate(\sH) \to \pstate(\sK)$, $\iota_*\omega = \omega \circ \alpha_\iota$, and the assignment $\iota \mapsto \iota_*$ is covariant; that is, $(\iota_2 \circ \iota_1)_* = (\iota_2)_* \circ (\iota_1)_*$ \cite[Prop.~2.9 \& Prop.~2.12]{HomotopicalFoundations}.

\begin{proposition}
    Let $\sH$ be a finite-dimensional Hilbert space with  representation $W:G \to \Unitary(\sH)$.
    Let $U$ be a representation of $G$ on $\bbC^\infty$ such that there exists an isometric intertwiner $\iota:\sH \to \bbC^\infty$ and let $d \in \bbN$ such that $\iota(\sH) \subset \bbC^d$. Regarding $\iota$ as an isometry $\sH \to \bbC^d$, the preimage
    \begin{equation}\label{eq:P_iMPS_def}
    \qstate_W(\sH) \defeq (\iota_*)^{-1}(\qstate_U(d))
    \end{equation}
    is independent of $d$, $\iota$, and $U$, as is the initial topology on $\qstate_W(\sH)$ obtained from the restricted map $\iota_*:\qstate_W(\sH) \to \qstate_U(d)$. With this topology, $\iota_*$ is a closed embedding, and a homeomorphism if $\iota(\sH) = \bbC^d$.
\end{proposition}

Of course, such a $U$, $d$, and $\iota$ always exist. For example, the representation $W$ is unitarily equivalent to a representation $\tilde W$ on $\bbC^{\dim \sH}$, and we could take $U$ to be the representation
\[
U(g) = \mqty{\tilde W(g) &  &  &  \\ &1&& \\ && 1& \\ &&&\ddots}.
\]
In this case, taking $d = \dim \sH$, we can construct an isometric intertwiner $\iota:\sH \to \bbC^\infty$ such that $\iota(\sH) = \bbC^d$.

\begin{proof}
    First, let $d' \geq d$ and let $\kappa:\bbC^d \to \bbC^{d'}$ be the inclusion.
    We claim that
    \begin{equation}\label{eq:easy_dimension_increase}
    (\iota_*)^{-1}(\qstate_U(d)) = (\kappa_* \circ \iota_*)^{-1}(\qstate_U(d')).
    \end{equation}
    Let $\omega$ be in the left hand side, so $\iota_*\omega = \omega_A$ for some $A \in \cE_U(d, D)$ for some $D$. It is shown in \cite[Prop.~2.19]{KZ3} that the diagram below commutes:
    \[
    \begin{tikzcd}
        \E(d, D) \arrow[d]\arrow[r, phantom, "\subset"] & \E(d', D)\arrow[d]\\
        \pstate(d) \arrow[r,"\kappa_*"]& \pstate(d')
    \end{tikzcd}
    \]
    Thus, $(\kappa \circ \iota)_*\omega = \kappa_*\omega_A = \omega_A$ as well. Thus, $\omega$ is in the right hand side of \eqref{eq:easy_dimension_increase}.

    Now suppose $\omega$ is in the right hand side of \eqref{eq:easy_dimension_increase}, so $(\kappa \circ \iota)_*\omega = \omega_A$ for some $A \in \cE_U(d', D)$ for some $D$. Hence $(\kappa \circ \iota)_*\omega = \omega_\MPSk$ for some $\MPSk \in \cI_U(d',\chi)$ where $\chi \leq D$. Then for any $C = C_1 \otimes \cdots \otimes C_n \in \fA(d)$ we note that $\kappa C_m \kappa^*$ is the $d' \times d'$ matrix with $C_m$ in the upper left block and zeros everywhere else, hence
    \begin{align*}
    \iota_*\omega(C) &= (\kappa \circ \iota)_*\omega(\kappa C_1 \kappa^* \otimes \cdots \otimes \kappa C_n \kappa^*)\\
    &= \underset{j_1,\ldots, j_n \leq d}{\sum_{i_1,\ldots, i_n \leq d}} C_{1,i_1j_1} \cdots C_{n,i_nj_n} \Tr[\MPSk^{i_n*} \cdots \MPSk^{i_1*}\targ{\MPSk} \MPSk^{j_1}\cdots \MPSk^{j_n}]
    \end{align*}
    This will prove that $\omega$ is in the left hand side of \eqref{eq:easy_dimension_increase} if we can show that $\MPSl = (\MPSk^1,\ldots, \MPSk^d)$ forms an element of $\cI_U(d, \chi)$ with $\targ{\MPSl} = \targ{\MPSk}$. We prove this by showing that $\MPSk^j = 0$ if $j > d$. Given $j > d$, let $|j \rangle$ be the $j$th standard basis vector of $\bbC^{d'}$ and let $|v\rangle \in \bbC^{d'}$ be arbitrary. Then $\alpha_\kappa(\ketbra{v}{j}) = 0$, hence
    \[
    0 = (\kappa \circ \iota)_*\omega(\ketbra{v}{j}) = \sum_{i} v_i \Tr[\MPSk^{i*} \targ{\MPSk}\MPSk^j]
    \]
    Since $\ket{v} \in \bbC^{d'}$ is arbitrary and $\MPSk$ is injective, this proves that $\MPSk^j = 0$.

    Let $\sT_\iota$ be the initial topology obtained from $\iota_*$ and let $\sT_{\kappa \circ \iota}$ be the initial topology obtained from $(\kappa \circ \iota)_*$. Consider the identity map $\id:(\qstate_W(\sH), \sT_\iota) \to (\qstate_W(\sH), \sT_{\kappa \circ \iota})$. We have the following commutative diagram:
    \[
    \begin{tikzcd}
        \qstate_W(\sH) \arrow[r,"\id"] \arrow[d,"\iota_*",->] & \qstate_W(\sH) \arrow[r,"(\kappa \circ \iota)_*",->] & \qstate_U(d') \arrow[d,<->,"\cong"] \\
        \qstate_U(d) \arrow[r,<->,"\cong"] & \cB_U(d) \arrow[r,"\subset"] & \cB_U(d')
    \end{tikzcd}
    \]
    The double-sided arrows in the diagram are homeomorphisms and the single-sided arrows are embeddings, so we can conclude that $\id$ is an embedding. Since $\id$ is bijective, it is a homeomorphism. Thus, $\sT_\iota = \sT_{\kappa \circ \iota}$.

    Now we show that \eqref{eq:P_iMPS_def} is independent of the choice of isometry in general. Let $U$ and $U'$ be representations of $G$ on $\bbC^\infty$, let $\iota:\sH \to \bbC^\infty$ and $\kappa:\sH \to \bbC^{\infty}$ to be any two isometric intertwiners for $U$ and $U'$, respectively, and let $d, d' \in \bbN$ such that $\iota(\sH) \subset \bbC^d$ and $\kappa(\sH) \subset \bbC^{d'}$. By the previous arguments we may assume $d = d'$. Let $\omega \in (\iota_*)^{-1}(\qstate_U(d))$ and let $\MPSk \in \cI_U(d, \chi)$ be an injective MPS such that $\iota_*\omega = \omega_\MPSk$. Then for any $C = C_1 \otimes \cdots \otimes C_n \in \fA(d)$ we have
    \begin{align*}
        \kappa_*\omega(C) &= \iota_*\omega(\iota\kappa^*C_1 \kappa \iota^* \otimes \cdots \otimes \iota \kappa^* C_n \kappa \iota^*)\\
        &= \underset{j_1,\ldots, j_n}{\sum_{i_1,\ldots, i_n}} C_{1,i_1j_1} \cdots C_{n,i_nj_n}\Tr[\MPSl^{i_n*} \cdots \MPSl^{i_1*}\targ{\MPSk} \MPSl^{j_1}\cdots \MPSl^{j_n}]
    \end{align*}
    where
    \[
    \MPSl^i = \sum_{j} (\kappa \iota^*)_{ij}\MPSk^j.
    \]
    We want to show that $\MPSl^i$ is a symmetric, injective, right-normalized, MPS tensor with $\targ{\MPSl} = \targ{\MPSk}$, for this will imply $\kappa_*\omega = \omega_\MPSl \in \qstate_{U'}(d)$.
    But first observe that for any $C \in \mat{d}$, we have
    \[
    \sum_{i,j} C_{ij}\Tr[\MPSk^{i*}\targ{\MPSk}\MPSk^j] = \iota_*\omega(C) = \iota_*\omega(C \iota\iota^*) = \sum_{i,j} C_{ij}\Tr\left[ \MPSk^{i*} \targ{\MPSk} \tilde \MPSk^j\right],
    \]
    where $\tilde \MPSk^i = \sum_j (\iota \iota^*)_{ij} \MPSk^j$. Since this holds for all $C \in \mat{d}$, we have
    \[
    \Tr[\MPSk^{i*}\targ{\MPSk}\MPSk^j] = \Tr[\MPSk^{i*}\targ{\MPSk}\tilde \MPSk^j]
    \]
    for all $i$ and $j$. Since $\MPSk$ is injective, this implies that $\MPSk^j = \tilde \MPSk^j$ for all $j$.

    Now we can show that $\MPSl$ is a symmetric, injective, right-normalized MPS tensor. Observe that
    \begin{align*}
    \sum_i \MPSl^i\MPSl^{i*} &=  \sum_{i,a,b} (\kappa \iota^*)_{ia} \overline{(\kappa \iota^*)_{ib}} \MPSk^a \MPSk^{b*}\\
    &= \sum_{i,a,b} (\iota \kappa^*)_{bi}(\kappa \iota^*)_{ia} \MPSk^a \MPSk^{b*}\\
    &= \sum_{a,b} (\iota\iota^*)_{ba}\MPSk^a \MPSk^{b*}\\
    &= \sum_b \MPSk^b \MPSk^{b*} \\
    &= \1.
    \end{align*}
    Next, if $h \in \mat{\chi}$, then there exists $\mu \in \bbC^d$ such that $\sum_i \mu_i \MPSk^i = h$, hence
    \begin{align*}
    \sum_i \left(\sum_a (\iota \kappa^*)_{ai}\mu_a\right) \MPSl^i &= \sum_{i,a,b} (\iota \kappa^*)_{ai} (\kappa \iota^*)_{ib} \mu_a \MPSk^b\\
    &= \sum_{a,b} (\iota \iota^*)_{ab} \mu_a \MPSk^b\\
    &= \sum_a \mu_a \MPSk^a\\
    &= h.
    \end{align*}
    This proves that $\MPSl$ is injective.

    We show that $\MPSl$ is respects the symmetry. Observe that
    \begin{align*}
        \sum_{j} U'(g)_{ij} \MPSl^j &= \sum_{j,a} U'(g)_{ij} (\kappa \iota^*)_{ja} \MPSk^a\\
        &= \sum_{a} (U'(g)\kappa \iota^*)_{ia} \MPSk^a \\
        &= \sum_{a} (\kappa W(g) \iota^*)_{ia} \MPSk^a\\
        &= \sum_{a} (\kappa \iota^* U(g))_{ia} \MPSk^a\\
        &= \sum_{j} (\kappa \iota^*)_{ij} (g \act \MPSk)^j\\
        &= \sum_j (\kappa \iota^*)_{ij} \phasearg{g,\MPSk}\bbvarg{g,\MPSk}^*\MPSk^j \bbvarg{g,\MPSk}\\
        &= \phasearg{g,\MPSk} \bbvarg{g,\MPSk}^* \MPSl^i \bbvarg{g,\MPSk}.
    \end{align*}
    Finally, observe that
    \begin{align*}
        \sum_i \MPSl^{i*} \targ{\MPSk} \MPSl^i &= \sum_{i,a,b} \overline{(\kappa \iota^*)_{ia}}(\kappa \iota^*)_{ib}  \MPSk^{a*} \targ{\MPSk}\MPSk^b\\
        &= \sum_{a,b}(\iota \iota^*)_{ab} \MPSk^{a*}\targ{\MPSk} \MPSk^b\\
        &= \sum_{a} \MPSk^{a*} \targ{\MPSk} \MPSk^a\\
        &= \targ{\MPSk}.
    \end{align*}
    We have now shown that $\MPSl$ is a right-normalized injective symmetric tensor satisfying $\targ{\MPSl} = \targ{\MPSk}$. Thus, $\kappa_*\omega \in \qstate_{U'}(d)$, as desired. This shows that \eqref{eq:P_iMPS_def} is independent of the choice of isometry.

    We show that the topology induced by $\iota_*$ coincides with the topology induced by $\kappa_*$. Let $\sT_\iota$ and $\sT_\kappa$ be the topologies induced by $\iota_*$ and $\kappa_*$, respectively. Let $\id:(\qstate_W(\sH),\sT_\iota) \to (\qstate_W(\sH),\sT_\kappa)$ be the identity. Then we have a commutative diagram:
    \begin{equation}\label{eq:id_continuous_phi}
    \begin{tikzcd}
        \qstate_W(\sH) \arrow[r,"\id"] \arrow[d,"\iota_*",->] & \qstate_W(\sH) \arrow[r,"\kappa_*",->] & \kappa_*(\qstate_W(\sH)) \arrow[d,<->,"\cong"] \\
        \iota_*(\qstate_W(\sH)) \arrow[r,<->,"\cong"] &  \cB_U^\iota(d) \arrow[r,"\phi"] & \cB_{U'}^\kappa(d)
    \end{tikzcd}
    \end{equation}
    where $\cB_U^\iota(d)$ is the image of $\iota_*(\qstate_W(\sH))$ in $\cB_U(d)$ and $\cB_{U'}^\kappa(d)$ is the image of $\kappa_*(\qstate_W(\sH))$ in $\cB_{U'}(d)$. As we saw from the proof above that obtained $\MPSl$ from $\MPSk$, the map $\phi$ is obtained from a commutative diagram
    \[
    \begin{tikzcd}
        p^{-1}(\cB_U^\iota(d)) \arrow[r,"\Phi"]\arrow[d,"p"] & p^{-1}(\cB_{U'}^\kappa(d)) \arrow[d,"p"]\\
        \cB_U^\iota(d) \arrow[r,"\phi"]& \cB_{U'}^\kappa(d)
    \end{tikzcd}
    \]
    where
    \[
    \Phi(A)^i = \sum_j (\kappa \iota^*)_{ij} A^j.
    \]
    From this it is clear that $\phi$ is continuous with respect to the quotient topologies on $\cB_U^\iota(d)$ and $\cB_{U'}^\kappa(d)$ obtained from $p^{-1}(\cB_U^\iota(d))$ and $p^{-1}(\cB_{U'}^\kappa(d))$. To show that these coincide with the subspace topologies obtained from $\cB_U(d)$ and $\cB_{U'}(d)$, it suffices to show that $\cB_U^\iota(d)$ and $\cB_{U'}^\kappa(d)$ are closed in $\cB_U(d)$ and $\cB_{U'}(d)$. We claim that $\cB_U^\iota(d)$ is the preimage under the  weak*-continuous map $\cB_U(d) \to \pstate(d)$  of the weak*-closed set
    \begin{equation}\label{eq:iotastar_image}
    \left\{ \psi \in \pstate(d) : \psi\left( \bigotimes_{i \in \Lambda} \iota \iota^* \right) = 1 \tn{ for all finite $\Lambda \subset \bbZ$}\right\}
    \end{equation}
    We note that the above set \eqref{eq:iotastar_image} is exactly $\iota_*(\pstate(\sH))$ by \cite[Prop.~2.12]{HomotopicalFoundations}. That $\cB_U^\iota$ is contained in the preimage of $\iota_*(\pstate(\sH))$ follows from the fact that
    \[
    \iota_*\omega\left( \bigotimes_{i \in \Lambda} \iota \iota^* \right) = \omega(\1) = 1
    \]
    for all $\omega \in \qstate(\sH)$. Having identified \eqref{eq:iotastar_image} as $\iota_*(\pstate(\sH))$, the reverse inclusion is easily verified from the definitions. Thus, $\cB_U^\iota(d)$ is closed in $\cB_U(d)$, and likewise $\cB_{U'}^\kappa(d)$ is closed in $\cB_{U'}(d)$.

    We conclude that $\phi$ is continuous. Since $\iota_*:\qstate_W(\sH) \to \iota_*(\qstate_W(\sH))$ and $\kappa_*:\qstate_W(\sH) \to \kappa_*(\qstate_W(\sH))$ are homeomorphisms, we see that $\id$ is continuous. This proves that $\sT_\kappa \subset \sT_\iota$, and the reverse inclusion follows since $\iota$ and $\kappa$ were arbitrary.

    The initial topology endowed on a space by a single injective function makes that function an embedding. Thus, $\iota_*:\qstate_W(\sH) \to \qstate_U(d)$ is an embedding. Furthermore, the fact that $\cB^\iota_U(d)$ is closed in $\cB_U(d)$ implies that $\iota_*(\qstate_W(\sH))$ is closed in $\qstate_U(d)$, so $\iota_*$ is a closed embedding. If $\iota(\sH) = \bbC^d$, then $\iota$ is a unitary map $\iota:\sH \to \bbC^d$ and $\iota_* : \pstate(\sH) \to \pstate(d)$ is a bijection. Thus, $\iota_*:\qstate_W(\sH) \to \qstate_U(d)$ is bijective by definition of $\qstate_W(\sH)$ and therefore it is a homeomorphism.
\end{proof}

\begin{proposition}\label{prop:iota_H1H2}
    Let $\sH_1$ and $\sH_2$ be finite-dimensional Hilbert spaces with unitary representations $W_1:G \to \Unitary(\sH_1)$ and $W_2:G \to \Unitary(\sH_2)$. If $\iota:\sH_1 \to \sH_2$ is an isometric intertwiner, then $\iota_*(\qstate_{W_1}(\sH_1)) \subset \qstate_{W_2}(\sH_2)$ and $\iota_*$ is a closed embedding $\qstate_{W_1}(\sH_1) \to \qstate_{W_2}(\sH_2)$.
\end{proposition}

\begin{proof}
    Let $U$ be a representation on $\bbC^\infty$, let $\kappa:\sH_2 \to \bbC^\infty$ be an isometric intertwiner, and let $d \in \bbN$ such that $\kappa(\sH_2) \subset \bbC^d$. Then $\kappa \circ \iota$ is an isometric intertwiner. Therefore, if $\omega \in \qstate_{W_1}(\sH_1)$, then $(\kappa \circ \iota)_*\omega = \kappa_*(\iota_*\omega) \in \qstate_{U}(d)$, which implies that $\iota_*\omega \in \qstate_{W_2}(\sH_2)$ by definition. By definition of the topology on $\qstate_{W_2}(\sH_2)$, we know that $\iota_*$ is continuous if and only if $\kappa_*\iota_* = (\kappa \circ \iota)_*$ is continuous, and it is by definition of the topology on $\qstate_{W_1}(\sH_1)$. Since $\kappa_*$ and $\kappa_*\iota_*$ are closed embeddings, we may conclude that $\iota_*$ is a closed embedding.
\end{proof}

Now that we have a nice topological space of states of interest for a given on-site Hilbert space $\sH$ and unitary representation $W$, we can define a notion of parametrized system and the corresponding notion of phase.

\begin{definition}\label{def:equivalence_relations}
    Let $X$ be a topological space and let $G$ be a compact Hausdorff topological group.
    An $X$-parametrized family of translation-invariant, injective, $G$-symmetric MPS consists of a finite-dimensional Hilbert space $\sH$ with a continuous representation $W:G \to \Unitary(\sH)$ and a continuous function $\psi:X \to \qstate_W(\sH)$. We will call this an \emph{parametrized family} for short. Below, we define an equivalence relation on the class of parametrized families.

    Let $\psi_1:X \to \qstate_{W_1}(\sH_1)$ and $\psi_2:X \to \qstate_{W_2}(\sH_2)$ be two parametrized families. We write $\psi_1 R \psi_2$ if any of the following hold:
    \begin{enumerate}
        \item\label{ite:finite_stage_homotopic} $\sH_1 = \sH_2$, $W_1 = W_2$, and $\psi_1$ and $\psi_2$ are homotopic maps,
        \item\label{ite:stacking} there exist a finite-dimensional Hilbert space $\sH_0$, a representation $W_0:G \to \Unitary(\sH_0)$, and $v \in \bbS \sH_0$ such that $W_0(g)v = v$ for all $g \in G$ and such that $\sH_2 = \sH_1 \otimes \sH_0$, $W_2 = W_1 \otimes W_0$, and $\psi_{2} = \iota_*\psi_{1}$, where $\iota:\sH_1 \mapsto \sH_2$ is the isometric intertwiner $\iota(u) = u \otimes v$,
        \item\label{ite:isomorphism} there exists a unitary intertwiner $V:\sH_1 \to \sH_2$  such that $V_*\psi_{1} = \psi_{2}$.
    \end{enumerate}
    The operations above are colloquially known as \emph{deformation}, \emph{stacking}, and \emph{isomorphism}, respectively. We let ${\sim}$ denote the equivalence relation generated by $R$. That is, $\psi_1 \sim \psi_2$ if $\psi_1$ can be related to $\psi_2$ by a finite sequence of deformations, stackings, and isomorphisms. The \emph{phase} of a parametrized family is its equivalence class with respect to ${\sim}$.
\end{definition}

\begin{lemma}\label{lem:direct_sum}
    Let $\psi_1:X \to \qstate_{W_1}(\sH_1)$ and $\psi_2:X \to \qstate_{W_2}(\sH_2)$ be two parametrized families. If there exists a Hilbert space $\sH_0$ and a representation $W_0$ on $\sH_0$ such that $\sH_2 = \sH_1 \oplus \sH_0$, $W_2 = W_1 \oplus W_0$, and $\psi_{2} = \iota_*\psi_{1}$, where $\iota$ is the isometric intertwiner $\iota:\sH_1 \to \sH_2$ defined by $\iota(u) = (u,0)$, then $\psi_1 \sim \psi_2$.
\end{lemma}

\begin{proof}
    Consider Hilbert spaces $\sK_1 \defeq \sH_1 \otimes ((\sH_1 \otimes \sH_1^*) \oplus (\sH_0 \otimes \sH_1^*))$ and $\sK_2 \defeq \sH_2 \otimes (\sH_1 \otimes \sH_1^*)$, where $\sH_1^*$ is regarded as a Hilbert space with inner product
    \[
    \langle f_1, f_2 \rangle = \langle f_2^\sharp, f_1^\sharp \rangle
    \]
    and ${}^\sharp : \sH^*_1 \to \sH_1$ is the inverse of the antilinear isomorphism ${}^\flat : \sH_1 \to \sH_1^*$ from the Riesz representation theorem. Define $\overline{W_1}(g) \in \Unitary(\sH_1^*)$ by $\overline{W_1}(g)f = f \circ W_1(g)$. Equip $\sK_1$ with the representation $T_1 \defeq W_1 \otimes ((W_1 \otimes \overline{W_1}) \oplus (W_0 \otimes \overline{W_1}))$ and equip $\sK_2$ with the representation $T_2 \defeq W_2 \otimes (W_1 \otimes \overline{W_1})$. Define
    \[
    v = \frac{1}{\sqrt{\dim \sH_1}}\sum_i e_i \otimes e_i^\flat
    \]
    where $(e_i)_i$ is an orthonormal basis for $\sH_1$ and observe that $v$ is invariant under $W_1 \otimes \overline{W_1}$.

    By item (\ref{ite:stacking}) of \zcref{def:equivalence_relations} we know $\psi_1 \sim \omega_1$ where $\omega_1 : X \to \qstate_{T_1}(\sK_1)$ is the parametrized family obtained from $\psi_1$ and the embedding $\sH_1 \to \sK_1$, $u \mapsto u \otimes (v, 0)$. Likewise, $\psi_2 \sim \omega_2$ where $\omega_2:X \to \qstate_{T_2}(\sK_2)$ is the parametrized family obtained from $\psi_2$ and the embedding $\sH_2 \to \sK_2$, $u \mapsto u \otimes v$.

    Finally, we have a natural way of identifying
    \[
    \sH_1 \otimes ((\sH_1 \otimes \sH_1^*) \oplus (\sH_0 \otimes \sH_1^*)) \cong (\sH_1 \oplus \sH_0) \otimes (\sH_1 \otimes \sH_1^*)
    \]
    and this is a unitary intertwiner $V:\sK_1 \to \sK_2$, so $\omega_1 \sim V_*\omega_1$. We have a commutative diagram:
    \[
    \begin{tikzcd}
        \sH_1 \arrow[d, "u \mapsto {(u,0)}"' ]\arrow[rr, "u \mapsto {u \otimes (v, 0)}"] && \sK_1 \arrow[d,"V"]\\
        \sH_2 \arrow[rr,"u \mapsto u \otimes v"]&& \sK_2
    \end{tikzcd}
    \]
    Thus, $V_*\omega_1 = \omega_2$, so $\psi_1 \sim \omega_1 \sim \omega_2 \sim \psi_2$, as desired.
\end{proof}

\begin{lemma}\label{lem:isometric_intertwiner_same_phase}
    Let $\psi_1:X \to \qstate_{W_1}(\sH_1)$ and $\psi_2:X \to \qstate_{W_2}(\sH_2)$ be two parametrized families. If there exists an isometric intertwiner $\iota:\sH_1 \to \sH_2$ such that $\iota_*\psi_1 = \psi_2$, then $\psi_1 \sim \psi_2$.
\end{lemma}

\begin{proof}
    The image $\iota(\sH_1)$ is invariant under $W_2$ and therefore so is the orthogonal complement $\iota(\sH_1)^\perp$. If $W_2^\iota$ and $W_2^{\perp}$ are the subrepresentations of $W_2$ on $\iota(\sH_1)$ and $\iota(\sH_1)^\perp$, then we have a natural unitary intertwiner $V:\sH_2 \to \iota(\sH_1) \oplus \iota(\sH_1)^\perp$, where the codomain is equipped with $W_2^\iota \oplus W_2^{\perp}$. Thus, $\psi_2 \sim V_*\psi_2$ by item (\ref{ite:isomorphism}) of \zcref{def:equivalence_relations}. With $W:\sH_1 \to \iota(\sH_1)$ defined as $Wu = \iota(u)$, we see that $W$ is a unitary intertwiner and therefore $\psi_1 \sim W_*\psi_1$ by item (\ref{ite:isomorphism}) of \zcref{def:equivalence_relations}. Finally, with $\kappa:\iota(\sH_1) \to \iota(\sH_1) \oplus \iota(\sH_1)^\perp$ defined as $\kappa(u) = (u, 0)$, we know that $\kappa_*(W_*\psi_1) \sim W_*\psi_1$ by \zcref{lem:direct_sum}. Since $V\iota = \kappa W$, we see that $\psi_1 \sim V_*\iota_*\psi_1 = V_*\psi_2 \sim \psi_2$, as desired.
\end{proof}

To relate the notion of phase back to $\cB_U$, let us first remark how to embed $\qstate_W(\sH)$ into $\cB_U$. If $U:G \to \Unitary(\bbC^\infty)$ is a representation as in \zcref{def:representation} and $\iota:\sH \to \bbC^\infty$ is an isometric intertwiner such that $\iota(\sH) \subset \bbC^d$, then we may compose the embedding $\iota_*:\qstate_W(\sH) \to \qstate_U(d)$ with the homeomorphism $\qstate_U(d) \to \cB_U(d)$ and the inclusion $\cB_U(d) \hookrightarrow \cB_U$. The resulting map $\qstate_W(\sH) \to \cB_U$, which by slight abuse of notation we will also call $\iota_*$, is independent of the choice of $d$ such that $\iota(\sH) \subset \bbC^d$. Thus, given an isometric intertwiner $\iota:\sH \to \bbC^\infty$, we get an embedding $\iota_*:\qstate_W(\sH) \to \cB_U$. Given a parametrized family $\psi:X \to \qstate_W(\sH)$, we may compose with $\iota_*$ to get a map $\iota_*\psi:X \to \cB_U$.

We would like to show that if $U$ is a $G$-universe, then the homotopy class of $\iota_*\psi$ is independent of the choice of embedding $\iota:\sH \to \bbC^\infty$. For this, we require the following proposition. Recall that $\cJ(W, U)$ is the set of isometric intertwiners $\iota:\sH \to \bbC^\infty$, where $\sH$ is equipped with the representation $W$ and $\bbC^\infty$ is equipped with the representation $U$, and that $\cJ(W, U)$ is equipped with the subspace topology obtained from the compact-open topology on $\cont{\sH}{\bbC^\infty}$.

\begin{proposition}\label{prop:continuous_isometry_action}
    Let $\sH$ be a finite-dimensional Hilbert space with unitary representation $W:G \to \Unitary(\sH)$ and let $U:G \to \Unitary(\bbC^\infty)$ be a representation on $\bbC^\infty$. If $Z$ is a compact Hausdorff space and $Z \to \cJ(W, U)$, $z \mapsto \iota(z)$ is continuous, then the map $\qstate_W(\sH) \times Z \to \cB_U$, $(\psi, z) \mapsto \iota(z)_*\psi$ is continuous.
\end{proposition}

\begin{proof}
    Since $Z$ is compact, there exists $d'$ such that $\iota(z)(\sH) \subset \bbC^{d'}$ for all $z \in Z$. Thus, the image of the map $\qstate_W(\sH) \times Z \to \cB_U$ is contained in $\cB_U(d')$ and it suffices to show continuity of the map $\qstate_{W}(\sH) \times Z \to \cB_U(d')$ with its codomain restricted.

    Let $d = \dim \sH$ and let $V$ be a representation on $\bbC^\infty$ that admits an isometric intertwiner $\kappa:\sH \to \bbC^\infty$ such that $\kappa(\sH) = \bbC^d$. Then $\kappa_*:\qstate_W(\sH) \to \qstate_{V}(d)$ is a homeomorphism and we also have the homeomorphism $\qstate_{V}(d) \cong \cB_{V}(d)$. Thus it suffices to show continuity of the map $\cB_V(d) \times Z \to \cB_U(d')$ obtained by composing the map $\qstate_W(\sH) \times Z \to \cB_U(d')$ with these homeomorphisms in the first component.

     Choose $D$ large enough so that $\cB_V(d) = \cB_{V}(d, D)$ and $\cB_U(d') = \cB_U(d', D)$. Again since $Z$ is compact and $\cB_V(d, D)$ is a quotient of $\cE_V(d, D)$, it suffices to prove continuity of the map $\cE_V(d, D) \times Z \to \cB_U(d', D)$ obtained by first projecting $\cE_V(d, D)$ onto $\cB_V(d, D)$ and then following the map $\cB_V(d, D) \times Z \to \cB_U(d', D)$. We then have a commutative diagram
     \[
     \begin{tikzcd}
         \cE_V(d, D) \times Y \arrow[r,"\Phi"] \arrow[d] & \cE_U(d', D) \arrow[d] \\
         \cB_V(d, D) \times Y \arrow[r] & \cB_U(d', D)
     \end{tikzcd}
     \]
     where the map $\Phi:\cE_V(d, D) \times Y \to \cE_U(d', D)$ is given by
     \[
     \Phi(A, z)^i = \sum_j (\iota(z)\kappa^{-1})_{ij}A^j
     \]
     and the notation $\kappa^{-1}$ is valid since $\kappa$ is being regarded as a unitary $\sH \to \bbC^d$. This function $\cE_V(d, D) \times Y \to \cE_U(d', D)$ is manifestly continuous since all matrix elements of $\iota(z)$ are continuous in $z$, so this concludes the proof.
\end{proof}

\begin{corollary}\label{cor:homotopic_embeddings_in_BU}
    Let $\psi:X \to \qstate_W(\sH)$ be a parametrized family. If $U:G \to \Unitary(\bbC^\infty)$ is a $G$-universe and $\iota_1, \iota_2 \in \cJ(W, U)$, then $(\iota_1)_*\psi$ is homotopic to $(\iota_2)_*\psi$.
\end{corollary}

\begin{proof}
    This is immediate from \zcref{thm:isometry_contractible} and \zcref{prop:continuous_isometry_action}.
\end{proof}

\begin{theorem}\label{thm:phase_equivalences}
    Let $\psi_1:X \to \qstate_{W_1}(\sH_1)$ and $\psi_2:X \to \qstate_{W_2}(\sH_2)$ be two parametrized families. If $\psi_1 \sim \psi_2$, then there exists a representation $U:G \to \Unitary(\bbC^\infty)$ and isometric intertwiners $\iota_1:\sH_1 \to \bbC^{\infty}$ and $\iota_2:\sH_2 \to \bbC^{\infty}$ such that $(\iota_1)_*\psi_1$ and $(\iota_2)_*\psi_2$ are homotopic as functions $X \to \cB_U$. The converse holds if $X$ is compact. If $\psi_1 \sim \psi_2$, then the representation $U$ may be chosen to be a $G$-universe. If $\psi_1 \sim \psi_2$ and $U$ is a complete $G$-universe, then for any isometric intertwiners $\iota_i:\sH_i \to \bbC^\infty$, the maps $(\iota_1)_*\psi_1$ and $(\iota_2)_*\psi_2$ are homotopic as function $X \to \cB_U$.
\end{theorem}

\begin{proof}
    Suppose $\psi_1 \sim \psi_2$. Then there exists a sequence of parametrized families $\omega_i:X \to \qstate_{V_i}(\sK_i)$, $i \in \{1,\ldots, n\}$ such that $\omega_1 = \psi_1$, $\omega_n = \psi_2$, and for all $i$, either $\omega_i R \omega_{i+1}$ or $\omega_{i+1}R \omega_i$. There are finitely many equivalence classes of irreducible subrepresentations among the $V_i$, so we may form a $G$-universe $U:G \to \Unitary(\bbC^\infty)$ such that each $V_i$ is unitarily equivalent to a subrepresentation of $U$. If $G$ is second-countable, then $U$ may be chosen to be any complete $G$-universe.

    Given parametrized families $\phi_1:X \to \qstate_{T_1}(\sL_1)$ and $\phi_2:X \to \qstate_{T_2}(\sL_2)$ such that $\cJ(T_i, U) \neq \varnothing$, let us write $\phi_1 \approx \phi_2$ if there exist isometric intertwiners $\kappa_i \in \cJ(T_i, U)$ such that $(\kappa_1)_*\phi_1$ and $(\kappa_2)_*\phi_2$ are homotopic as maps into $\cB_U$. Then $\approx$ is an equivalence relation on the class of parametrized families $\phi:X \to \qstate_T(\sL)$ with $\cJ(T, U) \neq \varnothing$, with transitivity of $\approx$ following from \zcref{cor:homotopic_embeddings_in_BU}.

    Let $\phi_1$ and $\phi_2$ be parametrized families as in the previous paragraph and let $\kappa_2 \in \cJ(T_2, U)$.
    We claim that $\phi_1 R \phi_2$ implies $\phi_1 \approx \phi_2$.
    Suppose item (\ref{ite:finite_stage_homotopic}) of \zcref{def:equivalence_relations} holds (with $\phi$, $T$, and $\sL$ replacing $\psi$, $W$, and $\sH$), so $\sL_1 = \sL_2 =\vcentcolon \sL$. Then $(\kappa_2)_*(\phi_1)$ is homotopic to $(\kappa_2)_*(\phi_2)$ since $\phi_1$ and $\phi_2$ are homotopic, so $\phi_1 \approx \phi_2$.

    Suppose item (\ref{ite:stacking}) of \zcref{def:equivalence_relations} holds. Let $\kappa_0:\sL_1 \to \sL_2$ be the embedding $\kappa_0(u) = u \otimes v$. Then $\kappa_1 \defeq \kappa_2 \circ \kappa_0$ is an isometric intertwiner and $(\kappa_1)_*(\phi_1) = (\kappa_2)_*(\phi_2)$, so $\phi_1 \approx \phi_2$.

    Suppose item (\ref{ite:isomorphism}) of \zcref{def:equivalence_relations} holds. Then we know $\kappa_1=\kappa_2 \circ V$ is an isometric intertwiner, and $(\kappa_1)_*(\phi_1) = (\kappa_2)_*(\phi_2)$, so $\phi_1 \approx \phi_2$.

    This proves that $\phi_1 R \phi_2$ implies $\phi_1 \approx \phi_2$. Since $\approx$ is an equivalence relation, we also know $\phi_2 R \phi_1$ implies $\phi_1 \approx \phi_2$. In particular, we see that $\omega_i \approx \omega_{i+1}$ for all $i$, so $\psi_1 \approx \psi_2$ since $\approx$ is transitive. This proves one direction of the theorem.

    For the other direction, we first make a general observation. Suppose $U:G \to \Unitary(\bbC^\infty)$ is a representation on $\bbC^\infty$. Then for any finite-dimensional invariant subspace $\sK \subset \bbC^\infty$, the inclusion $\kappa:\sK \to \bbC^\infty$ induces an embedding $\kappa_*:\qstate_U(\sK) \to \cB_U$. Suppose $\bbC^d \subset \sK \subset \bbC^{d'}$. We claim that $\cB_U(d) \subset \kappa_*(\qstate_U(\sK))$. Suppose $\omega \in \cB_U(d)$. Then $\omega \in \cB_U(d')$ since $\cB_U(d) \subset \cB_U(d')$. Furthermore, $\omega = \omega_\MPSk$ for some symmetric translation-invariant injective MPS tensor $\MPSk$ such that $\MPSk^i = 0$ for $i > d$. Regarding $\kappa$ as the inclusion $\sK \to \bbC^{d'}$, we may embed the projection $\kappa \kappa^*$ in any tensor component of $\fA(d')$, whence
    \begin{align*}
        \omega_\MPSk(\kappa \kappa^*) &= \sum_{i,j = 1}^{d'} (\kappa \kappa^*)_{ij} \Tr[\MPSk^{i*}\targ{\MPSk}\MPSk^j] = \sum_{i,j=1}^d (\kappa \kappa^*)_{ij} \Tr[\MPSk^{i*}\targ{\MPSk}\MPSk^j].
    \end{align*}
    Since $\kappa \kappa^*$ is the projection onto $\sK$ and $\bbC^d \subset \sK$, we know $(\kappa \kappa^*)_{ij} = \delta_{ij}$ for $i, j \leq d$. Thus, $\omega_\MPSk(\kappa \kappa^*) = 1$. This proves that $\omega_\MPSk \in \kappa_*(\qstate_U(\sK))$. We have now proven that $\cB_U(d) \subset \kappa_*(\qstate_U(\sK)) \subset \cB_U(d')$. Thus, $\cB_U$ has the final topology induced by the inclusions $\kappa_*(\qstate_U(\sK)) \to \cB_U$ across the finite-dimensional invariant subspaces $\sK \subset \bbC^\infty$.

    Now suppose there exists a representation $U:G \to \Unitary(\bbC^\infty)$ and isometric intertwiners $\iota_i \in \cJ(\sH_i, \bbC^\infty)$ such that $(\iota_1)_*\psi_1$ and $(\iota_2)_*\psi_2$ are homotopic as maps $X \to \cB_U$.
    Since $X$ is compact, there exists a finite-dimensional invariant subspace $\sK \subset \bbC^\infty$ such that $\iota_1(\sH_1) \cup \iota_2(\sH_2) \subset \sK$ and this homotopy has image in $\kappa_*(\qstate_U(\sK))$, where $\kappa:\sK \to \bbC^\infty$ is the inclusion. If $\tilde \iota_i:\sH_i \to \sK$ is $\iota_i$ with the codomain restricted, then the fact that $\kappa_*$ is an embedding implies that $(\tilde \iota_1)_*\psi_1$ is homotopic to $(\tilde \iota_2)_*\psi_2$ as parametrized families $X \to \qstate_U(\sK)$.  Thus, $\psi_1 \sim (\tilde \iota_1)_*\psi_1 \sim (\tilde \iota_2)_*\psi_2 \sim \psi_2$ by \zcref{lem:isometric_intertwiner_same_phase} and item (\ref{ite:finite_stage_homotopic}) of \zcref{def:equivalence_relations}.
\end{proof}

%!TEX root = MPS_Symmetry-master.tex

\section{Construction of the Invariant}\label{sec:general_construction}

\zcref{thm:phase_equivalences} states that if two parametrized families $\psi_1:X \to \qstate_{W_1}(\sH_1)$ and $\psi_2:X \to \qstate_{W_2}(\sH_2)$ are in the same phase, then for any complete $G$-universe $U$ and isometric embeddings $\iota_i:\sH_i \to \bbC^\infty$ for $i \in \{1,2\}$, the maps $(\iota_1)_*\psi_1$ and $(\iota_2)_*\psi_2$ are homotopic as maps $X \to \cB_U$. Thus, to construct a phase invariant for parametrized families, we may construct a cohomology class on $\cB_U$ and map it to a cohomology class on $X$ under the homomorphism induced by $(\iota_1)_*\psi_1$. In this section, we construct one such interesting class in the \v{C}ech cohomology group $\check{H}^1(\cB_U; \widehat{G})$, which then maps to a class in $\check{H}^1(X;\widehat{G})$. Here, $\widehat{G}$ is the Pontryagin dual of $G$, namely the group of continuous group homomorphisms \( G \to \Unitary(1) \).
Since $G$ is compact Hausdorff, the compact-open topology on $\widehat{G}$ coincides with the discrete topology. Thus, the sheaf of continuous functions into $\widehat{G}$ coincides with the sheaf of locally constant functions in $\widehat{G}$, and this is the sheaf we use to define the \v{C}ech cohomology groups.

An oversimplified idea of the construction of the $\check{H}^1(X;\widehat{G})$ invariant can be given as follows. With a parametrized family of tensors $\MPSk(x) \in \cI_U(\chi)$, $x \in X$, one can act with the symmetry for arbitrary $g$ and $x$ to obtain projective unitary matrices $\bbvarg{g, x}$ defined as
\[
g \act \MPSk(x) = \lambda(g, x)\bbvarg{g,x}^* \MPSk(x) \bbvarg{g,x}.
\]
One may try to find an open cover $(O_\alpha)_{\alpha}$ of $X$ on which the matrices $\bbvarg{g,x}$ have continuous lifts $\varg[\alpha]{g,x}$ for $g \in G$ and $x \in O_\alpha$. On an intersection of patches $O_\alpha \cap O_\beta$, the lifts may differ by a phase:
\[
\phi_{\alpha \beta}(g,x)\1 = \varg[\alpha]{g,x}^*\varg[\beta]{g,x},
\]
and these phases form a \v{C}ech 1-cocycle.

This is an oversimplification first and foremost because we are not given a parametrized family of tensors, rather we are given a continuous map into $\sQ_W(\sH)$ or $\cB_U$, from which we cannot in general extract a continuous family of tensors $\MPSk(x)$. There are other issues to be dealt with as well: a family of projective unitary representations on $G \times X$ cannot in general be lifted continuously on $G \times O$ for any open subset $O \subset X$, and $\phi_{\alpha \beta}(g,x)$ will not be a homomorphism in $g$ unless $v_\alpha(g,x)$ and $v_\beta(g,x)$ define the same 2-cocycle at every $x \in O_\alpha \cap O_\beta$.

Thus, we must show that with the given minimalistic hypotheses one may indeed understand something like $\phi_{\alpha \beta}$ as a well-defined \v{C}ech 1-cocycle. In \zcref{sec:construction_of_lifts} we find a well-defined and continuous replacement for $\bbv$ and discuss the way we may lift it. Then in \zcref{sec:invariant_def} we pull the results of \zcref{sec:construction_of_lifts} together and carefully define the \v{C}ech  invariant.

While it is helpful in the overarching classification of phases for $U$ to be a  $G$-universe or complete $G$-universe, the constructions in this section do not require this. Thus, we continue to let $G$ be a compact Hausdorff group and we let $U$ be an arbitrary representation of $G$ on $\bbC^\infty$.

\subsection{Projective Matrices and their Lifts}\label{sec:construction_of_lifts}
Projective matrix spaces like $\PU(\chi)$ and $\PM{D}$ are by definition given quotient topologies. This makes it easy to check whether a map out of a projective space is continuous.
It is useful for us to recharacterize these topologies as initial topologies so that we can easily check whether a map into such a space is continuous. Thus, we begin with the following lemma.

\begin{lemma}\label{lem:initial_top}
    Let $D \in \bbN$ and endow $\mat{D}$ with any norm. Let $\SM{D}$ be the unit sphere of $\mat{D}$ and let $\pi:\SM{D} \to \PM{D}$ be the projection. For each $B \in \mat{D}$, define
    \[
    \hat B : \PM{D} \to \mat{D}, \quad \hat B(\bbA) = A^* B A
    \]
    where $A$ is any matrix in $\SM{D}$ such that $\pi(A) = \bbA$. Then the quotient topology on $\PM{D}$ coincides with the initial topology induced by the maps $\hat B$ across any set of matrices $B$ that spans $\mat{D}$.
\end{lemma}

\begin{proof}
    The initial topology induced by $\hat B$ across all $B \in \mat{D}$ is equal to that induced by any spanning subset of $\mat{D}$ since $\hat B$ is linear in $B$.

    For each $B \in \mat{D}$, the composition $\hat B \circ \pi$ is continuous  and therefore $\hat B$ is continuous with respect to the quotient topology on $\PM{D}$. Thus, the quotient topology is finer than the initial topology.

    Let $\sT_i$ and $\sT_q$ be the initial and quotient topologies on $\PM{D}$, respectively. We will show that the identity $\id:(\PM{D}, \sT_i) \to (\PM{D}, \sT_q)$ is continuous at an arbitrary point $\bbA_0 \in \PM{D}$. Pick  $A_0 \in \SM{D}$ such that $\pi(A_0) = \bbA_0$. There exists a nonzero matrix element $A_{0,ab} \neq 0$ and we may multiply $A_0$ by an element of $\Unitary(1)$ if necessary to obtain $A_{0,ab} > 0$.

    Note that for all $i,j \in \left\{1,\ldots, D \right\}$ the function
    \[
        \PM{D} \to \bbC, \quad \pi(A) \mapsto \abs{A_{ij}}^2
    \]
    is continuous with respect to $\sT_i$ since it is the composition of $\hat B$ with $B = \ketbra{i}{i}$ together with taking the $(j,j)$ matrix entry of $A^*\ketbra{i}{i}A$. Thus, the set $O \subset \PM{D}$ consisting of those $\pi(A)$ such that $\abs{A_{ab}}^2 > 0$ is open with respect to $\sT_i$. It contains $\bbA_0$.

    We now define a map $F: O \to \SM{D}$ by mapping $\bbA$ to the unique matrix $F(\bbA) \in \SM{D}$ such that $\pi(F(\bbA)) = \bbA$ and $F(\bbA)_{ab} > 0$. Observe that for arbitrary $i$ and $j$ and $\bbA \in O$ we have
    \[
        \bra{b} \widehat{\ketbra{a}{i}}(\bbA) \ket{j} = \bra{b}F(\bbA)^*\ketbra{a}{i} F(\bbA) \ket{j} = \overline{F(\bbA)_{ab}}F(\bbA)_{ij}.
    \]
    This is a continuous function $O \to \bbC$ with respect to $\sT_i$. Setting $i = a$ and $j = b$ we see that $F(\bbA)_{ab}$ is continuous in $\bbA$. Then letting $i$ and $j$ be arbitrary it follows that each matrix element $F(\bbA)_{ij}$ is continuous with respect to $\sT_i$. Thus, $F$ is continuous with respect to $\sT_i$. Composing with $\pi$ yields $\id|_O$. Thus, $\id|_O$ is continuous. Since $\bbA_0$ was arbitrary, we see that $\id$ is continuous as a map from $\sT_i$ to $\sT_q$, as desired.
\end{proof}

The same proof applies to projective unitary matrices, yielding the following lemma. We will not repeat the proof.

\begin{lemma}\label{lem:initial_top_unitary}
    Let $D \in \bbN$. The quotient topology on $\PU(D)$ coincides with the initial topology induced by the maps
    \[
        \hat B:\PU(D) \to \mat{D}, \quad \hat B(\bbV) = \bbV^*B\bbV,
    \]
    across any set of matrices $B$ that spans $\mat{D}$, where in the equation above $\bbV^*B\bbV \defeq V^*BV$ for any unitary $V$ representing $\bbV$.
\end{lemma}

As mentioned at the start of the section, let $G$ be a compact Hausdorff topological group and let $U:G \to \Unitary(\bbC^\infty)$ be a representation. Recall that if $\MPSk \in \cI_U(\chi)$, then the $\lambda \in \Unitary(1)$ and $\bbv \in \PU(\chi)$ appearing in \eqref{eq:g_act_K_invariance} are unique. \zcref{lem:initial_top_unitary} now aids in the proof of the following  important proposition regarding these projective unitaries.

\begin{proposition}\label{prop:lambda_v_homomorphisms}
    Given $\MPSk \in \cI_U(\chi)$ and $g \in G$, let $\phasearg{g,\MPSk} \in \Unitary(1)$ and $\bbvarg{g,\MPSk} \in \PU(\chi)$ be the unique elements such that
    \[
        (g \act \MPSk)^i = \phasearg{g,\MPSk} \bbvarg{g,\MPSk}^* \MPSk^i \bbvarg{g,\MPSk}
    \]
    for all $i \in \bbN$. Then:
    \begin{enumerate}
    \item The maps $g \mapsto \phasearg{g,\MPSk}$ and $g \mapsto \bbvarg{g,\MPSk}$ are homomorphisms.
    \item The maps $\lambda:G \times \cI_U(\chi) \to \Unitary(1)$ and $\bbv:G \times \cI_U(\chi) \to \PU(\chi)$ are continuous.
    \end{enumerate}
\end{proposition}
\begin{proof}
\leavevmode
\begin{enumerate}
\item \textit{Homomorphism.} Below, we write $\phasearg{g} \defeq \phasearg{g,\MPSk}$ and $\bbvarg{g} \defeq \bbvarg{g,\MPSk}$ for ease of notation.
    Given $g, h \in G$ and $i \in \bbN$, we have
    \begin{align*}
    \phasearg{gh}\bbvarg{gh}^*\MPSk^i \bbvarg{gh} &= (gh \act \MPSk)^i\\
    &= (g \act (h \act \MPSk))^i\\
    &= \sum\nolimits_j U(g)_{ij} (h \act \MPSk)^j\\
    &= \sum\nolimits_j U(g)_{ij} \phasearg{h}\bbvarg{h}^* \MPSk^j \bbvarg{h}\\
    &= \phasearg{h}\bbvarg{h}^* (g \act \MPSk)^i \bbvarg{h}\\
    &= \phasearg{g}\phasearg{h}\bbvarg{h}^*\bbvarg{g}^* \MPSk^i \bbvarg{g}\bbvarg{h}.
    \end{align*}
    The uniqueness of $\lambda$ and $\bbv$ now implies that $g \mapsto \phasearg{g,\MPSk}$ and $g \mapsto \bbvarg{g,\MPSk}$ are homomorphisms.

\item \textit{Continuity.} To see that $\lambda$ and $\bbv$ are continuous, first note that since $G$ is  compact and $\cI_U(\chi)$ is the colimit of the spaces $\cI_U(d, \chi)$ across all $d \in \bbN$, the product topology on $G \times \cI_U(\chi)$ coincides with the colimit topology induced by the inclusions $G \times \cI_U(d, \chi) \hookrightarrow G \times \cI_U(\chi)$. Thus, it suffices to show continuity when restricted to $G \times \cI_U(d, \chi)$.

    Observe that the map $C:\cI_U(d, \chi) \to \cL(\bbC^{d}, \mat{\chi})$ defined by
    \[
    C_\MPSk(\mu) = \sum_i \mu_i \MPSk^i
    \]
    is continuous in $\MPSk$ and $C_\MPSk$ is surjective for every $k \in \cI_U(d, \chi)$. Thus, the Moore-Penrose inverse $C_\MPSk^+ : \mat{\chi} \to \bbC^{d}$ provides a continuous map $\cI_U(d, \chi) \to \cL(\mat{\chi}, \bbC^{d})$, $\MPSk \mapsto C_\MPSk^+$ such that $C_\MPSk C_\MPSk^+$ is the identity for all $\MPSk$.

    Thus, for every $B \in \mat{\chi}$ we have
    \[
    \sum_i C_\MPSk^+(B)_i \MPSk^i = B \qquad \forall\, \MPSk \in \cI_U(d, \chi).
    \]
    Multiplying by $\phasearg{g,\MPSk}$ and conjugating by $\bbvarg{g,\MPSk}^*$ yields
    \begin{equation}\label{eq:lambda_bbV_continuity}
    \sum_i C_\MPSk^+(B)_i (g \act \MPSk)^i = \phasearg{g,\MPSk}\bbvarg{g,\MPSk}^*B \bbvarg{g,\MPSk}.
    \end{equation}
    The left hand side is manifestly continuous in $g$ and $\MPSk$. Setting $B = \1_\chi$ proves that $\phasearg{g,\MPSk}$ is continuous. Continuity of $\bbvarg{g,\MPSk}$ then follows from \zcref{lem:initial_top_unitary}. \qedhere
\end{enumerate}
\end{proof}

\begin{remark}\label{rem:various_gauge_transforms}
    It is useful to see how various quantities transform under a gauge transformation  of an MPS tensor.
    Thus, let $\MPSl \in \cI_U(\chi)$, $\mu \in \Unitary(1)$, $\bbw \in \PU(\chi)$, and define
    \[
        \MPSk^i \defeq \mu \bbw^* \MPSl^i \bbw
    \]
    for all $i$. Then for all $g \in G$ and all $i \in \bbN$ we have
    \begin{align*}
        \targ{\MPSk} &= \bbw^* \targ{\MPSl} \bbw, \\
        \phasearg{g,\MPSk} &= \phasearg{g,\MPSl}, \\
        \bbvarg{g,\MPSk} &= \bbw^* \bbvarg{g,\MPSl}\bbw.
    \end{align*}
    These can all be easily checked using the uniqueness properties of these quantities.
\end{remark}

Next, we extend $\lambda$ and $\bbv$ to functions on $\cE_U$.

\begin{proposition}\label{prop:bbV}
    The map $\bbV: G \times \cE_U \to \PM{\infty}$ defined by
    \[
    \bbVarg{g,\MPSA} = Y\begin{pmatrix} \bbvarg{g,\MPSk} & 0 \\ 0 & 0 \end{pmatrix}Y^* \quad \textnormal{where $\MPSA^i = \mu Y\mqty{ \MPSk^i & 0 \\ \MPSm^i & 0 }Y^*$ as in \eqref{eq:A_form_simplified}}
    \]
    is well-defined and satisfies $\bbVarg{gh, \MPSA} = \bbVarg{g,\MPSA} \bbVarg{h,\MPSA}$ for all $g, h \in G$ and $\MPSA \in \cE$. The map $\altphase:G \times \cE_U \to \Unitary(1)$, $\altphasearg{g,\MPSA} = \phasearg{g,\MPSk}$ is also well-defined. Furthermore, $\altphase$ is continuous.
\end{proposition}
\begin{proof}
    If we can show that $\bbV$ is well-defined, then the identity $\bbVarg{gh,\MPSA} = \bbVarg{g,\MPSA}\bbVarg{h,\MPSA}$ will follow since $g \mapsto \bbvarg{g,\MPSk}$ is a group homomorphism.

    Let $\chi$ be the essential rank of $\MPSA$. Suppose $\MPSA$ has another representation as in \eqref{eq:A_form_simplified}:
    \[
    \MPSA = \mu' Y' \begin{pmatrix} \MPSl & 0 \\ \MPSm' & 0 \end{pmatrix}Y'.
    \]
    Since $\MPSk$ and $\MPSl$ define the same state, there exists $\nu \in \Unitary(1)$ and $\smallz \in \Unitary(\chi)$ such that $\MPSk^i = \nu \smallz \MPSl^i \smallz^*$ for all $i$. One can easily check that $\phasearg{g,\MPSk} = \phasearg{g,\MPSl}$ and $\bbvarg{g,\MPSk} = \smallz\bbvarg{g,\MPSl}\smallz^*$ for all $g \in G$. In particular, $\altphase$ is well-defined.

    Since $\vecspan_i \MPSl^i = \mat{\chi}$, there must be some $i$ such that $\Tr \MPSl^i \neq 0$. Then
    \[
    0 \neq \mu' \Tr \MPSl^i = \Tr \MPSA^i = \mu \Tr \MPSk^i = \nu \mu \Tr(\smallz\MPSl^i \smallz^*) =  \mu \nu \Tr \MPSl^i.
    \]
    We see that $\mu \nu = \mu'$.

    Note that
    \[
    Q(\MPSA) = Y\begin{pmatrix} \1_\chi & 0 \\0 & 0 \end{pmatrix}Y^* = Y' \begin{pmatrix} \1_\chi & 0 \\ 0 & 0 \end{pmatrix} Y'^*.
    \]
    Since $Y^*Y'$ commutes with $\begin{pmatrix} \1_\chi & 0 \\0 & 0 \end{pmatrix}$, it follows that $Y^*Y'$ is block diagonal:
    \[
    Y^*Y' = \begin{pmatrix} \smallw_1 & 0 \\ 0 & \smallw_2 \end{pmatrix}
    \]
    where $\smallw_1 \in \Unitary(\chi)$ and $\smallw_2 \in \Unitary(\infty)$.

    Next, computing $Q(\MPSA)\MPSA^i$ with the two different representations of $\MPSA$ yields
    \[
    \mu Y\begin{pmatrix} \MPSk^i & 0 \\ 0 & 0 \end{pmatrix} Y^* = \mu' Y' \begin{pmatrix} \MPSl^i & 0 \\ 0 & 0 \end{pmatrix} Y'^*.
    \]
    Acting with $g$ on both sides yields
    \[
    \mu Y\begin{pmatrix} \phasearg{g,\MPSk}\bbvarg{g,\MPSk}^* \MPSk^i \bbvarg{g, \MPSk} & 0 \\ 0 & 0 \end{pmatrix} Y^* = \mu'  Y'\begin{pmatrix} \phasearg{g,\MPSl}\bbvarg{g,\MPSl}^* \MPSl^i \bbvarg{g, \MPSl} & 0 \\ 0 & 0 \end{pmatrix} Y'^*.
    \]
    Since $\phasearg{g,\MPSk} = \phasearg{g,\MPSl}$ we may cancel this scalar factor from both sides. Choosing a representative $\varg{g,\MPSl}$ for $\bbvarg{g,\MPSl}$, we know that $\varg{g,\MPSk} \defeq \smallz \varg{g, \MPSl}\smallz^*$ is a representative for $\bbvarg{g,\MPSk}$. Then substituting $\MPSk^i = \nu \smallz \MPSl^i \smallz^*$ and $\varg{g,\MPSk} = \smallz \varg{g,\MPSl}\smallz^*$, canceling $\mu \nu = \mu'$ from both sides, and rearranging terms yields
    \[
    \begin{pmatrix} \MPSl^i \varg{g,\MPSl} \smallz^*\smallw_1 \varg{g,\MPSl}^* & 0 \\ 0 & 0 \end{pmatrix} =  \begin{pmatrix} \varg{g,\MPSl}\smallz^* \smallw_1 \varg{g,\MPSl}^* \MPSl^i & 0 \\0 & 0 \end{pmatrix}.
    \]
    Since this holds for all $i$, injectivity of $\MPSl$ implies that $\varg{g,\MPSl}\smallz^*\smallw_1 \varg{g,\MPSl}^*$ is a scalar multiple of the identity. Since $\varg{g,\MPSl}$ is unitary this implies that $\smallz^*\smallw_1$ is a scalar multiple of the identity. Using this, one easily checks that
    \[
    Y\begin{pmatrix} \bbvarg{g,\MPSk} & 0 \\ 0 & 0 \end{pmatrix} Y^* = Y'\begin{pmatrix} \bbvarg{g,\MPSl} & 0 \\ 0 & 0 \end{pmatrix} Y'^*,
    \]
    from which we see that $\bbVarg{g,\MPSA}$ is well-defined.

    To see that $\altphase$ is continuous, we observe that
\[
    \Tr[(g \act \MPSA)^i] = \altphasearg{g,\MPSA} \Tr[\MPSA^i].
\]
Given $\MPSA_0 \in \cE_U$, we may decompose $\MPSA_0$ as in \eqref{eq:A_form_simplified} with an injective component $\MPSk_0$. Then by injectivity of $\MPSk_0$ there exists $i$ such that $\Tr[\MPSk_0^i] = \Tr[\MPSA_0^i] \neq 0$. Then for every $g \in G$ and every $\MPSA$ such that $\Tr[\MPSA^i] \neq 0$, we have
\[
    \altphasearg{g,\MPSA} =\frac{\Tr[(g \act \MPSA)^i]}{\Tr[\MPSA^i]}.
\]
It follows that $\altphasearg{g, \MPSA}$ is continuous on a neighborhood of $G \times \{\MPSA_0\}$.
\end{proof}
The map $\bbVarg{g,A}$ is not continuous in $A$ since $\Tr[\bbVarg{g,A}\bbVarg{g,A}^*] = \chi$ and $\chi$ is not a continuous function of $A$. However, when multiplied by $\Targ{A}^{1/2}$ (see \zcref{def:tilde_T}), we can show that we get a continuous map $(g, A) \mapsto   \Targ{A}^{1/2}\bbVarg{g,A}$. Before showing this it is worth pointing out that $\bbVarg{g,A}$ commutes with $\Targ{A}$, hence also with $\Targ{A}^{1/2}$. We prove this below.

\begin{proposition}
If $\MPSk \in \cI_U(\chi)$, then $[\varg{g,\MPSk},\targ{\MPSk}] = 0$ for every $g \in G$ and every lift $\varg{g,\MPSk}$ of $\bbvarg{g,\MPSk}$. If $A \in \cE_U$, then $[\Varg{g,A}, \Targ{A}] = 0$ for every $g \in G$ and every lift $\Varg{g,A}$ of $\bbVarg{g,A}$.
\end{proposition}
\begin{proof}
Let $\MPSk \in \cI_U(\chi)$, let $g \in G$, and let $\varg{g,\MPSk}$ be a unitary lift of $\bbvarg{g,\MPSk}$. We observe that
\begin{align*}
\sum_i \MPSk^{i*}\varg{g,\MPSk}\targ{\MPSk}\varg{g,\MPSk}^* \MPSk^i &= \sum_i \varg{g,\MPSk}(g \act \MPSk)^{i*} \targ{\MPSk} (g \act \MPSk)^i \varg{g,\MPSk}^*\\
&= \sum_{i,j_1,j_2} \overline{U(g)_{ij_1}} U(g)_{ij_2} \varg{g,\MPSk} \MPSk^{j_1*} \targ{\MPSk} \MPSk^{j_2} \varg{g,\MPSk}^*\\
&= \sum_{j_1, j_2} \delta_{j_1j_2} \varg{g,\MPSk} \MPSk^{j_1*} \targ{\MPSk} \MPSk^{j_2} \varg{g,\MPSk}^*\\
&= \sum_{j} \varg{g,\MPSk} \MPSk^{j*} \targ{\MPSk} \MPSk^{j} \varg{g,\MPSk}^*\\
&= \varg{g,\MPSk}\targ{\MPSk}\varg{g,\MPSk}^*.
\end{align*}
This proves that $\targ{\MPSk} = \varg{g,\MPSk}\targ{\MPSk}\varg{g,\MPSk}^*$ by the uniqueness properties of $\targ{\MPSk}$. Thus, $[\varg{g,\MPSk}, \targ{\MPSk}] = 0$.

It is now easily checked that $[\Varg{g,A}, \Targ{A}] = 0$ using the definitions of $\Varg{g,A}$ and $\Targ{A}$.
\end{proof}

We now prepare for the proof that $(g, A) \mapsto \Targ{A}^{1/2}\bbVarg{g,A}$ is continuous with the following technical lemma.

\begin{lemma}\label{lem:Moore_Penrose_coordinates}
Let $d, D \in \bbN$.
Let $A_0 \in \cE(d, D)$ and represent $A_0$ as
\[
A_0 = Y\mqty{ \MPSl & 0 \\ \MPSn & 0 } Y^*,
\]
where $Y \in \Unitary(D)$,  $\MPSl \in \cI(d, \chi_0)$, and $\MPSn \in \mat{D- \chi_0, \chi_0}^{d}$.
For convenience, let
\[
    Q_0 \defeq Q(A_0) = Y\mqty{ \1_{\chi_0} & 0 \\ 0 & 0 } Y^*.
\]
There exists a neighborhood $O \subset \cE(d, D)$ containing $A_0$ such that for any $B \in Q_0\mat{D} Q_0$, there exists a continuous function $\mu:O \to \bbC^d$ such that
\[
    Q_0 \Targ{A}^{1/2}\left(\sum_i \mu_i(A)A^i \right)\Targ{A}^{1/2} Q_0 = Q_0 \Targ{A}^{1/2} B \Targ{A}^{1/2} Q_0
\]
for every $A \in O$.
\end{lemma}

\begin{proof}
Given $A \in \cE(d, D)$, define $S_A :\bbC^{d} \to Q_0 \mat{D} Q_0$ as
\[
    S_A(\mu) = Q_0 \Targ{A}^{1/2} \cdot\left(\sum_i \mu_i A^i\right) \cdot \Targ{A}^{1/2} Q_0.
\]
Observe that $\cE(d, D) \to  \cL(\bbC^{d}, Q_0 \mat{D} Q_0)$, $A \mapsto S_A$ is continuous. Furthermore, $S_{A_0}$ is surjective. The surjective linear maps form an open subset of $\cL(\bbC^d, Q_0 \mat{D} Q_0)$, so there exists a neighborhood $O$ of $A_0$ on which $S_{A}$ is surjective for all $A \in O$. Endowing $Q_0 \mat{D} Q_0$ with the Hilbert-Schmidt inner product, the Moore-Penrose inverse $S_A^+ = S_A^* (S_AS_A^*)^{-1}$ is a right inverse of $S_A$ for all $A \in O$, and the map $O \to \cL(Q_0 \mat{D} Q_0, \bbC^{d})$, $A \mapsto S_A^+$ is continuous. Setting
\[
    \mu(A) = S_A^+\left(Q_0\Targ{A}^{1/2} B \Targ{A}^{1/2} Q_0\right)
\]
then gives a continuous function with the desired property.
\end{proof}

Below we topologize $\PM{\infty}$ as a quotient of $\mat{\infty}$ which is equivalent to topologizing $\PM{\infty}$ as a colimit of its subspaces $\PM{D}$.

\begin{theorem}\label{thm:VT_continuity}
The map
\[
    G \times \cE_U \to \PM{\infty}, \quad (g, A) \mapsto \Targ{A}^{1/2}\bbVarg{g,A}
\]
is continuous.
\end{theorem}

\begin{proof}
Since $\cE_U$ and $\PM{\infty}$ are topologized as colimits and since $G$ is compact Hausdorff, it suffices to prove that the restriction of the above map to $G \times \cE_U(d, D) \to \PM{D}$ is continuous.

Endow $\mat{D}$ with the Hilbert-Schmidt norm $\norm{\cdot}_2$. By \zcref{lem:initial_top}, it suffices to show that the map $G \times \cE_U(d, D) \to  \mat{D}$ defined by
\begin{equation}\label{eq:continuous_conjugate_on_B}
    (g, A) \mapsto \Varg{g,A}^*\Targ{A}^{1/2} B \Targ{A}^{1/2} \Varg{g,A}
\end{equation}
is continuous for all $B \in \mat{D}$. Here $\Varg{g,A}$ may be taken to be any lift of $\bbVarg{g,A}$ such that $\Varg{g,A}^* \Varg{g,A} = Q(A)$.

We show continuity at an arbitrary point $(g_0, A_0) \in G \times \cE_U(d, D)$. Write
\[
    A_0 = Y\mqty{ \MPSl & 0 \\ \MPSn & 0 } Y^*
\]
with $\MPSl \in \cI_U(d, \chi_0)$, $\MPSn \in \mat{D - \chi_0, \chi_0}$, and $Y \in \Unitary(D)$. It suffices to show that \eqref{eq:continuous_conjugate_on_B} is continuous for all $B$ of the form $B = Y E_{\alpha \beta} Y^*$ where $E_{\alpha \beta}$ is the matrix with a $1$ in the $(\alpha, \beta)$ component and a $0$ everywhere else.

If $\alpha > \chi_0$ or $\beta > \chi_0$, then
\[
    \Targ{A_0}^{1/2} YE_{\alpha \beta} Y^* \Targ{A_0}^{1/2} = 0.
\]
Given $\varepsilon > 0$, continuity of $\bigT$ implies that there exists a neighborhood $O$ of $A_0$ such that
\[
    \norm{\Targ{A}^{1/2} YE_{\alpha \beta} Y^* \Targ{A}^{1/2}}_2 < \varepsilon
\]
for all $A \in O$. Thus, for all $(g, A) \in G \times O$, we have
\[
    \norm{\Varg{g,A}^*\Targ{A}^{1/2} YE_{\alpha \beta}Y^*  \Targ{A}^{1/2} \Varg{g,A}}_2 < \varepsilon.
\]
This proves continuity for $\alpha > \chi_0$ or $\beta > \chi_0$.

Now suppose $\alpha \leq \chi_0$ and $\beta \leq \chi_0$. Let $O$ be a neighborhood of $A_0$ as in \zcref{lem:Moore_Penrose_coordinates}, set $B = YE_{\alpha \beta}Y^*$, and let $\mu:O \to \bbC^{d}$ be the function obtained from \zcref{lem:Moore_Penrose_coordinates} applied to $B$. For ease of notation, given $A \in O \cap \cE_U(d, D)$ and $g \in G$, let
\[
    C(g,A) \defeq \sum_i \mu_i(A)(g \act A)^i.
\]

For every $A \in O \cap \cE_U(d, D)$ and $g \in G$, one can use the fact
that $[ \Varg{g,A} , \Targ{A} ] = 0$
to check that
\[
    \overline{\altphasearg{g,A}}\Ad\!\left[\Targ{A}^{1/2}\right]\left(C(g,A)\right) = \Ad\!\left[\Varg{g,A}^*\Targ{A}^{1/2}\right]\left(C(1,A)\right),
\]
where we use the notation $\Ad[R](S) = RSR^*$. Thus, with $Q_0 = Q(A_0)$, we have:
\begin{align*}
&\norm{\overline{\altphasearg{g,A}}\Ad\!\left[\Targ{A}^{1/2}\right]\left( C(g,A) \right) - \Ad\!\left[\Varg{g,A}^* \Targ{A}^{1/2}\right]\left(B \right)}_2\\
&\leq\norm{\Ad\!\left[\Varg{g,A}^* \Targ{A}^{1/2}\right]\left(C(1,A) \right) - \Ad\!\left[\Varg{g,A}^* Q_0\Targ{A}^{1/2} \right]\left(C(1,A)\right)}_2\\
&\qquad\qquad + \norm{ \Ad\!\left[\Varg{g,A}^*Q_0 \Targ{A}^{1/2}\right](B) - \Ad\!\left[\Varg{g,A}^* \Targ{A}^{1/2}\right](B)}_2\\
&\leq \left(2\norm{C(1,A)} + 2\right)\norm{\Targ{A}^{1/2} - Q_0\Targ{A}^{1/2}}
\end{align*}
where the norms in the last line are the supremum norms.
At $A_0$, we have $Q_0 \Targ{A_0} = \Targ{A_0}$ so that the bound above is zero. Furthermore, the above bound is a continuous function of $A \in O$. Finally, we observe that the function
\[
    \overline{\altphasearg{g,A}}\Ad\!\left[\Targ{A}^{1/2}\right]\left( C(g,A)\right)
\]
is a continuous function of $(g,A) \in G \times (O \cap \cE_U(d, D))$. Thus, given $\varepsilon > 0$, there exists a neighborhood $O'$ of $(g_0, A_0)$ such that
\[
    \norm{\Ad\left[\Varg{g,A}^*\Targ{A}^{1/2}\right](B) - \Ad\left[\Varg{g_0, A_0}^*\Targ{A_0}^{1/2}\right](B)}_2 < \varepsilon
\]
for all $(g,A) \in O'$. This concludes the proof.
\end{proof}

One easily obtains variations on \zcref{thm:VT_continuity} such as the following.

\begin{corollary}\label{cor:VT_variations}
    Each of the maps $G \times \cE_U \to \PM{\infty}$ below is continuous:
    \begin{align*}
        (g, A) &\mapsto \bbVarg{g, A}^*\Targ{A}^{1/2}, \\
        (g, A) &\mapsto \Targ{A} \bbVarg{g,A}.
    \end{align*}
\end{corollary}

\begin{proof}
    The first map is obtained by composing the map in \zcref{thm:VT_continuity} with Hermitian conjugation. The second is obtained by multiplying the map in \zcref{thm:VT_continuity} with $\Targ{A}^{1/2}$ on the left. Each of these operations is continuous. In particular, multiplying by $\Targ{A}^{1/2}$ on the left is a continuous operation since matrix multiplication is continuous as a map $\mat{\infty} \times \PM{\infty} \to \PM{\infty}$. This is true because the product topology on $\mat{\infty} \times \PM{\infty}$ coincides with the final topology induced by the inclusions $\mat{D} \times \PM{D} \hookrightarrow \mat{\infty} \times \PM{\infty}$ \cite[Thm.~4.1 \& Lem.~4.2]{Hirai_et_al}.
\end{proof}

We are now ready to define the main quantity used in the construction of our  invariant.
Note that if $(g,A) \in G \times \cE_U$, where $A^i = \mu Y\bigl(\smallmat{\MPSk^i & 0 \\ \MPSm^i & 0}\bigr) Y^*$ as in \eqref{eq:A_form_simplified},
and $\Varg{g,A}$ is a lift of $\bbVarg{g,A}$ to a partial isometry, then there exists a unique lift $\varg{g,\MPSk}$ of $\bbvarg{g,\MPSk}$ to a unitary such that
\begin{equation}\label{eq:bigV_smallv_relation}
    \Varg{g,A} = Y\mqty{\varg{g,\MPSk} & 0 \\ 0 & 0}Y^*.
\end{equation}
Below and for the remainder of the paper, whenever we refer to a ``lift'' or ``representative'' $\Varg{g,A}$ of $\bbVarg{g,A}$, it will always be understood that the lift $\Varg{g,A}$ is chosen to be a partial isometry. Likewise, a ``lift'' or ``representative'' $\varg{g,\MPSk}$ of $\bbvarg{g,\MPSk}$ is understood to be unitary.

\begin{proposition}\label{prop:GxBU_matrix}
    Define a function $\Xi: G \times \cE_U \to \PM{\infty}$ by setting $\Xi(g,A)$ to be the projectivization of the matrix whose $(i,j)$ entry is:
    \[
        \Tr\left[\Targ{A} \Varg{g,A} A^iA^{j*} \right]
    \]
    where $\Varg{g,A}$ is any lift of $\bbVarg{g,A}$ (the same choice of lift being used for all $i,j$). Then $\Xi$ is continuous and factors through a continuous function $\xi : G \times \cB_U \to \PM{\infty}$ so that the diagram below commutes.
    \[
        \begin{tikzcd}
            G \times \cE_U \arrow[r,"\Xi"]\arrow[d,"\id \times p"']& \PM{\infty}\\
            G \times \cB_U \arrow[ur,"\xi"']&
        \end{tikzcd}
    \]
\end{proposition}

\begin{proof}
    First we check that $\Xi$ is well-defined. A different choice of $\Varg{g,A}$ will lead to an overall phase on the matrix representing $\Xi(g,A)$, which does not change the projectivization of the matrix. To see that the matrix is nonzero (and therefore has a projectivization at all), let $A$ have a representation
    \[
    A^i = Y\mqty{ \MPSk^i & 0 \\ \MPSm^i & 0 } Y^*
    \]
    where $\MPSk$ is an injective right-normalized MPS tensor. Observe that
    \[
    \Tr[\Targ{A} \Varg{g,A} A^iA^{j*}] = \Tr[\targ{\MPSk} \varg{g,\MPSk} \MPSk^i\MPSk^{j*}]
    \]
    where $\Varg{g,A}$ and $\varg{g,\MPSk}$ are related as in \eqref{eq:bigV_smallv_relation}.
    Since $\targ{\MPSk} \varg{g,\MPSk} \neq 0$ and $\MPSk$ is an injective MPS tensor, the above must be nonzero for some $i$ and $j$.

    This also shows that $\Xi$ factors through $G \times \cB_U$. Indeed, if
    \[
    B = Z\mqty{ \MPSl^i & 0 \\ \MPSn^i & 0 } Z^*
    \]
    is any other element of $\cE_U$ representing the same state as $A$, then there exists $\mu \in \Unitary(1)$ and $\smallw \in \Unitary(\chi)$, where $\chi$ is the bond dimension of the state, such that $\MPSk^i = \mu \smallw^* \MPSl^i \smallw$. Then one easily sees from \zcref{rem:various_gauge_transforms} that we can choose $\varg{g,\MPSk} = \smallw^* \varg{g,\MPSl} \smallw$, hence
    \[
    \Tr[\targ{\MPSk} \varg{g,\MPSk} \MPSk^i\MPSk^{j*}] =  \Tr[\targ{\MPSl} \varg{g,\MPSl} \MPSl^i\MPSl^{j*}]
    \]
    so in particular the vectors obtained from $A$ and $B$ have the same projectivization.

    To prove continuity, it suffices to prove continuity of each of the restrictions $G \times \cE_U(d, D) \to \PM{d}$. Since $(g, A) \mapsto  \Targ{A}\bbVarg{g,A} \in \PM{D}$ is continuous by \zcref{cor:VT_variations}, it locally lifts to a continuous function $(g, A) \mapsto \Targ{A} \Varg{g,A}$, where $\Varg{g,A}$ is a lift of $\bbVarg{g,A}$. Then it is clear that $(g, A) \mapsto \Tr[\Targ{A} \Varg{g,A} A^iA^{j*}]$ is continuous on the domain of this local lift, hence $(g, A) \mapsto \Xi(g,A)$ is continuous on the domain of the local lift. Since these domains cover $G \times \cE_U(d, D)$, this proves global continuity of $\Xi$.

    Continuity of $\xi$ follows from continuity of $\Xi$ because the quotient topology on $G \times \cB_U$ obtained from $G \times \cE_U$ coincides with the product topology on $G \times \cB_U$, since $G$ is assumed to be compact Hausdorff \cite[Thm.~XII.4.1]{DugundjiTopology}.
\end{proof}

It will also be useful to keep track of the magnitudes of the individual components of the matrix defining $\Xi(g, A)$.

\begin{proposition}\label{prop:ij_abs_value}
    For each $i,j \in \bbN$, define $C_{ij} : G \times \cE_U \to [0,\infty)$  by
    \[
    C_{ij}(g, A) = \abs{\Tr[\Targ{A} \Varg{g,A} A^i A^{j*}]}
    \]
    where $\Varg{g,A}$ is any lift of $\bbVarg{g,A}$. Then $C_{ij}$ is continuous and factors through a continuous function $G \times \cB_U \to [0,\infty)$.
\end{proposition}

\begin{proof}
    It is clear that $C_{ij}$ is well-defined since different choices of $\Varg{g,A}$ will differ only by a phase, which is eliminated by the absolute value. Let $(g_0, A_0) \in G \times \cE_U$. Since the function $(g,A) \mapsto \Targ{A}^{1/2} \bbVarg{g,A} \in \PM{\infty}$ is continuous, it may locally be lifted on some domain $(g_0, A_0) \in O \subset G \times \cE_U$ to a function $(g, A) \mapsto R(g,A) \in \SM{\infty}$, where $\SM{\infty}$ is the unit sphere $\mat{\infty}$ with respect to the Hilbert-Schmidt norm. Since $R(g,A)$ and $\Targ{A}^{1/2}$ both have Hilbert-Schmidt norm equal to one, we see that $R(g,A) = \Targ{A}^{1/2}\Varg{g,A}$ where $\Varg{g,A}$ is a lift of $\bbVarg{g,A}$. Thus, we see that $C_{ij} = |\Tr[\Targ{A}^{1/2}R(g,A) A^i A^{j*}]|$ is continuous on $O$. Since $(g_0, A_0)$ was arbitrary, this proves global continuity of $C_{ij}$. As in \zcref{prop:GxBU_matrix},  $C_{ij}(g,A)$ is invariant under gauge transformations of $A$, so that $C_{ij}$ factors through a continuous function $G \times \cB_U \to [0,\infty)$.
\end{proof}

We will continue to use $\SM{\infty}$ to denote the unit sphere of $\mat{\infty}$ with respect to the Hilbert-Schmidt norm.

\begin{proposition}\label{prop:kappa_continuity}
    Define a function $\kappa:G \times \cE_U \to (0,\infty)$ by
    \[
    \kappa(g, A) = \sqrt{\sum_{i,j} \abs{\Tr[\Targ{A} \Varg{g,A} A^i A^{j*} ]}^2},
    \]
    where $\Varg{g,A}$ is any lift of $\bbVarg{g,A}$.
    Then $\kappa$ is continuous and factors through a continuous function $G \times \cB_U \to (0,\infty)$.
\end{proposition}

\begin{proof}
    Fix $(g_0, A_0) \in G \times \cE_U$.
    Let $\Xi:G \times \cE_U \to \PM{\infty}$ be the map from \zcref{prop:GxBU_matrix}. Since $\SM{\infty} \to \PM{\infty}$ is a fiber bundle, there exists a neighborhood $O$ of $(g_0, A_0)$ and a continuous lift $\hat \Xi : O \to \SM{\infty}$ of $\Xi$. Given $(g,A) \in O$, there exists a unique lift of $\bbVarg{g,A}$ to a partial isometry $\Varg{g,A}$ such that
    \[
    \kappa(g,A) \hat \Xi(g,A)_{ij} = \Tr[\Targ{A} \Varg{g,A} A^iA^{j*}]
    \]
    for all $i$ and $j$. Moreover, there exists some $i$ and $j$ such that $\hat \Xi(g_0, A_0)_{ij} \neq 0$, and we may shrink $O$ if necessary so that $\hat \Xi(g, A)_{ij} \neq 0$ for all $(g, A) \in O$. Thus, on $O$ we see that
    \[
    \kappa(g,A) = \frac{C_{ij}(g,A)}{\abs{\hat \Xi(g,A)_{ij}}},
    \]
    where $C_{ij}$ is as in \zcref{prop:ij_abs_value}. Since the numerator and denominator are nonzero, we see that $\kappa$ is continuous on $O$. Since $O$ was a neighborhood of the arbitrary point $(g_0, A_0)$, we see that $\kappa$ is continuous.

    Since $\kappa(g,A) = \sqrt{\sum_{i,j} C_{ij}(g,A)^2}$ and each $C_{ij}$ factors through $G \times \cB_U$, we see that $\kappa$ factors through $G \times \cB_U$.
\end{proof}

By slight abuse of notation, we will also denote the factored function from \zcref{prop:kappa_continuity} by $\kappa:G \times \cB_U \to (0,\infty)$.

Suppose $\omega \in \cB_U$ and $\hat \xi:G \times \{\omega\} \to \SM{\infty}$ is a (not necessarily continuous) lift of the function $\xi$ from \zcref{prop:GxBU_matrix} on $G \times \{\omega\}$. Choose $A \in \cE_U$ such that $p(A) = \omega$. Then for every $g \in G$, the lift $\hat \xi(g, \omega)$ determines a unique lift $\Varg{g, A}$ of $\bbVarg{g, A}$ such that
\begin{equation}\label{eq:V_from_hatxi}
\kappa(g, \omega) \hat \xi(g, \omega)_{ij} = \Tr[\Targ{A} \Varg{g,A} A^i A^{j*}].
\end{equation}
for all $i$ and $j$. Furthermore, since $\bbVarg{g,A}\bbVarg{h,A} = \bbVarg{gh,A}$ for all $g,h \in G$, we know
\[
\Varg{g,A}\Varg{h,A} = \eta_A(g,h) \Varg{gh,A}
\]
for some function  $\eta_A : G \times G \to \Unitary(1)$ satisfying
\begin{equation}\label{eq:2cocycle_condition}
\eta_A(f,gh)\eta_A(g,h) = \eta_A(f,g)\eta_A(fg,h)
\end{equation}
for all $f,g,h \in G$. Had we chosen a different tensor $B \in \cE_U$ such that $p(B) = \omega$, then we would find that there exists $Z \in \Unitary(\infty)$ such that $\Varg{g,B} = Z \Varg{g,A} Z^*$ for every $g \in G$.
In particular this implies that $\eta_A(g,h) = \eta_B(g,h)$ for all $g,h \in G$. Thus, the function $\eta = \eta_A = \eta_B$ depends only on $\omega$ and the lift $\hat \xi$.

In this paper, we will consider the Borel cohomology of the compact Hausdorff group $G$ \cite{MooreI}.\footnote{In \cite{MooreI}, the group $G$ is required to be second-countable. However, if $G$ is any topological group, the Borel cohomology groups may still be defined without difficulty using the usual differential given in \cite{MooreI} if an $n$-cochain is defined in the way we prescribe.} In Borel cohomology, an $n$-cochain is a function $\eta:G^n \to \Unitary(1)$ such that $\eta(g_1,\ldots, g_n) = 1$ if any $g_i$ is the identity and such that $\eta$ is measurable with respect to the Borel $\sigma$-algebra on $G^n$ coming from the product topology. Thus, it is important for us to identify conditions under which the function $\eta$ from \eqref{eq:2cocycle_condition} is Borel measurable. The normalization condition---that $\eta(g_1,\ldots, g_n) = 1$ when any $g_i$ is the identity---will be dealt with in \zcref{prop:lift_with_normalized_2cocycle}.

\begin{proposition}\label{prop:pointwise_Borel_measurable_cocycle}
    Let $\omega \in \cB_U$. If $\hat \xi:G \times \{\omega\}  \to \SM{\infty}$ is a lift of $\xi$ on $G \times \{\omega\}$ and each matrix element of $\hat \xi$ is a Borel measurable function $G \to \bbC$, then the 2-cocycle $\eta:G \times G \to \Unitary(1)$ induced by $\hat \xi$ is Borel measurable.
\end{proposition}

\begin{proof}
    Let $\MPSA \in \cE_U$ be an MPS tensor representing $\omega$.
    Then for each $g \in G$, there exists a unique $\Varg{g,\MPSA} \in \Unitary(\chi)$ such that
    \[
        \kappa(g,\omega) \hat \xi(g, \omega)_{ij} = \Tr[\Targ{\MPSA} \Varg{g,\MPSA} \MPSA^i \MPSA^{j*}]
    \]
    for all $i$ and $j$. If we decompose $\MPSA$ as $\MPSA^i = Y\bigl(\smallmat{\MPSk^i & 0 \\ \MPSm^i & 0}\bigr)Y^*$ as in \eqref{eq:A_form_simplified}, then there exists a unique lift $\varg{g,\MPSk}$ of $\bbvarg{g,\MPSk}$ such that $\Varg{g,\MPSA} = Y\bigl(\smallmat{\varg{g,\MPSk} & 0 \\ 0 & 0}\bigr)Y^*$.
    We then have
    \[
    \kappa(g,\omega) \hat \xi(g, \omega)_{ij} = \Tr[\targ{\MPSk}\varg{g,\MPSk}\MPSk^i\MPSk^{j*}]
    \]
    The 2-cocycle induced by $\hat \xi$ is then the 2-cocycle induced by $\varg{g,\MPSk}$.

    By our hypothesis, the above is a Borel measurable function of $g$.
    Thanks to the injectivity of \( k \), one can replace \( k^{i}k^{j*} \) with \( \ket{\beta}\bra{\alpha} \), where \( \{\ket{\alpha}\}_{\alpha=1, \ldots, \chi} \) is an orthonormal basis, and still get a Borel measurable function of \( g \).
    In this way, the trace reduces to the entry \( (\alpha,\beta) \) of $\targ{\MPSk} \varg{g,\MPSk}$, which is again a measurable function.
    From this it follows that $\targ{\MPSk} \varg{g,\MPSk}$ itself is a Borel measurable function of $g$.
    Since $\targ{\MPSk}$ is invertible and left multiplication by $\targ{\MPSk}^{-1}$ is continuous, we know $\varg{g,\MPSk}$ is a Borel measurable function of $g$.
    We then see that the 2-cocycle
    \[
        \eta(g,h) = \frac{1}{\chi} \Tr[\varg{gh,\MPSk}^*\varg{g,\MPSk}\varg{h,\MPSk}]
    \]
    is Borel measurable because it can be viewed as a composition
    \[
    G \times G \to G \times G \times G \to \Unitary(\chi) \times \Unitary(\chi) \times \Unitary(\chi) \to \bbC,
    \]
    where the first map is $(g,h) \mapsto (gh, g, h)$, the second maps applies $\varg{\,\cdot\,, \MPSk}$ to each component and the third map is $(u, v, w) \mapsto \Tr[u^*vw]/\chi$. The first and third map are continuous. To see that the second map is Borel measurable, note that $\Unitary(\chi)$ is second-countable, so that the Borel $\sigma$-algebra on $\Unitary(\chi) \times \Unitary(\chi) \times \Unitary(\chi)$ coincides with the $\sigma$-algebra generated by measurable rectangles. A function into a product with the latter $\sigma$-algebra is measurable if and only if each component is measurable, and indeed each component of the second map is Borel measurable.
\end{proof}

In the above proposition, we lifted a map $G \times \cB_U \to \PM{\infty}$ to $\SM{\infty}$ at a single state $\omega \in \cB_U$. In the following, we will be lifting such maps on open sets $O \subset \cB_U$. Thus, we will obtain functions $F:G \times O \to \SM{\infty}$. These functions will not in general be continuous, but will be Borel measurable with respect to the product topology on $G \times O$ and the Hilbert-Schmidt topology on $\SM{\infty}$. More importantly, the function $F^\wedge:O \to \SM{\infty}^G$, $F^\wedge(\omega)(g) = F(g,\omega)$ will be continuous with respect to the uniform topology on $\SM{\infty}^G$. Following \cite[Prop.~2.4.2]{Dieck2008}, we call $F^\wedge$ the \emph{adjoint} of $F$.

\begin{proposition}\label{prop:TV_lifting}
    Let $O \subset \cE_U$ and let $F:G \times O \to \SM{\infty}$ be a (not necessarily measurable) lift of the continuous function $G \times O \mapsto \PM{\infty}$, $(g, A) \mapsto \Targ{A}^{1/2} \bbVarg{g,A}$. For each $(g, A) \in G \times O$, let $\Varg{g,A}$ be the unique lift of $\bbVarg{g,A}$ such that
    \[
    F(g,A) = \Targ{A}^{1/2} \Varg{g,A}.
    \]
    If the map $O \to \SM{\infty}^G$, $A \mapsto F(\,\cdot \, , A)$ is continuous with respect to the uniform topology on $\SM{\infty}^G$, then the 2-cocycle $\eta_A(g, h)$ associated to $\Varg{g,A}$ is continuous as a map
    \begin{equation}\label{eq:eta_adjoint}
    O \to \Unitary(1)^{G \times G}, \quad A \mapsto \eta_A,
    \end{equation}
    where $\Unitary(1)^{G \times G}$ is given the uniform topology.
\end{proposition}

\begin{proof}
    Given $A \in O$ and $g,h \in G$, the fact that $[\Targ{A}^{1/2},\Varg{g,A}] =0$ implies
    \[
    F(g,A)F(h,A) = \eta_A(g,h)\Targ{A}^{1/2}F(gh,A).
    \]
    Note also that
    \[
    \norm{F(g,A)F(h,A)}_2 = \norm{\Targ{A}}_2 \leq 1,
    \]
    where the norms are Hilbert-Schmidt norms.
    Thus,
    \begin{equation}\label{eq:eta_F}
    \eta_A(g,h) = \frac{\ev{\Targ{A}^{1/2}F(gh,A), F(g,A)F(h,A)}_2}{\norm{\Targ{A}}_2},
    \end{equation}
    where the inner product in the numerator is a Hilbert-Schmidt inner product.
    Using the facts that $\Targ{A}^{1/2}$ and $\norm{\Targ{A}}^{-1}_2$ are continuous in $A$ and $F(g,A)$ is continuous as a function $O \to \SM{\infty}^G$, one can check that $\eta$ is a continuous function $O \to \Unitary(1)^{G \times G}$ by making routine estimates of $\abs{\eta_A(g, h) - \eta_{A_0}(g,h)}$ for $A, A_0 \in O$ and $g, h \in G$.
\end{proof}

\begin{proposition}\label{prop:Borel_lifting}
    Let $X$, $Y$, $E$, and $B$ be topological spaces, let $X$ be compact Hausdorff, and let $E$ be a metric space with bounded metric $d$. Let $p:E \to B$ be a function that admits a continuous local sections in a neighborhood of every point of $B$. If $f:X \times Y \to B$ is continuous, then for every $y_0 \in Y$, there exists a neighborhood $y_0 \in O \subset Y$ and a Borel measurable lift $F:X \times O \to E$ of $f|_{X \times O}$ such that the adjoint function
    \begin{equation}\label{eq:adjoint_function}
    F^\wedge: O \to E^X, \quad F^\wedge(y)(x) = F(x,y)
    \end{equation}
    is continuous in the uniform topology on $E^X$.
\end{proposition}

\begin{proof}
    Fix $y_0 \in Y$. For each $x \in X$, there exists an open set $W_x \subset B$ containing $f(x,y_0)$ and a continuous local section $s_x:W_x \to E$ of $p$. Since $f$ is continuous, there exist open sets  $x \in U_x \subset X$ and $y_0 \in O_x \subset X$ such that $f(U_x \times O_x) \subset W_x$. There exists an open set $V_x \subset X$ such that $x \in V_x \subset \overline{V_x} \subset U_x$. Since $X$ is compact, there exist $x_1,\ldots, x_n \in X$ such that $X = V_{x_1} \cup \cdots \cup V_{x_n}$. Define $O = O_{x_1} \cap \cdots \cap O_{x_n} \ni y_{0}$.

    Let $V_1' = V_{x_1}$ and for $1 < i \leq n$ let $V_{i}' = V_{x_i} \setminus \bigcup_{j < i} V_{x_j}$. Then the sets $V_1' \times O,\ldots, V_n' \times O$ form a partition of $X \times O$ into Borel measurable sets. We define $F:X \times O \to E$ piecewise by setting
    \[
    F(x,y) = s_{x_i}(f(x, y))
    \]
    where $i \in \{1,\ldots, n\}$ is the unique number such that $x \in V_i'$. Clearly $F$ is continuous on each set $V_i' \times O$, so $F$ is Borel measurable. It is also clear that $F$ is a lift of $f|_{X \times O}$.

    It remains to show that \eqref{eq:adjoint_function} is continuous. First, define $K_i = \overline{V_i'}$ and observe that each function $F_i: K_i \times O \to E$, $F_i(x,y) = s_{x_i}(f(x,y))$ is well-defined and continuous, hence each adjoint function
    \[
    F_i^\wedge:O \to C(K_i, E), \quad F_i^\wedge(y)(x) = F_i(x,y)
    \]
    is continuous with respect to the compact-open topology on $C(K_i, E)$. Since $K_i$ is compact Hausdorff, the compact-open topology coincides with the uniform topology on $C(K_i, E)$.

    Thus, given $y \in O$ and $\varepsilon > 0$, there exists a neighborhood $y \in O' \subset O$ such that for all $z \in O'$ and $i \in \{1,\ldots, n\}$ we have
    \[
    \sup_{x \in K_i}  d(F_i^\wedge(y)(x), F_i^\wedge(z)(x)) < \varepsilon.
    \]
    It follows that for all $z \in O'$ and $x \in X$, we have
    \[
    d(F^\wedge(y)(x), F^\wedge(z)(x)) \leq \max_{i} \sup_{x \in K_i} d(F_i^\wedge(y)(x),  F_i^\wedge(z)(x)) < \varepsilon
    \]
    Taking a supremum of the right hand side over all $x \in X$ yields
    \[
    \sup_{x \in X} d(F^\wedge(y)(x), F^\wedge(z)(x))  < \varepsilon.
    \]
    This proves that $F^\wedge$ is continuous at the arbitrary point $y$.
\end{proof}

\begin{proposition}\label{prop:xi_lift_continuous_cocycle}
    Let $O \subset \cB_U$ be open and let $\hat \xi:G \times O \to \SM{\infty}$ be a lift of $\xi$ such that the associated function $O \to \SM{\infty}^G$ is continuous with respect to the uniform topology on $\SM{\infty}^G$. Then the 2-cocycle induced by $\hat \xi$ is continuous as a function $\eta:O \to \Unitary(1)^{G \times G}$
\end{proposition}

\begin{proof}
    Let $O$ and $\hat \xi$ satisfy the hypotheses of the theorem. Since $O$ is open, the subspace topology on $O$ obtained from $\cB_U$ coincides with the quotient topology obtained from the projection $p^{-1}(O) \to O$. Thus, it suffices to show that the map $O \to \Unitary(1)^{G\times G}$ is continuous when composed with this projection.

    Given $(g, A) \in G \times p^{-1}(O)$, let $\Varg{g,A}$ be the unique lift of $\bbVarg{g,A}$ such that
    \[
    \kappa(g, p(A))\hat \xi(g, p(A)) = \Tr[\Targ{A} \Varg{g,A} A^iA^{j*}].
    \]
    We claim that
    \begin{equation}\label{eq:adjoint_function_TV}
    p^{-1}(O) \to \SM{\infty}^G, \quad A \mapsto \left( g \mapsto \Targ{A}^{1/2} \Varg{g,A} \right)
    \end{equation}
    is continuous with respect to the uniform topology on $\SM{\infty}^G$. The result will then follow immediately from \zcref{prop:TV_lifting}.

    Fix $A_0 \in p^{-1}(O)$; we will prove continuity of \eqref{eq:adjoint_function_TV} at $A_0$. By \zcref{prop:Borel_lifting} applied to $X = G$, $Y = \cE_U$, $B = \PM{\infty}$, $E = \SM{\infty}$, and $f$ as the map $(g, A) \mapsto \Targ{A}^{1/2} \bbVarg{g,A}$, we know there exists a neighborhood $A_0 \in O' \subset p^{-1}(O)$ and a Borel measurable lift $F:G \times O' \to \SM{\infty}$ of $(g,A) \mapsto \Targ{A}^{1/2} \bbVarg{g,A}$ such that the adjoint function $F^\wedge:O' \to \SM{\infty}^G$, $F^\wedge(A)(g) = F(g,A)$ is continuous with respect to the uniform topology. For each $(g, A) \in G \times O'$, there exists a unique lift $\Varg{g,A}'$ of $\bbVarg{g,A}$  such that
    \[
    F(g,A) = \Targ{A}^{1/2} \Varg{g,A}'.
    \]
    Since $\Varg{g,A}$ and $\Varg{g,A}'$ both represent $\bbVarg{g,A}$, there exists a unique function $\gamma: G \times O' \to \Unitary(1)$ such that
    \[
    \Varg{g,A} = \gamma(g, A) \Varg{g,A}'
    \]
    for all $(g, A) \in G \times O'$. Since
    \[
    \Targ{A}^{1/2} \Varg{g,A} = \gamma(g,A)F(g,A),
    \]
    it suffices to prove that $\gamma^\wedge : O' \to \Unitary(1)^G$, $\gamma^\wedge(A)(g) = \gamma(g,A)$ is continuous in the uniform topology.

    To do this, define a function $\zeta : G \times O' \to \SM{\infty}$ by setting
    \[
    \zeta(g, A) = \Tr[\Targ{A}\Varg{g,A}'A^i A^{j*} ] = \overline{\gamma(g, A)} \kappa(g, p(A)) \hat \xi(g, p(A)).
    \]
    Then
    \[
    \gamma(g,A) = \frac{\ev{\zeta(g,A), \hat \xi(g,p(A))}_2}{\kappa(g, p(A))}.
    \]
    Thus, if we can show that $\zeta^\wedge :O' \to \SM{\infty}^G$, $\zeta^\wedge(A)(g) = \zeta(g,A)$ is continuous, then we are done.

    To see that $\zeta^\wedge$ is continuous, given $g \in G$ and $A_1, A_2 \in O'$, we estimate the difference $\Delta \defeq \norm{\zeta(g,A_1) - \zeta(g, A_2)}_2^2$ as:
    \begin{align*}
        \Delta &= \sum_{i,j} \abs{\Tr[\Targ{A_1}\Varg{g,A_1}' A_1^iA_1^{j*}] - \Tr[\Targ{A_2} \Varg{g,A_2}' A_2^iA_2^{j*}]}^2\\
        &= \sum_{i,j} \abs{\ev{\Targ{A_1}^{1/2}A_1^j A_1^{i*}, F(g, A_1)} - \ev{\Targ{A_2}^{1/2}A_2^j A_2^{i*}, F(g, A_2)} }^2\\
        &\leq \sum_{i,j} \left( \norm{\Targ{A_2}^{1/2}A_2^jA_2^{i*}}_2 \norm{F(g, A_1) - F(g, A_2)}_2\right.\\
        &\qquad \qquad \qquad\qquad \qquad \left. +  \norm{\Targ{A_1}^{1/2}A_1^jA_1^{i*} - \Targ{A_2}^{1/2}A^j_2 A_2^{i*}}_2 \right)^2
    \end{align*}
    Expanding out the square, there are three terms to consider. The term
    \[
    \sum_{i,j} \norm{\Targ{A_1}^{1/2}A_1^j A_1^{i*} - \Targ{A_2}^{1/2} A_2^j A_2^{i*}}^2_2
    \]
    may be made arbitrarily small as $A_1 \to A_2$ since the function
    \[
    \cE_U \to [0,\infty), \quad A \mapsto \sum_{i,j} \norm{\Targ{A}^{1/2}A^j A^{i*} -\Targ{A_2}^{1/2} A_2^{j}A_2^{i*}}_2^2
    \]
    is continuous. The term
    \[
    \norm{F(g, A_1) - F(g, A_2)}_2^2 \sum_{i,j} \norm{\Targ{A_2}^{1/2}A_2^j A_2^{i*}}_2^2
    \]
    may be made arbitrarily small as $A_1 \to A_2$, uniformly in $g$, since $F^\wedge$ is continuous. By similar arguments, the final term
    \begin{align*}
    2 \norm{F(g, A_1) - F(g, A_2)}_2 \sum_{i,j} &\norm{\Targ{A_2}^{1/2}A_2^j A_2^{i*}}_2 \\
    &\qquad \cdot \norm{\Targ{A_1}^{1/2} A_1^j A_1^{i*} - \Targ{A_2}^{1/2}A_2^j A_2^{i*}}_2
    \end{align*}
    can be made arbitrarily small as $A_2 \to A_1$. Thus, for arbitrary $A_2 \in O'$ and $\varepsilon > 0$, there exists a neighborhood of $A_2$ on which $\norm{\zeta(g, A_1) - \zeta(g, A_2)}_2 < \varepsilon$ for all $A_1$ in that neighborhood. This proves that $\hat \zeta$ is continuous, concluding the proof.
\end{proof}

A function $\gamma:G^n \to \Unitary(1)$ is \emph{normalized} if $\gamma(g_1,\ldots, g_n) = 1$ whenever $g_i = 1$ for at least one $i$.
Thus, a 2-cocycle $\eta:G \times G \to \Unitary(1)$ is \emph{normalized} if $\eta(g, 1) = \eta(1, g) = 1$ for all $g \in G$.
In the theorem below, $\Zgrp^2(G;\Unitary(1))$ denotes the set of  normalized 2-cocycles that are measurable with respect to the Borel $\sigma$-algebra of the product topology on $G \times G$ and the Borel $\sigma$-algebra on $\Unitary(1)$.
We equip $\Zgrp^2(G;\Unitary(1))$ with the uniform topology.

\begin{proposition}\label{prop:lift_with_normalized_2cocycle}
    Let $\omega_0 \in \cB_U$. There exists an open set $\omega_0 \in O \subset \cB_U$ and a Borel measurable (relative to the Hilbert-Schmidt topology) lift $\hat \xi:G \times O \to \SM{\infty}$ of $\xi$ such that the adjoint function $\hat \xi^\wedge:O \to \SM{\infty}^G$ is continuous and such that each associated 2-cocycle $\eta_\omega$, $\omega \in O$, is normalized and Borel measurable, and has continuous adjoint function $\eta:O \to \Zgrp^2(G;\Unitary(1))$.
\end{proposition}

\begin{proof}
    Given $\omega_0 \in \cB_U$, \zcref{prop:Borel_lifting} (applied to $X = G$, $Y = \cB_U$, $E = \SM{\infty}$ with the Hilbert-Schmidt topology, $B = \PM{\infty}$ with the quotient of the Hilbert-Schmidt topology, and $f = \xi$) yields an open set $\omega_0 \in O \subset \cB_U$ and a Borel measurable lift $\hat \xi:G \times O \to \SM{\infty}$ such that the adjoint map $O \to \SM{\infty}^G$ is continuous with respect to the uniform topology on $\SM{\infty}^G$. Each 2-cocycle $\eta_\omega$ obtained from $\hat \xi$ is Borel measurable by \zcref{prop:pointwise_Borel_measurable_cocycle}. Furthermore, $\eta$ is continuous as a map $O \to \Unitary(1)^{G \times G}$ by \zcref{prop:xi_lift_continuous_cocycle}.

    The only problem is that $\eta$ might not be normalized. To rectify this, observe that $\Varg{1,A} = \eta_\omega(1, 1)Q(A)$ for all $\omega \in O$ and all $A \in \cE_U$ such that $p(A) = \omega$, where $\Varg{1,A}$ is the unique lift of $\bbVarg{1,A}$ such that \eqref{eq:V_from_hatxi} holds. The function $\omega \mapsto \eta_\omega(1, 1)$ is continuous on $O$ and therefore the lift
    \[
    \hat \xi'(g, \omega) \defeq \overline{\eta_\omega(1,1)}\hat \xi(g, \omega)
    \]
    is a Borel measurable function $\hat \xi':G \times O \to \SM{\infty}$ such that the adjoint function $O \to \SM{\infty}^G$, $\omega \mapsto (g \mapsto \hat \xi'(g, \omega))$ is continuous. The 2-cocycle $\eta_\omega'$ associated to $\hat \xi'$ is
    \[
    \eta_\omega'(g,h) = \overline{\eta_\omega(1,1)} \eta_\omega(g,h).
    \]
    This is normalized, so $\hat \xi'$ satisfies the requirements of the proposition.
\end{proof}

In the theorem below, we let $\Cgrp^1(G;\Unitary(1))$ denote the set of normalized 1-cochains, i.e., normalized Borel measurable functions $G \to \Unitary(1)$.
We equip $\Cgrp^1(G;\Unitary(1))$ with the uniform topology.

\begin{theorem}\label{thm:adjust_cocycle_to_constant}
    Let $\eta_0 \in \Zgrp^2(G;\Unitary(1))$ and let $B_{\sqrt{2}}(\eta_0)$ be the open ball of radius $\sqrt{2}$ centered on $\eta_0$ with respect to the uniform metric. There exists a continuous map $B_{\sqrt{2}}(\eta_0) \to \Cgrp^1(G;\Unitary(1))$, $\eta \mapsto \beta_\eta$ such that
    \[
    \eta(g, h) = \frac{\beta_\eta(g)\beta_\eta(h)}{\beta_\eta(gh)} \eta_0(g,h)
    \]
    for all $\eta \in B_{\sqrt{2}}(\eta_0)$ and all $g, h \in G$.
\end{theorem}

\begin{proof}
    If $\eta \in B_{\sqrt{2}}(\eta_0)$, then for all $g, h \in G$ we have
    \[
    \abs{1 - \eta(g,h)\eta_0(g,h)^{-1}} < \sqrt{2},
    \]
    which implies that $\Re \eta(g,h)\eta_0(g,h)^{-1} > 0$. Thus, we may define a function $\theta: B_{\sqrt{2}}(\eta_0) \to (-\pi/2, \pi/2)^{G \times G}$, $\eta \mapsto \theta_\eta$ by
    \[
    \eta(g,h)\eta_0(g,h)^{-1} = e^{i\theta_\eta(g,h)}.
    \]
    Furthermore, it is elementary to check that $\theta_\eta$ is Borel measurable for each $\eta$ and $\theta$ is continuous when $(-\pi/2, \pi/2)^{G \times G}$ is given the uniform metric. Note that the cocycle condition on $\eta$ and $\eta_0$ implies
    \[
    \theta_\eta(f,g) + \theta_\eta(fg,h) = \theta_\eta(f, gh) + \theta_\eta(g,h)
    \]
    for all $\eta \in B_{\sqrt{2}}(\eta_0)$ and $g, h \in G$, and the normalization condition on $\eta$ and $\eta_0$ implies
    \[
    \theta_\eta(g,1) = \theta_\eta(1, g) = 0
    \]
    for all $\eta \in B_{\sqrt{2}}(\eta_0)$ and $g \in G$.

    Next we integrate out one of the variables of $\theta$ using the normalized Haar measure on $G$. We define $\alpha:B_{\sqrt{2}}(\eta_0) \to (-\pi/2, \pi/2)^G$, $\eta \mapsto \alpha_\eta$ by
    \[
    \alpha_\eta(g) = \int \theta_\eta(g,h) \dd{h}.
    \]
    Then $\alpha_\eta$ is measurable\footnote{Note there is a subtlety if $G$ is not second-countable. In this case, we have only argued that $\theta_\eta$ is measurable with respect to the Borel $\sigma$-algebra on $G \times G$, not the $\sigma$-algebra generated by measurable rectangles, so the ordinary Fubini  theorem (e.g, \cite[Thm.~8.8]{paparudin}) cannot be used. The cited theorem \cite[Thm.~7.6.7]{Cohn_Measure_Theory} provides a sufficient replacement, fortunately.} by  \cite[Thm.~7.6.7]{Cohn_Measure_Theory} and $\alpha$ is continuous when $(-\pi/2, \pi/2)^G$ is given the uniform metric. Also note that $\alpha_\eta(1) = 0$ since $\theta_\eta(1, h) = 0$ for all $h \in G$.

    We observe that
    \begin{align*}
        \alpha_\eta(f) + \alpha_\eta(g) - \alpha_\eta(fg) &= \int \left[\theta_\eta(f,h) + \theta_\eta(g,h) - \theta_\eta(fg,h)\right] \dd{h}\\
        &= \int \left[ \theta_\eta(f,h) + \theta_\eta(f, g) - \theta_\eta(f, gh) \right] \dd{h}\\
        &= \theta_\eta(f,g),
    \end{align*}
    where we have used the invariance and normalization of the Haar measure in the last step.

    Finally, define $\beta:B_{\sqrt{2}}(\eta_0) \to \Cgrp^1(G; \Unitary(1))$ by
    \[
    \beta_\eta(g) = e^{i\alpha_\eta(g)},
    \]
    Observe that $\beta_\eta$ is Borel measurable for each $\eta \in B_{\sqrt{2}}(\eta_0)$ and $\beta$ is continuous with respect to the uniform topology on $\Cgrp^1(G;\Unitary(1))$. Moreover, we have
    \begin{align*}
        \frac{\beta_\eta(g)\beta_\eta(h)}{\beta_\eta(gh)} &= e^{i\theta_\eta(g,h)} = \eta(g,h)\eta_0(g,h)^{-1},
    \end{align*}
    as desired.
\end{proof}

\begin{proposition}\label{prop:flatten}
    Let $\omega_0 \in \cB_U$. There exists an open set $\omega_0 \in O \subset \cB_U$ and a Borel measurable (relative to the Hilbert-Schmidt topology) lift $\hat \xi:G \times O \to \SM{\infty}$ of $\xi$ such that the associated function $O \to \SM{\infty}^G$ is continuous and the associated 2-cocycle $\eta_\omega$ is normalized, Borel measurable, and constant as a function of $\omega$.
\end{proposition}

\begin{proof}
    Let $O$ be an open set containing $\omega_0$ and let $\hat \xi:G \times O \to \SM{\infty}$ be as in \zcref{prop:lift_with_normalized_2cocycle}. Since the associated 2-cocycle is continuous as a function $O \to \Zgrp^2(G;\Unitary(1))$, we may shrink $O$ if necessary so that for all $\omega \in O$, we have that $\eta_\omega$ is in the open ball of radius $\sqrt{2}$ centered on $\eta_{\omega_0}$.
    By \zcref{thm:adjust_cocycle_to_constant}, there exists a continuous map $\beta:O \to \Cgrp^1(G;\Unitary(1))$, $\omega \mapsto \beta_\omega$ such that
    \[
    \eta_\omega(g, h) = \frac{\beta_\omega(g)\beta_\omega(h)}{\beta_\omega(gh)} \eta_{\omega_0}(g,h)
    \]
    for all $\omega \in O$ and $g,h \in G$. Now we may define a lift $\hat \xi':G \times O \to \SM{\infty}$  by
    \[
    \hat \xi'(g, \omega) = \overline{\beta_\omega(g)} \hat \xi(g, \omega).
    \]
    One can then check that the adjoint function $O \to \SM{\infty}^G$, $\omega \mapsto (g \mapsto \hat \xi'(g, \omega))$ is continuous and the associated 2-cocycle is $\eta'_\omega(g,h) = \eta_{\omega_0}(g,h)$ for all $\omega \in O$.

    It remains to argue that $\hat \xi'$ is Borel measurable. If we can show that the associated function $O \times G \to \Unitary(1)$, $(\omega, g) \mapsto \beta_\omega(g)$ is Borel measurable, then the result will follow since $\hat \xi$ is Borel measurable and the product of these measurable functions is Borel measurable by an argument similar to that at the end of the proof of \zcref{prop:pointwise_Borel_measurable_cocycle}.

    To see that $(\omega, g) \mapsto \beta_\omega(g)$ is Borel measurable, we refer the reader to \zcref{cor:adjoint_evaluation_measurable}. The only thing that remains to be checked before applying \zcref{cor:adjoint_evaluation_measurable} is that $O$ is Lindel\"of. Since continuous surjective images of Lindel\"of spaces are Lindel\"of, it suffices to check that $p^{-1}(O)$ is Lindel\"of. We may write $p^{-1}(O)$ as a countable union $p^{-1}(O) = \bigcup_{d,D} p^{-1}(O) \cap \cE_U(d,D)$. Each space $p^{-1}(O) \cap \cE_U(d, D)$ is a subspace of the finite-dimensional topological vector space $\cM(d, D)$ is therefore metrizable and second-countable, hence Lindel\"of. Thus, $p^{-1}(O)$ is Lindel\"of as it is the countable union of Lindel\"of spaces. Thus, $O$ is Lindel\"of, and we may now apply \zcref{cor:adjoint_evaluation_measurable} to conclude the proof.
\end{proof}

Let $\omega \in \cB_U$ and let $\hat \xi_1:G \times \{\omega\} \to \SM{\infty}$ and $\hat \xi_2:G \times \{\omega\} \to \SM{\infty}$ be two Borel measurable lifts of $\xi$ that yield normalized 2-cocycles $\eta_1$ and $\eta_2$. Choose $A \in \cE_U$ such that $p(A) = \omega$ and let $\Varg[1]{g,A}$ and $\Varg[2]{g, A}$ be the lifts of $\bbVarg{g,A}$ associated to $\hat \xi_1$ and $\hat \xi_2$. Since $\Varg[1]{g,A}$ and $\Varg[2]{g,A}$ are both lifts of $\bbVarg{g,A}$, we know that for each $g \in G$ there exists $\beta(g) \in \Unitary(1)$ such that $\Varg[1]{g,A} = \beta(g)\Varg[2]{g,A}$. This implies that
\[
    \eta_1(g,h) = \frac{\beta(g)\beta(h)}{\beta(gh)} \eta_2(g,h).
\]
Furthermore, since
\[
    \beta(g) = \ev{\hat \xi_2(g, \omega), \hat \xi_1(g,\omega)},
\]
we see that $\beta$ is Borel measurable. Thus, $\eta_1$ and $\eta_2$ are in the same cohomology class in $\Hgrp^2(G;\Unitary(1))$. This cohomology class is therefore a well-defined function $\cB_U \to \Hgrp^2(G;\Unitary(1))$, i.e., it is independent of the chosen lift $\hat \xi$.

This cohomology class is the well-known \emph{SPT index} for 1d phases, adapted to the classifying space $\cB_U$, and we can now easily recover the well-known fact that it is a phase invariant of systems parametrized by a point $X = \{*\}$. This result was obtained for matrix product states in~\cite{PollmannEntanglementSpectrum,PollmannBergTurnerOshikawa12,ChenGuWen2011_1,ChenGuWen2011_2,SchuchClassifyingQuantumPhasesMPS}, and in the operator-algebraic setting, for finite $G$ and pure split states, by Ogata~\cite{Ogata20,Ogata21}.
In our context, this takes the form of the following corollary.

\begin{corollary}[{cf.~\cite{SchuchClassifyingQuantumPhasesMPS}}]\label{cor:SPT_locally_constant}
    The cohomology class $\cB_U \to \Hgrp^2(G;\Unitary(1))$ is constant on each connected component of $\cB_U$.
    In particular, states with different SPT indices lie in different connected components of $\cB_U$.
\end{corollary}
\begin{proof}
    \zcref{prop:flatten} shows that the cohomology class is locally constant.
    Locally constant functions are constant on connected components.
\end{proof}

\begin{remark}[SPT index]\label{rmk:SPT_phase}
    \zcref{cor:SPT_locally_constant} provides a well-defined function $\pi_0(\cB_U)\to \Hgrp^2(G;\Unitary(1))$.
    We do not address here whether this map is injective or surjective, although we expect it to be a bijection when $U$ is a complete $G$-universe, and it has been shown to be a bijection in the contexts considered in other works \cite{SchuchClassifyingQuantumPhasesMPS,Ogata21}.

    In \zcref{sec:invariant_def} we construct a degree-one invariant which lives independently of the SPT index: a class $\Phi\in\check H^1(\cB_U;\widehat G)$.
    Pulling back along loops yields a  homomorphism $\Lambda_\Phi : \pi_1(\cB_U)\to\widehat G$, $[\gamma]\mapsto\gamma^*\Phi$, which is well defined since homotopy invariance of \v{C}ech cohomology only requires paracompactness of the domain.
    For a parametrized family $\psi:X \to \cB_U$ on a connected CW complex $X$, the parametric invariant $\psi^*\Phi$ corresponds to $\Lambda_\Phi\circ\psi_*$ under the natural isomorphism $\check H^1(X;\widehat G)\cong\Hom(\pi_1(X),\widehat G)$.
    Morally, the SPT index sees $\pi_0$ and the Thouless invariant sees $\pi_1$ of the space of $G$-symmetric states.
\end{remark}

\subsection{Definition of the \v{C}ech Cocycle}\label{sec:invariant_def}
We now use the results from \zcref{sec:construction_of_lifts} to construct an element of the \v{C}ech cohomology group $\check{H}^1(X;\widehat{G})$ from a parametrized family $\psi:X \to \sQ_W(\sH)$, where $X$ is an arbitrary topological space. This class will then be shown to depend only on the phase of $\psi$. If we choose a representation $U$ on $\bbC^\infty$ that admits an isometric intertwiner $\kappa:\sH \to \bbC^\infty$, then we get an embedding $\kappa_*:\sQ_W(\sH) \to \cB_U$ that we may compose with $\psi$ to get a continuous map $\kappa_*\psi:X \to \cB_U$. A continuous map $X \to \cB_U$ is the only data needed to define the class in $\check{H}^1(X;\widehat{G})$, so will temporarily forget about the intermediate space $\sQ_W(\sH)$ and just assume we are given a topological space $X$ and a continuous function $f:X \to \cB_U$ for any arbitrary representation $U$ on $\bbC^\infty$. In particular, we could take $X = \cB_U$ and $f = \id$.

We have so far used $\hat \xi$ to denote a function $G \times O \to \SM{\infty}$, where $O \subset \cB_U$, such that $\hat \xi(g, x)$ is a lift of $\xi(g,x)$. In what follows, it will be useful to use $\hat \xi$ to denote a function $G \times O \to \SM{\infty}$, where $O \subset X$, and $\hat \xi(g, x)$ is a lift of $\xi(g,f(x))$. Observe that such a $\hat \xi$ defined on $G \times O$, where $O \subset X$, still determines a 2-cocycle $\eta_x$ for each $x \in O$ as well as partial isometries $\Varg{g,A_x}$ for any choice of lift $A_x \in \cE_U$ such that $p(A_x) = f(x)$.

For each element of $[\eta] \in \Hgrp^2(G;\Unitary(1))$, we pick once and for all a representative $\eta \in \Zgrp^2(G;\Unitary(1))$.
Given $x \in X$, it follows by taking preimages of the open sets from \zcref{prop:flatten} that there exists a neighborhood $x \in O \subset X$ and a Borel measurable (with respect to the Hilbert-Schmidt topology on $\SM{\infty}$) function $\hat \xi:G \times O \to \SM{\infty}$ such that $\hat \xi(g, x)$ is a lift of $\xi(g, f(x))$ for every $x \in O$,
such that the associated function $O \to \SM{\infty}^G$ is continuous, and such that the associated 2-cocycle is normalized, Borel measurable, and constant as a function of $x \in O$. Multiplying $\hat \xi$ with a Borel measurable function $\beta :G \to \Unitary(1)$ if necessary, we can enforce that the associated 2-cocycle $\eta_x$ is the chosen representative of the class $[\eta_x]$.

Thus, $X$ admits an open cover $\cO = (O_\alpha)_\alpha$ and Borel measurable functions $\hat \xi_\alpha : G \times O_\alpha \to \SM{\infty}$ such that $\hat \xi_\alpha(g,x)$ is a lift of $\xi(g,f(x))$ for every $x \in O_\alpha$,
such that the associated function $O_\alpha \to \SM{\infty}^G$ is continuous, and such that the associated 2-cocycle is normalized, Borel measurable, constant and equal to one of the chosen representatives.

\begin{definition}\label{def:phi}
    In the above setting, consider domains $O_\alpha$ and $O_\beta$ such that $O_\alpha \cap O_\beta \neq \varnothing$.
    We define
    \begin{equation}\label{eq:def-phi}
        \phi_{\alpha \beta} : O_\alpha \cap O_\beta \to \widehat{G},
        \quad
        \phi_{\alpha \beta}(x)(g)
        = \ev{\hat \xi_\alpha(g, x), \hat \xi_\beta(g, x)},
    \end{equation}
    where \( \widehat{G} \) denotes continuous group homomorphisms \( G \to \Unitary(1) \).
    More transparently, if $A \in \cE_U$ such that $p(A) = f(x)$ and $\Varg[\alpha]{g, A}$ and $\Varg[\beta]{g, A}$ are the partial isometries associated to $\hat \xi_\alpha$ and $\hat \xi_\beta$, respectively, then
    \begin{equation}\label{eq:more_explicit_phi}
        \phi_{\alpha \beta}(x)(g)Q(A) = \Varg[\alpha]{g, A}^* \Varg[\beta]{g, A}.
    \end{equation}
    This equation is useful for obtaining the invariant in practical computations, but~\eqref{eq:def-phi} makes it easier to infer good properties of \( \phi \).

    Equation~\eqref{eq:more_explicit_phi} follows from the fact that \( V_{\alpha}(g,A) \) and \( V_{\beta}(g,A) \), being lifts of the same matrix to a partial isometry, must differ by a phase.
    Using~\eqref{eq:def-phi}, one sees that \( V_{\beta}(g,A) = V_{\alpha}(g,A) \phi_{\alpha \beta}(x)(g) \), from which~\eqref{eq:more_explicit_phi} follows using the general form of \( V \) as in~\eqref{eq:bigV_smallv_relation}.
\end{definition}

\begin{remark}\label{rmk:residual_freedom}
    The 2-cocycle that we fix on each patch is a representative of the SPT index $[\eta_\omega]$ discussed at the end of \zcref{sec:construction_of_lifts}.
    Once it is fixed, the only remaining freedom in the choice of a lift is a character: if $\hat\xi(\cdot, \omega)$ and $\hat\xi'(\cdot, \omega)$ are Borel measurable lifts of $\xi(\cdot,\omega)$ with the same
    associated 2-cocycle, then $\hat\xi'(g,\omega)=\nu(g)\hat\xi(g,\omega)$ for some $\nu\in\widehat G$.
    The functions $\phi_{\alpha\beta}$ record precisely how this residual $\widehat G$-ambiguity varies between patches.
\end{remark}

\begin{theorem}\label{thm:universal-class}
    The collection of all $(\phi_{\alpha \beta})_{\alpha,\beta}$ in \zcref{def:phi} defines a Čech class
    \( \Phi \defeq i([\phi]) \in \check{H}^{1}(X;\widehat{G}) \),
    depending only on the continuous function $f:X \to \cB_U$ and independent of all the choices made for the 2-cocycles $\eta$, the open cover $\cO = (O_\alpha)_\alpha$, and the lifts $\hat \xi_\alpha$.
    Here, \( i \) is the canonical inclusion $\check{H}^1(\cO;\widehat{G}) \to \check{H}^1(X;\widehat{G})$.
\end{theorem}

\begin{proof}
    \emph{Continuous Homomorphism.}
    First we check that  \( g \mapsto \phi_{\alpha \beta}(x)(g) \) is a continuous group homomorphism.
    Using~\eqref{eq:more_explicit_phi} and the fact that the 2-cocycles associated to $\Varg[\alpha]{g, A}$ and $\Varg[\beta]{g, A}$ are the same, we compute
    \begin{align*}
        \phi_{\alpha \beta}(x)(gh) Q(A)
        &= V_{\alpha}(gh,A)^{*} V_{\beta}(gh,A) \\
        &=
        \left(
            \overline{\eta_{x}(g,h)}
            V_{\alpha}(g,A)V_{\alpha}(h,A)
        \right)^{*}
        \overline{\eta_{x}(g,h)}
        V_{\beta}(g,A)V_{\beta}(h,A) \\
        &= V_{\alpha}(h,A)^{*} \left(V_{\alpha}(g,A)^{*} V_{\beta}(g,A)\right) V_{\beta}(h,A) \\
        &= \phi_{\alpha \beta}(x)(g)
        V_{\alpha}(h,A)^{*} Q(A) V_{\beta}(h,A) \\
        &= \phi_{\alpha \beta}(x)(g)
        Q(A) V_{\alpha}(h,A)^{*} V_{\beta}(h,A) \\
        &= \phi_{\alpha \beta}(x)(g) \phi_{\alpha \beta}(x)(h) Q(A).
    \end{align*}
    In the second-to-last line we have furthermore used that \( V_{\alpha}(h,A) \) and \( Q(A) \) commute, and in the last line the fact that \( Q(A) \) is idempotent.
    This shows that $\phi_{\alpha\beta}(x)$ is indeed a homomorphism.

    Since $\hat \xi_\alpha$ and $\hat \xi_\beta$ are Borel measurable, we see that $\phi_{\alpha \beta}(x)$ is also Borel measurable.
    Indeed this can be seen by observing that $\phi_{\alpha \beta}(x)$ is the composition of the map
    $G \to \SM{\infty} \times \SM{\infty}$, $g \mapsto (\hat \xi_\alpha(g,x), \hat \xi_\beta(g,x))$
    with the inner product $\SM{\infty} \times \SM{\infty} \to \bbC$. The former is measurable with respect to the $\sigma$-algebra generated by measurable rectangles, and the inner product $\SM{\infty} \times \SM{\infty} \to \bbC$ is continuous, hence measurable with respect to the Borel $\sigma$-algebra induced by the product topology on $\SM{\infty} \times \SM{\infty}$, where $\SM{\infty}$ is given the Hilbert-Schmidt topology. Since $\SM{\infty}$ is second-countable with the Hilbert-Schmidt topology, these $\sigma$-algebras on $\SM{\infty} \times \SM{\infty}$ coincide,
    so the composition $\phi_{\alpha \beta}(x)$ is measurable.
    Since Borel measurable homomorphisms between locally compact Hausdorff groups are continuous~\cite{Sasvari}, we know $\phi_{\alpha \beta}(x) \in \widehat{G}$.

    \medskip

    \emph{Cocycle condition.} Since $\hat \xi_\alpha$ and $\hat \xi_\beta$ are continuous when regarded as functions into $\SM{\infty}^G$, we see that $\phi_{\alpha \beta} : O_\alpha \cap O_\beta \to \widehat{G}$ is continuous.\footnote{
      This can be proven using the estimate
      \begin{multline*}
          \abs{
            \phi_{\alpha \beta}(x)(g) - \phi_{\alpha \beta}(x_{0})(g)
          } \\
          \leq
          \norm{
            \hat\xi_{\alpha}(g,x) - \hat\xi_{\alpha}(g,x_{0})
          } \norm{
            \hat \xi_{\beta}(g,x)
          }
          +
          \norm{
            \hat\xi_{\beta}(g,x) - \hat\xi_{\beta}(g,x_{0})
          } \norm{
            \hat \xi_{\alpha}(g,x)
          }
      \end{multline*}
      and the fact that the \( \hat \xi \) are valued in the unit sphere.
      This allows one to show continuity in the uniform topology on \( \Unitary(1)^{G} \).
      By the point above, the image of \( \phi_{\alpha \beta} \) is contained in the continuous functions \( G \to \Unitary(1) \).
      Because \( G \) is compact, the uniform topology coincides with the compact-open topology.
    }
    This shows that \( \phi \in \check{C}^{1}(\mathcal{O}; \widehat{G}) \). We remark that since $\widehat{G}$ is discrete, this implies that $\phi_{\alpha \beta}$ is locally constant.

    Finally, it is obvious from~\eqref{eq:more_explicit_phi} that
    \[
        \phi_{\alpha \beta}(x)\phi_{\beta \gamma}(x) = \phi_{\alpha \gamma}(x)
    \]
    whenever $x \in O_\alpha \cap O_\beta \cap O_\gamma$. Thus, $\phi$ is a \v{C}ech 1-cocycle $\phi \in \check{Z}^1(\cO; \widehat{G})$. Modding out by coboundaries we get an element of the degree one \v{C}ech cohomology group $\check{H}^1(\cO;\widehat{G})$.
    We may then apply the canonical inclusion $i : \check{H}^1(\cO;\widehat{G}) \to \check{H}^1(X;\widehat{G})$ to get a \v{C}ech cohomology class $ \Phi \in \check{H}^1(X;\widehat{G})$.
    It is worth noting that \( i \) is injective~\cite[Lem.~12.4]{Foster1981}.

    \medskip

    \emph{Independence from cover and lifts.} It remains to show that the class $\Phi \in \check{H}^1(X;\widehat{G})$ is independent of all choices.

    Let us first prove -- for a fixed choice of representative of each class \( \Hgrp^{2}(G;\Unitary(1)) \) -- that any two pairs \( (\mathcal{O}, \{\hat\xi_{\alpha}\}) \) and \( (\mathcal{O}', \{\hat\xi_{\alpha'}'\}) \) of open covers \( \mathcal{O} = (O_{\alpha})_{\alpha} \) and \( \mathcal{O}' = (O'_{\alpha'})_{\alpha'} \) of \( X \) and of local lifts of \( \xi \) define the same \( \Phi \in \check{H}^1(X;\widehat{G}) \).
    Let \( \phi \) and \( \phi' \) be the Čech cocycles associated to the two pairs above.
    Consider the smallest common refinement of the two open covers
    \begin{equation*}
        \mathcal{O} \vee \mathcal{O}' \defeq ( O_{\alpha \alpha'} = O_{\alpha} \cap O'_{\alpha'} )_{\alpha, \alpha'}.
    \end{equation*}
    The restrictions \( \hat \zeta_{\alpha \alpha'} \defeq \hat \xi_{\alpha}|_{G \times O_{\alpha \alpha'}} \)
    and \( \hat \zeta_{\alpha \alpha'}' \defeq \hat \xi_{\alpha'}'|_{G \times O_{\alpha \alpha'}} \)
    provide two sets of local lifts adapted to the same open cover \( \mathcal{O} \vee \mathcal{O}' \), both satisfying the thesis of \zcref{prop:flatten}
    (with 2-cocycles constantly equal to the same chosen representative for both of them) and hence define, via~\eqref{eq:def-phi}, Čech cocycles \( \varphi_{\alpha \alpha', \beta \beta'} \) and \( \varphi_{\alpha \alpha', \beta \beta'}' \) in \( \check Z^{1}(\mathcal{O} \vee \mathcal{O}'; \widehat{G}) \).
    We claim that these are cohomologous, namely that there exists \( \theta \in \check C^{0}(\mathcal{O} \vee \mathcal{O}';\widehat{G}) \) such that
    \begin{equation}\label{eq:phis-coboundary}
        \varphi'_{\alpha \alpha', \beta\beta'} = \varphi_{\alpha \alpha', \beta\beta'} \overline{\theta_{\alpha \alpha'}} \theta_{\beta \beta'}.
    \end{equation}
    Indeed, since \( \hat \zeta(g,x) \) and \( \hat \zeta'(g,x) \) are lifts of  \( \xi(g, f(x)) \), they must differ locally by a phase factor:
    \begin{equation}\label{eq:zetas-relation}
        \hat \zeta'_{\alpha \alpha'}(g,x) = \hat \zeta_{\alpha \alpha'}(g, x) \theta_{\alpha \alpha'}(g,x),
        \qquad
        \theta_{\alpha \alpha'}(g,x) \in \Unitary(1)
    \end{equation}
    and thus
    \begin{equation}\label{eq:zeta_inner_product}
        \theta_{\alpha \alpha'}(g, x) = \ev{\hat \zeta_{\alpha \alpha'}(g, x), \hat \zeta'_{\alpha \alpha'}(g, x)}.
    \end{equation}
    From \eqref{eq:zeta_inner_product} and the fact that $\hat \zeta$ and $\hat \zeta'$ are Borel measurable and have continuous adjoints, we see that the function \( \theta_{\alpha \alpha'}: G \times O_{\alpha \alpha'} \to \Unitary(1) \) is Borel measurable and has continuous adjoint $O_{\alpha \alpha'} \to \Unitary(1)^G$.
    If $O_{\alpha \alpha'} \neq \varnothing$, then the 2-cocycles associated to $\hat \zeta_{\alpha \alpha'}$ and $\hat \zeta_{\alpha \alpha'}'$ must both be equal to the 2-cocycle $\eta_x$ for any $x \in O_{\alpha \alpha'}$. From \eqref{eq:zetas-relation} it follows that
    \begin{equation*}
        \eta_{x}(g,h) = \eta_{x}(g,h) \delta\theta_{\alpha \alpha'}(g,h,x),
    \end{equation*}
    where $\delta$ is the differential in group cohomology.
    In other words, \( \theta_{\alpha \alpha'}(\cdot, x) \) is a group homomorphism and thus, being also Borel measurable, a continuous one.
    We conclude that \( \theta \in \check C^{0}(\mathcal{O} \vee \mathcal{O}'; \widehat{G}) \).
    Using~\eqref{eq:def-phi}, it is immediate to see that such a \( \theta \) satisfies the relation~\eqref{eq:phis-coboundary}.

    Finally, recall that Čech cohomologies form a directed system (over the directed set of open covers) of abelian groups with the refinement homomorphisms, and thus the following diagram commutes
    
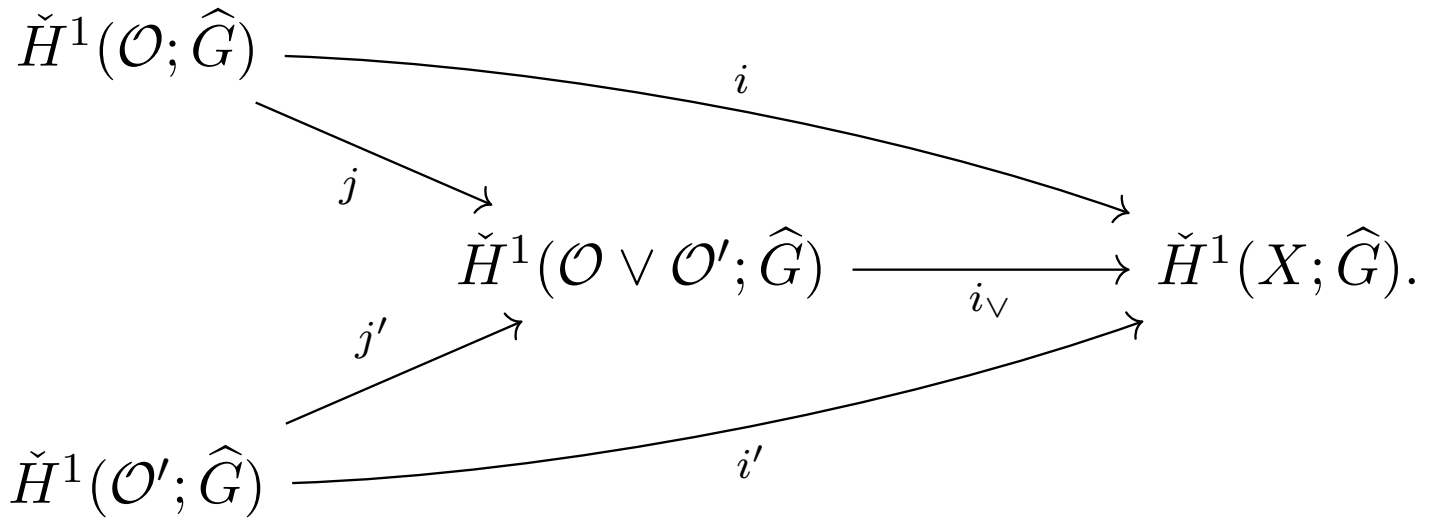
\[
        \begin{tikzcd}
            {\check H^1(\mathcal{O};\widehat{G})} &&& \\
            & {\check H^1(\mathcal{O}\vee \mathcal{O}';\widehat{G})} && {\check H^1(X;\widehat{G}).} \\
            {\check H^1(\mathcal{O}';\widehat{G})}
            \arrow["j"', from=1-1, to=2-2]
            \arrow["i", curve={height=-12pt}, from=1-1, to=2-4]
            \arrow["{i_\vee}"', from=2-2, to=2-4]
            \arrow["{j'}", from=3-1, to=2-2]
            \arrow["{i'}"', curve={height=12pt}, from=3-1, to=2-4]
        \end{tikzcd}
    \]
    Here, \( j,j' \) are the canonical maps induced by restriction, and \( i,i',i_{\vee} \) are the canonical inclusions into the direct limit.
    Observe that
    \( \hat \xi_{\alpha}|_{O_{\alpha \alpha'}\cap O_{\beta \beta'}} = \hat \zeta_{\alpha \alpha'}|_{O_{\alpha \alpha'}\cap O_{\beta \beta'}} \),
    \( \hat \xi_{\beta}|_{O_{\alpha \alpha'}\cap O_{\beta \beta'}} = \hat \zeta_{\beta \beta'}|_{O_{\alpha \alpha'}\cap O_{\beta \beta'}} \),
    and thus
    \begin{equation*}
        \phi_{\alpha \beta}|_{O_{\alpha \alpha'}\cap O_{\beta \beta'}}
        = \varphi_{\alpha \alpha', \beta \beta'},
    \end{equation*}
    with analogous statements for the primed functions.
    Together with~\eqref{eq:phis-coboundary}, this shows that \( j([\phi]) = j'([\phi']) \).
    It then follows from the commutativity of the diagram above that
    \begin{equation*}
        i([\phi]) = i_{\vee}(j([\phi])) = i_{\vee}(j'([\phi'])) = i'([\phi']),
    \end{equation*}
    thus making \( \Phi \) independent of the cover and of the lift chosen.

    \medskip

    \emph{Independence from 2-cocycles.} To conclude, let us show that \( \phi_{\alpha \beta} \) (and thus \( \Phi \)) is independent of the representative $\eta_c$ chosen for each cohomology class \(c \vcentcolon= [\eta_c] \in \Hgrp^2(G;\Unitary(1)) \). Suppose we choose a different representative $\eta'_c$ for each class $c$. Then for each class $c$ there exists a Borel measurable $\nu_c:G \to \Unitary(1)$ such that $\eta'_c = \eta_c \delta \nu_c$.
    For a given lift $\hat \xi_\alpha$ with associated class $c(\alpha)$ and 2-cocycle $\eta_{c(\alpha)}$ from the original choice of representative 2-cocycles, we can achieve a change of representative \( \eta'_{c(\alpha)} = \eta_{c(\alpha)} \delta\nu_{c(\alpha)} \) by defining
    \( \hat \xi'_{\alpha}(g,x) \vcentcolon= \hat \xi_{\alpha}(g, x) \nu_{c(\alpha)}(g) \).
    Since $\nu_{c(\alpha)}$ is a Borel-measurable function, \( \hat \xi'_{\alpha} \) is also Borel-measurable;
    it satisfies the thesis of \zcref{prop:flatten}, and the associated cocycle is equal to \( \eta'_{c(\alpha)} \) and still constant on connected components.
    If $O_\alpha \cap O_\beta \neq \varnothing$, then the classes $c(\alpha)$ and $c(\beta)$ associated to $\hat \xi_\alpha$ and $\hat \xi_\beta$ are the same and thus $\hat \xi_\alpha'$ and $\hat \xi_\beta'$ have been modified by the same phase $\nu_{c(\alpha)} = \nu_{c(\beta)}$. Using~\eqref{eq:def-phi} we see that the associated \v{C}ech cocycle is unchanged $\phi'_{\alpha \beta} = \phi_{\alpha \beta}$, as desired.
\end{proof}

\begin{definition}[Universal Class]
    We call the $\check{H}^1(\cB_U;\widehat{G})$ class coming from $X = \cB_U$ and $f = \id_{X}$ the \emph{universal class} on $\cB_U$.
    It is clear from the construction that for arbitrary $X$ and continuous $f:X \to \cB_U$, the $\check{H}^1(X;\widehat{G})$ class constructed from $f$ in \zcref{def:phi} coincides with the image of the universal class under the induced map $f^*:\check{H}^1(\cB_U; \widehat{G}) \to \check{H}^1(X; \widehat{G})$.
\end{definition}

We would like to show that the universal classes for different representations on $\bbC^\infty$ are in some sense compatible. Suppose $U$ and $U'$ are two representations on $\bbC^\infty$ and $\gamma \in \cJ(U, U')$, where we recall that $\cJ(U, U')$ is the set of isometric intertwiners between $U$ and $U'$. We leave it to the reader to check that for any $\MPSA \in \cE_U$,
\[
	\Gamma(\MPSA)^i = \sum_j \gamma_{ij}\MPSA^j
\]
is a well-defined element of $\cE_{U'}$ with $\Targ{\Gamma(\MPSA)} = \Targ{A}$ and $\bbVarg{g,\Gamma(\MPSA)} = \bbVarg{g, \MPSA}$. It is also routine to check that $\Gamma$ is continuous and factors through a well-defined continuous map $\underline{\Gamma}:\cB_U \to \cB_{U'}$.

\begin{proposition}\label{prop:universal_to_universal}
    Let $U$ and $U'$ be two representation on $\bbC^\infty$, let $\gamma \in \cJ(U, U')$, and let $\underline{\Gamma}: \cB_U \to \cB_{U'}$ be the continuous function described above. On cohomology, the induced map $\underline{\Gamma}^*:\check{H}^1(\cB_{U'}; \widehat G) \to \check{H}^1(\cB_U; \widehat G)$ maps the universal class on $\cB_{U'}$ to the universal class on $\cB_{U}$.
\end{proposition}

\begin{proof}
Fix representative 2-cocycles $\eta$ for every class $[\eta] \in \Hgrp^2(G;\Unitary(1))$, let $\cO' = (O_\alpha')_\alpha$ be an open cover of $\cB_{U'}$, and let $\hat \xi_\alpha': G \times O_\alpha' \to \SM{\infty}$ be lifts of $\xi':G \times \cB_{U'} \to \PM{\infty}$ satisfying the conditions of \zcref{prop:flatten} with associated 2-cocycles equal to one of the chosen representatives.  Let $V_\alpha'$ be the partial isometry associated to $\hat \xi_\alpha'$ and let $\phi'$ be the \v{C}ech cocycle induced by this data. Under $\underline{\Gamma}^*$, the universal class on $\cB_{U'}$ maps to the \v{C}ech class represented by the open cover $\cO = (O_\alpha \vcentcolon= \underline{\Gamma}^{-1}(O_\alpha'))_\alpha$ and the cocycle
\[
	\phi_{\alpha \beta}(\omega)(g) \vcentcolon= \phi'_{\alpha \beta}(\underline{\Gamma}(\omega))(g) = \ev{\hat \xi_\alpha'(g, \underline{\Gamma}(\omega)), \hat \xi_\beta'(g, \underline{\Gamma}(\omega))}.
\]
Choose a representative $A \in \cE_U$ such that $p(A) = \omega$. Then $p'(\Gamma(A)) = \underline{\Gamma}(\omega)$, where $p':\cE_{U'} \to \cB_{U'}$ is the projection. From the definitions of $\xi'$, $\xi$, and $\Gamma(\MPSA)$, and the fact that $\gamma$ is an isometry, one can show that
\[
	\hat \xi_\alpha(g, \omega)_{ij} \vcentcolon= \sum_{k,\ell} \hat\xi'_\alpha(g, \underline \Gamma(\omega))_{k\ell} \overline{\gamma_{ki}} \gamma_{\ell j}
\]
is a lift of $\xi(g,\omega)$ for $\omega \in O_\alpha$. Furthermore, the partial isometry associated to this lift is $V_\alpha(g,\omega) = V_\alpha'(g, \underline{\Gamma}(\omega))$. Thus, the maps $\hat \xi_\alpha:G \times O_\alpha \to \SM{\infty}$ represent valid data for constructing the universal class on $\cB_U$, and they yield the same \v{C}ech cocycle $\phi$ as above.
\end{proof}

\begin{proposition}\label{prop:parametric_invariant_independent_of_embedding}
    Let $X$ be paracompact Hausdorff and let $\psi:X \to \qstate_W(\sH)$ be a parametrized family. Choose a representation $U$ on $\bbC^\infty$ with an isometric intertwiner $\kappa:\sH \to \bbC^\infty$ and consider the map $\kappa_*\psi:X \to \cB_U$. The class in $\check{H}^1(X;\widehat{G})$ constructed from $\kappa_*\psi$ as in \zcref{def:phi} is independent of the choice of $U$ and $\kappa$. In particular, this class depends only on $\psi$.
\end{proposition}

\begin{proof}
    Suppose $U'$ is another representation on $\bbC^\infty$ with an isometric intertwiner $\kappa':\sH \to \bbC^\infty$. There exists a $G$-universe $U''$ that admits isometric intertwiners $\gamma \in \cJ(U, U'')$ and $\gamma' \in \cJ(U', U'')$. Then $\gamma \kappa, \gamma'\kappa' \in \cJ(W, U'')$ and moreover one can check that the embedding $(\gamma \kappa)_* : \qstate_W(\sH) \to \cB_{U''}$ is the composition $\underline{\Gamma} \circ \kappa_*$, and likewise $(\gamma' \kappa')_* = \underline{\Gamma}' \circ \kappa'_*$, where $\underline{\Gamma}$ and $\underline{\Gamma}'$ are the maps described above.

    For ease of notation, let $f:\check{H}^1(\cB_{U};\widehat{G}) \to \check{H}^1(X;\widehat{G})$ and $f':\check{H}^1(\cB_{U'};\widehat{G}) \to \check{H}^1(X;\widehat{G})$ be the homomorphisms induced by $\kappa_*\psi$ and $\kappa'_*\psi$, respectively.
    Since $U''$ is a $G$-universe, we know $\gamma \kappa$ and $\gamma'\kappa'$ are connected by a path in $\cJ(W, U'')$, hence $(\gamma \kappa)_*\psi$ is homotopic to $(\gamma' \kappa')_*\psi$ by \zcref{prop:continuous_isometry_action}. Since $X$ is paracompact Hausdorff, $(\gamma \kappa)_*\psi$ and $(\gamma' \kappa')_*\psi$ induce the same homomorphism $\check{H}^1(\cB_{U''};\widehat{G}) \to \check{H}^1(X;\widehat{G})$. By the previous paragraph, we know $f \circ \underline{\Gamma}^* =  f' \circ \underline{\Gamma}'^*$.

    Let $\theta$ be the universal class on $\cB_{U''}$. Then by \zcref{prop:universal_to_universal}, we know $\underline{\Gamma}^*(\theta)$ and $\underline{\Gamma}'^*(\theta)$ are the universal classes on $\cB_{U}$ and $\cB_{U'}$ respectively. Thus, the \v{C}ech class on $X$ obtained from $\kappa_*\psi$ is $f(\underline{\Gamma}^*(\theta)) = f'(\underline{\Gamma}'^*(\theta))$, which is the class obtained from $\kappa_*'\psi$, as desired.
\end{proof}

\begin{definition}
    Let $X$ be a paracompact Hausdorff space and let $\psi:X \to \qstate_W(\sH)$ be a parametrized family. The \emph{parametric Thouless invariant} of $\psi$ is the $\check{H}^1(X;\widehat{G})$ class on $X$ obtained from the map $\iota_*\psi$ for any representation $U$ on $\bbC^\infty$ and any isometric intertwiner $\iota \in \cJ(W, U)$.
\end{definition}

\begin{theorem}
    If $X$ is paracompact Hausdorff and $\psi_1:X \to \qstate_{W_1}(\sH_1)$ and $\psi_2:X \to \qstate_{W_2}(\sH_2)$ are two parametrized families in the same phase, then they have the same parametric Thouless invariant on $X$.
\end{theorem}

\begin{proof}
    By \zcref{thm:phase_equivalences}, there exists a $G$-universe $U$ on $\bbC^\infty$ and $\iota_1 \in \cJ(W_1, U)$ and $\iota_2 \in \cJ(W_2, U)$ such that $(\iota_1)_*\psi_1$ is homotopic to $(\iota_2)_*\psi_2$. It follows that $(\iota_1)_*\psi_1$ and $(\iota_2)_*\psi_2$ induce the same homomorphism $\check{H}^1(\cB_U;\widehat{G}) \to \check{H}^1(X;\widehat{G})$, so they have the same parametric Thouless invariant.
\end{proof}

\begin{remark}\label{rem:pullback-parameter}
    Suppose $\psi:X \to \qstate_W(\sH)$ is a parametrized family and $f:Y \to X$ is a continuous function. It is worth noting, and easily checked, that the parametric Thouless invariant associated to the family $\psi \circ f$ is the same as the image of the parametric Thouless invariant associated to $\psi$ under the induced homomorphism $f^*:\check{H}^1(X; \widehat{G}) \to \check{H}^1(Y;\widehat{G})$. This will be used in \zcref{sec:examples} where we use a circle $Y = S^1$ in the parameter space $X$ to detect the parametric Thouless invariant of the parametrized family $\psi$.
\end{remark}

\begin{remark}
    Let us now explain in which sense this construction generalizes the usual Thouless pump. The latter corresponds to a family of states parametrized by the circle, namely to the choice $X=S^1$. In this case, the first \v{C}ech cohomology group can be computed explicitly:
    \[
        \check H^1(S^1; \widehat{G})
        \cong \widehat{G}
        \cong \Hgrp^1(G; U(1)).
    \]
    For the second isomorphism, note that by definition of the differential in group cohomology, a $1$-cocycle in group cohomology is a Borel measurable homomorphism $G \to \Unitary(1)$. Since $G$ and $\Unitary(1)$ are compact Hausdorff, Borel measurable homomorphisms are continuous \cite{Sasvari}, so the group of $1$-cocycles coincides with $\widehat{G}$. The group of $1$-coboundaries is trivial, hence $\Hgrp^1(G;\Unitary(1))$ is the group of $1$-cocycles.

    Thus, for one-parameter cyclic families, the generalized invariant reduces precisely to the familiar character-valued invariant classifying \(G\)-Thouless pumps (see for example~\cite{Bachmann2024} and references therein).
\end{remark}

%!TEX root = MPS_Symmetry-master.tex

\section{Examples}\label{sec:examples}

In many concrete examples, the parametrized family $\psi:X \to \qstate_W(\sH)$  possesses certain nice properties that simplify the computation of the parametric Thouless invariant. Indeed, all of the examples in this section satisfy the following additional hypotheses.

\begin{itemize}
    \item There exists a representation $U$ on $\bbC^n$, a unitary intertwiner $\iota:\sH \to \bbC^n$, and a continuous function $A:X \to \cE_U(n,D)$ for some $D$ such that $p(A(x)) = \iota_*\psi(x)$ for all $x \in X$.
    \item There exists an open cover $(O_\alpha)_{\alpha \in I}$ of $X$ and (not necessarily continuous) lifts $V_\alpha :G \times O_\alpha \to \mat{D}$ of $\bbV \circ (\id_G \times A):G \times X \to \PM{D}$, $(g,x) \mapsto \bbV(g,A(x))$ such that each function $G \times O_\alpha \to \mat{D}$, $(g,x) \to T(A(x))V_\alpha(g,x)$ is continuous.
    \item For each $\alpha \in I$ and $x \in O_\alpha$ the function $g \mapsto V_\alpha(g,x)$ is continuous and the associated 2-cocycle is normalized and is the same across all $x \in O_\alpha$ and $\alpha \in I$. Note that the associated 2-cocycle is automatically Borel measurable by the continuity assumption on $V_\alpha(g,x)$.
\end{itemize}

We will typically leave the representation $U$ on $\bbC^n$ and the isometric intertwiner $\iota$ implicit. For example, if $\sH = \bbC^{n_1} \otimes \bbC^{n_2}$ and $W = W_1 \otimes W_2$, then we may label the matrices of the tensor $A$ as $A^{ij}$ with $i \in \{1,\ldots, n_1\}$ and $j \in \{1,\ldots, n_2\}$, with the understanding that the role of $U$ and $\iota$ would essentially be to relabel the $A^{ij}$ with a single index that runs from $1$ to $n_1n_2$.

Under the above hypotheses, we may form continuous lifts $\hat \xi_\alpha:G \times O_\alpha \to \SM{n} \subset \SM{\infty}$ defined by
\[
    \kappa(g, A(x))\hat \xi_\alpha(g,x)_{ij} = \Tr[T(A(x))V_\alpha(g,x)A(x)^i A(y)^{j*}].
\]
These lifts satisfy the requirements needed to construct the \v{C}ech cocycle of \zcref{def:phi}, and the lift of $\bbV(g, A(x))$ associated to $\hat \xi_\alpha$ is exactly $V_\alpha(g,x)$. Thus, the \v{C}ech cocycle may be constructed as
\begin{equation}\label{eq:phi-from-V}
    \phi_{\alpha \beta}(x)(g)Q(A(x)) = V_\alpha(g, x)^* V_\beta(g,x).
\end{equation}
Note that one does not need to explicitly compute the matrix $\hat \xi_\alpha$ (though we will do so in the first example as a sanity check);
it is sufficient to work with the $V_\alpha$ matrices.

\medskip

Alternatively, one can instead directly consider the function \( \mathbb{M} : G \times X \to \PM{D} \), \( (g,x) \mapsto T(A(x)) \bbVarg{g,A(x)} \) -- which is continuous by \zcref{cor:VT_variations} -- and find continuous local lifts \( M_{\alpha} : G \times O_{\alpha} \to \mat{D} \) such that the associated 2-cocycle is normalized, Borel measurable and constant.
Then, the lifts \( \kappa(g, A(x))\hat \xi_\alpha(g,x)_{ij} = \Tr[M_{\alpha}(g,x) A(x)^i A(y)^{j*}] \) satisfy the required properties, and the Čech cocycle may be extracted as:
\begin{equation}\label{eq:phi-from-M-alpha}
    M_{\beta}(g,x) = M_{\alpha}(g,x) \phi_{\alpha \beta}(x)(g).
\end{equation}

\medskip

In all cases, note that \zcref{thm:universal-class} guarantees that the choice of open cover of \( X \) is immaterial.
Hence, we may choose a (good) cover convenient for the computation, even if this does not come from the pullback of an open cover of \( \mathcal{B}_{U} \).

In the following subsections we will carry out the computation of this \v{C}ech cocycle in four examples.

\subsection{Thouless Pumps}\label{sec:example-sven}
In this section we aim to give an MPS description of the Thouless pump (namely the parameter space is \( X = S^{1} \)) in~\cite[\S 4.2]{Bachmann2024}.
We then compute its invariant defined in \zcref{thm:universal-class} and compare it to the one of \cite{Bachmann2024}.
Indeed, note that it is \( \check{H}^1(S^1; \widehat{G}) \cong \widehat{G} = \Hgrp^{1}(G;\Unitary(1)) \) so the comparison is meaningful.

Consider a compact Hausdorff topological group \( G \) and a continuous group homomorphism \( h : G \to \R/2\pi\Z \), so that \( g \mapsto e^{ih(g)} \) defines an element of \( \widehat{G} \).
Furthermore, let \( 0 \in \widehat{G} \) denote the trivial homomorphism.
We consider a three-dimensional on-site Hilbert space spanned by the orthonormal basis \( \ket{0}, \ket{h}, \ket{-h} \), where the label denotes the representation in which each state transforms.

The loop of states is defined as follows.
We start from the product state given by \( \ket{0} \) on each site; then, each even-odd pair is transformed into the state \( \ket{-h}_{\mathrm{e}} \otimes \ket{h}_{\mathrm{o}} \).
Finally, the sites are regrouped in odd-even pairs, and each state \( \ket{h}_{\mathrm{o}} \otimes \ket{-h}_{\mathrm{e}} \) is transformed back into the original product state.
Pictorially:
\begin{gather*}
    \cdots\ \ket{0}_{-1} \big] \otimes
    \big[ \ket{0}_{0} \otimes \ket{0}_{1} \big] \otimes
    \big[ \ket{0}_{2} \otimes \ket{0}_{3} \big] \otimes
    \big[ \ket{0}_{4} \ \cdots \\
    \downarrow
    \\
    \cdots \ \ket{h}_{-1} \big] \otimes
    \big[ \ket{-h}_{0} \otimes \ket{h}_{1} \big] \otimes
    \big[ \ket{-h}_{2} \otimes \ket{h}_{3} \big] \otimes
    \big[ \ket{-h}_{4} \ \cdots \\
    \rotatebox{90}{$=$} \\
    \cdots \ \big[ \ket{h}_{-1} \otimes \ket{-h}_{0} \big] \otimes
    \big[ \ket{h}_{1} \otimes \ket{-h}_{2} \big] \otimes
    \big[ \ket{h}_{3} \otimes \ket{-h}_{4} \big] \ \cdots \\
    \downarrow
    \\
    \cdots \ \big[ \ket{0}_{-1} \otimes \ket{0}_{0} \big] \otimes
    \big[ \ket{0}_{1} \otimes \ket{0}_{2} \big] \otimes
    \big[ \ket{0}_{3} \otimes \ket{0}_{4} \big] \ \cdots
\end{gather*}
We refer to~\cite[\S 4.2]{Bachmann2024} for details about the Hamiltonian generating the specific path of states.

In order to describe this process with a translation-invariant injective MPS tensor, we let the physical Hilbert space be associated to an even-odd pair.
On such a pair of sites, it is enough to consider a four-dimensional space, since only the following basis vectors appear:
\begin{equation*}
    \ket{1} \defeq \ket{0,0}, \quad
    \ket{2} \defeq \ket{-h,h}, \quad
    \ket{3} \defeq \ket{-h,0}, \quad
    \ket{4} \defeq \ket{0,h}.
\end{equation*}
As we now describe, this Thouless pump can be described by a loop in the space \( \mathcal{E}(d=4,D=2) \).
The essential rank of the MPS tensor will take values 1 or 2.
We parametrize \( S^{1} \) with \( t \in [0,1]/(0 \sim 1) \), though the quotient will be left implicit in what follows.
The MPS description of the path described in~\cite{Bachmann2024} is then:
\begin{align*}
    A^{1}(t) &=
    \begin{cases}
      \begin{psmallmatrix}
        c & 0 \\
        0 & 0
      \end{psmallmatrix}
      & t \in [0,1/2], \\
      F_{t}\bigl(
      \begin{psmallmatrix}
        0 & 0 \\
        0 & c
      \end{psmallmatrix}
      \bigr)
      & t \in (1/2,1),
    \end{cases}
    &&
    A^{2}(t) &=
    \begin{cases}
      \begin{psmallmatrix}
        -is & 0 \\
        0 & 0
      \end{psmallmatrix}
      & t \in [0,1/2], \\
      F_{t}\bigl(
      \begin{psmallmatrix}
        -is & 0 \\
        0 & 0
      \end{psmallmatrix}
      \bigr)
      & t \in (1/2,1),
    \end{cases}
    \\
    A^{3}(t) &=
    \begin{cases}
      \begin{psmallmatrix}
        0 & 0 \\
        c & 0
      \end{psmallmatrix}
      & t \in [0,1/2], \\
      F_{t}\bigl(
      \begin{psmallmatrix}
        0 & c \\
        0 & 0
      \end{psmallmatrix}
      \bigr)
      & t \in (1/2,1),
    \end{cases}
    &&
    A^{4}(t) &=
    \begin{cases}
      \begin{psmallmatrix}
        0 & 0 \\
        -is & 0
      \end{psmallmatrix}
      & t \in [0,1/2], \\
      F_{t}\bigl(
      \begin{psmallmatrix}
        0 & 0 \\
        -is & 0
      \end{psmallmatrix}
      \bigr)
      & t \in (1/2,1),
    \end{cases}
\end{align*}
where we have set
\begin{equation}\label{eq:sven-pump-aux-def}
    \begin{alignedat}{2}
      c &= \cos(\pi t),
      &\qquad\quad X(t)
      &= \begin{psmallmatrix}
        -is & c \\
        c & -is
      \end{psmallmatrix}, \\
      s &= \sin(\pi t),
      & F_{t}(M) &= -e^{2\pi i t}X(t) M X(t)^{*}.
    \end{alignedat}
\end{equation}
One can check that all the \( A^{i} \) are continuous functions of \( S^{1} \).
Note that the essential rank is 1 for \( t \in [0,1/2] \), and 2 for \( t \in (1/2,1) \).
Furthermore, notice that the purpose of the time-dependent gauge transformation \( F_{t} \) is to make the path closed in \( \mathcal{E}(4,2) \), as opposed to closed up to gauge transformations.

We can now write the projective matrices defined in \zcref{prop:bbV}:
\begin{equation}\label{eq:sven-pump-v-matrices}
    \bbVarg{g,A(t)} =
    \begin{cases}
      \biggl[
      \begin{matrix}
        1 & 0 \\
        0 & 0
      \end{matrix}
      \biggr]
      & t \in [0,1/2], \\
      X(t)
      \biggl[
      \begin{matrix}
        e^{ih(g)} & 0 \\
        0 & 1
      \end{matrix}
      \biggr]
      X(t)^{*}
      & t \in (1/2,1),
    \end{cases}
\end{equation}
where the square brackets denote a class in \( \PM{\infty} \).
Moreover, \( \altphasearg{g,A(t)} = 1 \).
One can verify that these reproduce the correct charges for the four basis states above, which are \( 0,0,-h,h \) respectively.
Instead, the matrices of~\eqref{eq:def-tilde-T} are given by
\begin{equation}\label{eq:sven-pump-t-matrices}
    \Targ{A(t)} =
    \begin{cases}
      \biggl(
      \begin{matrix}
        1 & 0 \\
        0 & 0
      \end{matrix}
      \biggr)
      &\quad t \in [0,1/2], \\
      X(t)
      \biggl(
      \begin{matrix}
        s^{2} & 0 \\
        0 & c^{2}
      \end{matrix}
      \biggr)
      X(t)^{*}
      &\quad t \in (1/2,1).
    \end{cases}
\end{equation}

In order to compute the Čech cocycle in \zcref{def:phi}, let us choose the following good cover \( \mathcal{O} = \{O_{1},O_{2},O_{3}\} \) of \( S^{1} = [0,1] / (0 \sim 1) \), for \( \epsilon \in (0, 1/6) \):
\begin{equation}\label{eq:circle-patches}
    \begin{aligned}
        O_{1} &= [0,1/6+\epsilon) \cup (5/6-\epsilon, 1], \\
        O_{2} &= (1/6-\epsilon, 1/2+\epsilon), \\
        O_{3} &= (1/2-\epsilon, 5/6+\epsilon).
    \end{aligned}
\end{equation}
Note that each open set and each pairwise intersection is contractible, while $O_1\cap O_2\cap O_3 = \varnothing$.
Then, consider the map
\begin{equation}\label{eq:class-from-cyclic-product}
    \check{Z}^{1}(\mathcal{O};\widehat{G}) \to \widehat{G},
    \qquad
    \phi \mapsto \phi_{12}\phi_{23}\phi_{31}.
\end{equation}
One can verify that this is a surjective group homomorphism.
Moreover, its kernel is precisely given by the Čech 1-coboundaries, and hence it descends to an isomorphism \( \check{H}^{1}(\mathcal{O};\widehat{G}) \overset{\sim}{\to} \widehat{G} \).
Furthermore, the canonical map \( \check{H}^{1}(\mathcal{O}; \widehat G) \to \check{H}^{1}(S^{1}; \widehat G) \) is also an isomorphism.
This gives an operative way of computing the invariant in the case \( X = S^{1} \).

\subsubsection{Computation with lifts of \texorpdfstring{\( \mathbb{M} \)}{M}}
We can choose the following continuous local lifts \( M_{\alpha} \) of \( \mathbb{M} \) (see around~\eqref{eq:phi-from-M-alpha}):
\begin{align*}
    M_{1}(g,t) &=
    \begin{cases}
      \begin{psmallmatrix}
        1 & 0 \\ 0 & 0
      \end{psmallmatrix}
      & t \in [0,\frac{1}{6}+\epsilon), \\
      X(t)
      \begin{psmallmatrix}
        e^{ih(g)}s^{2} & 0 \\ 0 & c^{2}
      \end{psmallmatrix}
      X(t)^{*}
      & t \in (\frac{5}{6}-\epsilon,1],
    \end{cases} \\
    M_{2}(g,t) &=
    \begin{cases}
      \begin{psmallmatrix}
        e^{ih(g)} & 0 \\ 0 & 0
      \end{psmallmatrix}
      & t \in (\frac{1}{6}-\epsilon, \frac{1}{2}], \\
      X(t)
      \begin{psmallmatrix}
        e^{ih(g)}s^{2} & 0 \\ 0 & c^{2}
      \end{psmallmatrix}
      X(t)^{*}
      & t \in [\frac{1}{2},\frac{1}{2}+\epsilon),
    \end{cases} \\
    M_{3}(g,t) &=
    \begin{cases}
      \begin{psmallmatrix}
        e^{ih(g)} & 0 \\ 0 & 0
      \end{psmallmatrix}
      & t \in (\frac{1}{2}-\epsilon, \frac{1}{2}], \\
      X(t)
      \begin{psmallmatrix}
        e^{ih(g)}s^{2} & 0 \\ 0 & c^{2}
      \end{psmallmatrix}
      X(t)^{*}
      & t \in [\frac{1}{2}, \frac{5}{6}+\epsilon).
    \end{cases}
\end{align*}
The associated matrices \( V_{\alpha} \) are given by
\begin{align*}
    V_{1}(g,t) &=
    \begin{cases}
      \begin{psmallmatrix}
        1 & 0 \\ 0 & 0
      \end{psmallmatrix}
      & t \in [0,\frac{1}{6}+\epsilon), \\
      X(t)
      \begin{psmallmatrix}
        e^{ih(g)} & 0 \\ 0 & 1
      \end{psmallmatrix}
      X(t)^{*} & t \in (\frac{5}{6}-\epsilon,1],
    \end{cases} \\
    V_{2}(g,t) &=
    \begin{cases}
      \begin{psmallmatrix}
        e^{ih(g)} & 0 \\ 0 & 0
      \end{psmallmatrix}
      & t \in (\frac{1}{6}-\epsilon, \frac{1}{2}], \\
      X(t)
      \begin{psmallmatrix}
        e^{ih(g)} & 0 \\ 0 & 1
      \end{psmallmatrix}
      X(t)^{*} & t \in [\frac{1}{2},\frac{1}{2}+\epsilon),
    \end{cases} \\
    V_{3}(g,t) &=
    \begin{cases}
      \begin{psmallmatrix}
        e^{ih(g)} & 0 \\ 0 & 0
      \end{psmallmatrix}
      & t \in (\frac{1}{2}-\epsilon, \frac{1}{2}], \\
      X(t)
      \begin{psmallmatrix}
        e^{ih(g)} & 0 \\ 0 & 1
      \end{psmallmatrix}
      X(t)^{*} & t \in [\frac{1}{2}, \frac{5}{6}+\epsilon),
    \end{cases}
\end{align*}
from which it follows that the 2-cocycle \( \eta \) is identically 1.
This confirms that the pump is contained in the trivial \( G \)-SPT, as it should be since it is based at a product state.
Using~\eqref{eq:phi-from-M-alpha} we obtain:
\begin{equation*}
    \phi_{12}(t) = e^{ih}, \qquad
    \phi_{23}(t) = 1, \qquad
    \phi_{31}(t) = 1.
\end{equation*}
Following the discussion after~\eqref{eq:class-from-cyclic-product}, the invariant is then \( e^{ih} \in \widehat{G} \cong \check{H}^{1}(S^{1};\widehat{G}) \).
This result agrees with the one of \cite[\S4.2]{Bachmann2024}.

\subsubsection{Computation with lifts of \texorpdfstring{\( \xi \)}{ξ}}
In more general settings were the previous procedure cannot be carried out, the only way to proceed may be to explicitly computing local lifts of \( \xi \).
Here, these will be continuous functions, though in general they will only be Borel measurable, as discussed in \zcref{sec:general_construction}.
Denoting classes in \( \PM{\infty} \) with square brackets, we have:
\begin{equation*}
    \Xi(g,A(t)) =
    \begin{cases}
      \left[
      \begin{smallmatrix}
        \cos(\pi t)^{2} & \frac{i}{2} \sin(2\pi t) \\
        -\frac{i}{2} \sin(2\pi t) & \sin(\pi t)^{2}
      \end{smallmatrix}
      \right]
      & t \in [0,1/2], \\[3ex]
      \left[
      \begin{smallmatrix}
        \cos(\pi t)^{4} & 0 \\
        0 & e^{i h(g)} \sin(\pi t)^{4}
      \end{smallmatrix}
      \right]
      & t \in (1/2,1).
    \end{cases}
\end{equation*}
One verifies that this is continuous on \( G \times S^{1} \).
Using the open cover~\eqref{eq:circle-patches}, we can choose the following normalized and continuous lifts
\begin{align*}
    \hat\xi_{1}(g,t) &=
    \begin{cases}
      \mathmakebox[12em][l]{
      \begin{psmallmatrix}
        \cos(\pi t)^{2} & \frac{i}{2} \sin(2\pi t) \\
        -\frac{i}{2} \sin(2\pi t) & \sin(\pi t)^{2}
      \end{psmallmatrix}
      }
      & t \in [0,\frac{1}{6}+\epsilon), \\[2ex]
      \frac{1}{\mathcal{N}(t)}
      \begin{psmallmatrix}
        \cos(\pi t)^{4} & 0 \\
        0 & e^{i h(g)} \sin(\pi t)^{4}
      \end{psmallmatrix}
      & t \in (\frac{5}{6}-\epsilon,1],
    \end{cases} \\
    \hat\xi_{2}(g,t) &=
    \begin{cases}
      \mathmakebox[12em][l]{
      \begin{psmallmatrix}
        \cos(\pi t)^{2} & \frac{i}{2} \sin(2\pi t) \\
        -\frac{i}{2} \sin(2\pi t) & \sin(\pi t)^{2}
      \end{psmallmatrix}
      }
      & t \in (\frac{1}{6}-\epsilon, \frac{1}{2}], \\[2ex]
      \frac{1}{\mathcal{N}(t)}
      \begin{psmallmatrix}
        e^{-i h(g)} \cos(\pi t)^{4} & 0 \\
        0 & \sin(\pi t)^{4}
      \end{psmallmatrix}
      & t \in [\frac{1}{2},\frac{1}{2}+\epsilon),
    \end{cases} \\
    \hat\xi_{3}(g,t) &=
    \begin{cases}
      \mathmakebox[12em][l]{
      \begin{psmallmatrix}
        \cos(\pi t)^{2} & \frac{i}{2} \sin(2\pi t) \\
        -\frac{i}{2} \sin(2\pi t) & \sin(\pi t)^{2}
      \end{psmallmatrix}
      }
      & t \in (\frac{1}{2}-\epsilon, \frac{1}{2}], \\[2ex]
      \frac{1}{\mathcal{N}(t)}
      \begin{psmallmatrix}
        e^{-ih(g)} \cos(\pi t)^{4} & 0 \\
        0 & \sin(\pi t)^{4}
      \end{psmallmatrix}
      & t \in [\frac{1}{2}, \frac{5}{6}+\epsilon),
    \end{cases}
\end{align*}
where \( \mathcal{N}(t) = \sqrt{\sin(\pi t)^{8} + \cos(\pi t)^{8}} \).
Continuity implies that the adjoint functions \( \hat \xi^{\wedge}_{\alpha} : O_{\alpha} \to \SM{\infty}^{G} \) are continuous when \( \SM{\infty}^{G} \) is given the compact open topology~\cite[Prop.~2.4.2]{Dieck2008}.
However, because \( G \) is compact, this coincides with the uniform topology.
The associated \( V_{\alpha} \) matrices are
\begin{align*}
    V_{1}(g,t) &=
    \begin{cases}
      \mathmakebox[9em][l]{
      \begin{psmallmatrix}
        1 & 0 \\ 0 & 0
      \end{psmallmatrix}
      }
      & t \in [0,\frac{1}{6}+\epsilon), \\
      X(t)
      \begin{psmallmatrix}
        e^{-ih(g)} & 0 \\ 0 & 1
      \end{psmallmatrix}
      X(t)^{*} & t \in (\frac{5}{6}-\epsilon,1],
    \end{cases} \\
    V_{2}(g,t) &=
    \begin{cases}
      \mathmakebox[9em][l]{
      \begin{psmallmatrix}
        1 & 0 \\ 0 & 0
      \end{psmallmatrix}
      }
      & t \in (\frac{1}{6}-\epsilon, \frac{1}{2}], \\
      X(t)
      \begin{psmallmatrix}
        1 & 0 \\ 0 & e^{ih(g)}
      \end{psmallmatrix}
      X(t)^{*} & t \in [\frac{1}{2},\frac{1}{2}+\epsilon),
    \end{cases} \\
    V_{3}(g,t) &=
    \begin{cases}
      \mathmakebox[9em][l]{
      \begin{psmallmatrix}
        1 & 0 \\ 0 & 0
      \end{psmallmatrix}
      }
      & t \in (\frac{1}{2}-\epsilon, \frac{1}{2}], \\
      X(t)
      \begin{psmallmatrix}
        1 & 0 \\ 0 & e^{ih(g)}
      \end{psmallmatrix}
      X(t)^{*} & t \in [\frac{1}{2}, \frac{5}{6}+\epsilon).
    \end{cases}
\end{align*}
Since \( h \) is a group homomorphism, one verifies that the group 2-cocycle is identically 1.
Then, using~\eqref{eq:def-phi}, one obtains:
\begin{equation*}
    \phi_{12}(t) = 1, \qquad
    \phi_{23}(t) = 1, \qquad
    \phi_{31}(t) = e^{ih}
\end{equation*}
and thus the class in \( \check{H}^{1}(S^{1};\widehat{G}) \) is again \( e^{ih} \).

\subsection{\texorpdfstring{$\RP^2$}{ℝℙ²}-Parametrized Phase}\label{sec:RP2_example}
Let us consider now the parameter space $X=\mathbb{RP}^2$.
The model described here is based on a model over the parameter space $\mathbb{RP}^2 \times S^1$ considered in~\cite{PhysRevB.110.035114} in the context of nontrivial Dixmier--Douady--Kapustin--Spodyneiko (DDKS) invariants. The $\mathbb{RP}^2$ model we consider is obtained by fixing a point on $S^1$ in the $\mathbb{RP}^2 \times S^1$ family. For the convenience of the reader, we will review the construction of the model on $\mathbb{RP}^2$ and the MPS representation of the ground states given in \cite{PhysRevB.110.035114}. While the DDKS invariant of the full $\mathbb{RP}^2 \times S^1$ model is nontrivial, the DDKS invariant of the restricted model on $\mathbb{RP}^2$ is trivial. On the other hand, the family on the $\mathbb{RP}^2$ subspace possess a $\mathbb{Z}_2$ symmetry not possessed by the full $\mathbb{RP}^2 \times S^1$ model, and this will lead to a nontrivial parametric Thouless invariant.

Consider a lattice labeled by integer sites (called \(\sigma\) sites) and half integer sites (called \(\tau\) sites). They admit a two and three-dimensional onsite Hilbert space, respectively.

Consider a unit vector $\mathbf n=(n_1,n_2,n_3)\in  S^2$.
At each $\tau$ site introduce the orthonormal basis
\[
    \ket{u_+(\mathbf n)}=
    \begin{pmatrix}1\\0\\0\end{pmatrix},
    \quad
    \ket{u_-(\mathbf n)}=
    \begin{pmatrix}0\\n_3\\n_1+i n_2\end{pmatrix},
    \quad
    \ket{u_-^\perp(\mathbf n)}=
    \begin{pmatrix}0\\-n_1+i n_2\\n_3\end{pmatrix}.
\]
Define operators acting on the \(\tau\) sites
\begin{align*}
    \tau^x(\mathbf n)
    &=
    \ket{u_+(\mathbf n)} \bra{u_+(\mathbf n)}
    -
    \ket{u_-(\mathbf n)}\bra{u_-(\mathbf n)},
    \\
    \tau^z(\mathbf n)
    &=
    \ket{u_+(\mathbf n)}\bra{u_-(\mathbf n)}
    +
    \ket{u_-(\mathbf n)}\bra{u_+(\mathbf n)}.
\end{align*}
These act as Pauli matrices on $\vecspan\{\ket{u_+(\mathbf{n})}, \ket{u_-(\mathbf{n})}\}$ and annihilate $\ket{u_-^\perp(\mathbf{n})}$.
We can write a family of Hamiltonians parametrized by $\mathbf n$ as follows:
\begin{equation}\label{eq:RP2_Hamiltonian}
    H(\mathbf n)=
    -\sum_{j \in \Z}
    \tau^z_{j-\frac12}(\mathbf n)\sigma_j^x
    \tau^z_{j+\frac12}(\mathbf n)
    -\sum_{j \in \Z}
    \sigma_j^z\tau^x_{j+\frac12}(\mathbf n)\sigma_{j+1}^z,
\end{equation}
where \( \sigma^{x,z}_{j} \) are the standard Pauli matrices on the \( j \)-th site.
Since
\[
    \tau^x(-\mathbf n) = \tau^x(\mathbf n),
    \qquad
    \tau^z(-\mathbf n) = -\tau^z(\mathbf n),
\]
we have $H(-\mathbf n)=H(\mathbf n)$.
Thus, this family of Hamiltonians actually descends to a family over $\mathbb R\mathbb P^2$ instead of $S^2$, since
\[
    \RP^2=S^2/(\mathbf{n} \sim -\mathbf{n}).
\]
The same is true for the (unique) ground state of \( H(\mathbf{n}) \), which will be given an MPS representation shortly.

\subsubsection{Symmetries}
Consider the unitary extension of $\tau^x(\mathbf n)$
\[
    \hat{\tau}^x
    =
    \tau^x(\mathbf n)
    -
    \ket{u_-^\perp(\mathbf n)}
    \bra{u_-^\perp(\mathbf n)}.
\]
One can check that this is independent of $\mathbf n$ and equal to \( \diag(1,-1,-1) \).
Moreover, it agrees with $\tau^x(\mathbf n)$ on $\vecspan\{\ket{u_+(\mathbf n)}, \ket{u_-(\mathbf n)}\}$.
Now consider the operators
\[
    G_\sigma = \prod_{j \in \Z} \sigma_j^x,
    \qquad
    G_\tau = \prod_{j \in \Z} \hat{\tau}_{j-\frac12}^x.
\]
They satisfy
\[
    G_\sigma^2 = G_\tau^2 = \1,
    \qquad
    [G_\sigma,G_\tau] = 0,
    \qquad
    [H(\mathbf{n}), G_\sigma] = [H(\mathbf{n}), G_\tau] = 0,
\]
and therefore generate a symmetry
\[
    G = \mathbb Z_2^\sigma \times \mathbb Z_2^\tau
\]
at all points on the parameter space \( X = \RP^{2} \).

\subsubsection{MPS representation and symmetry action}

For practical computations, it is convenient to derive the MPS tensors by considering a finite periodic chain, here of $N$ cells, where a cell consists of a half-integer site paired with an integer site. Once the tensors are obtained, we know they represent a well-defined family of states on the infinite lattice. We will provide tensors that represent the ground state of the Hamiltonian \eqref{eq:RP2_Hamiltonian} on a ring with $N$ cells. It is then possible to use \eqref{eq:state_def} to show that these MPS tensors also give ground states of the Hamiltonian in the thermodynamic limit. However, only the tensors are needed for the computation of the parametric Thouless invariant, so we leave the verification that these represent a ground state on an infinite chain to the interested reader.

On a periodic chain of $N$ cells,
the ground state of this Hamiltonian can be written as
\begin{equation}\label{eq:finite-chain-Psi}
    \ket{\Psi(\mathbf n)}
    = 2^{-N/2} \sum_{s_1,\ldots,s_N=\pm1}
    \bigotimes_{j=1}^N
    \left(
        \ket{u_{s_{j-1} s_{j}}(\mathbf n)}_{j-\frac{1}{2}}
        \otimes
        \ket{s_j}_{j}
    \right),
\end{equation}
where \( s_0=s_N \), $\sigma_j^z \ket{s_j} = s_j \ket{s_j}$, and $\ket{u_{\pm1}} = \ket{u_\pm}$.

The operation $G_\sigma$ simultaneously flips all $s_j$ and leaves the $\tau$ sites unchanged.
It permutes the terms in the equal-weight sum.
The operation $G_\tau$ multiplies each term in the sum by \( \prod_{j=1}^{N} s_{j-1}s_j = 1 \).
Therefore the ground state is symmetric under $G = \mathbb Z_2^\sigma\times \mathbb Z_2^\tau$.

In what follows, we will focus on the symmetry acting on the integer sites $\sigma$, i.e., $\mathbb Z_2^\sigma$.
Then, the family of ground states parametrized by \( \RP^{2} \) necessarily has a trivial SPT index at every point since $\Hgrp^2(\bbZ_2;\Unitary(1))$ is trivial.

Following~\cite{PhysRevB.110.035114}, we introduce the parameter-dependent physical basis for the \( \tau \) sites:
\begin{equation}\label{eq:n-dependent-basis}
    \begin{aligned}
        \ket{\tau_\uparrow(\mathbf n)} &= \bigl[ \ket{u_+(\mathbf n)} + \ket{u_-(\mathbf n)} \bigr] / \sqrt{2}, \\
        \ket{\tau_\downarrow(\mathbf n)} &= \bigl[ \ket{u_+(\mathbf n)} - \ket{u_-(\mathbf n)} \bigr] / \sqrt{2}, \\
        \ket{\tau_0(\mathbf n)} &= \ket{u_-^\perp(\mathbf n)}.
    \end{aligned}
\end{equation}
Then, the following MPS tensors:
\[
    A_{\tau(\mathbf n)}^\uparrow=
    \begin{pmatrix}1&1\\0&0\end{pmatrix},
    \qquad
    A_{\tau(\mathbf n)}^\downarrow=
    \begin{pmatrix}0&0\\1&-1\end{pmatrix},
    \qquad
    A_{\tau(\mathbf n)}^0=0,
\]
and
\[
    A_{\sigma}^{1}=
    \begin{pmatrix}1&1\\0&0\end{pmatrix},
    \qquad
    A_{\sigma}^{-1}=
    \begin{pmatrix}0&0\\1&-1\end{pmatrix}
\]
provide an MPS presentation of~\eqref{eq:finite-chain-Psi} as:
\[
    \ket{\Psi(\mathbf n)}
    = 2^{-N}
    \sum_{\{i_j, s_j\}}
    \Tr\sbr{
      A_{\tau(\mathbf{n})}^{i_{1}} A_{\sigma}^{s_{1}} \cdots
      A_{\tau(\mathbf{n})}^{i_{N}} A_{\sigma}^{s_{N}}
    }
    \bigotimes_{j=1}^N \ket{\tau_{i_j}(\mathbf{n})}_{j-\frac{1}{2}}\ket{s_j}_{j},
\]
These MPS tensors do not depend on the parameter \( \mathbf{n} \), however the basis vectors~\eqref{eq:n-dependent-basis} do.
In order to get an MPS presentation suited to computing our invariant, we should move the parameter dependence to the MPS tensors instead.
Let \( \ket{\tau_i} = \ket{\tau_i(\mathbf{n} = (0,0,1))} \).
Then we have:
\[
    \ket{\tau_k(\mathbf n)}
    =
    \sum_i K_\tau(\mathbf n)_{ik} \ket{\tau_i},
    \qquad
    K_\tau(\mathbf n) =
    \begin{pmatrix}
      \frac{1+n_3}{2} & \frac{1-n_3}{2} & -\frac{n_1-i n_2}{\sqrt2} \\
      \frac{1-n_3}{2} & \frac{1+n_3}{2} & \frac{n_1-i n_2}{\sqrt2} \\
      \frac{n_1+i n_2}{\sqrt2} & -\frac{n_1+i n_2}{\sqrt2} & n_3
    \end{pmatrix},
\]
where the entries of the matrix are ordered as in~\eqref{eq:n-dependent-basis}.
The tensors at the $\tau$ sites in this fixed physical basis are then parametrized by $\mathbf n$ and read
\begin{equation*}
    A_\tau^i(\mathbf n)
    =
    \sum_k K_\tau(\mathbf n)_{ik}
    A_{\tau(\mathbf n)}^k.
\end{equation*}
In order to obtain a translation-invariant MPS tensor, we should pair together sites \( (\tau_{j-\frac12}, \sigma_j) \):
\begin{equation}\label{eq:rp2-blocked}
    A_{\tau\sigma}^{i,s}(\mathbf n) \defeq \frac{1}{2} A_\tau^i(\mathbf n)A_\sigma^s.
\end{equation}
Then, the ground state can be written as
\[
    \ket{\Psi(\mathbf n)}
    =
    \sum_{\{i_j,s_j\}}
    \Tr\left[
        A_{\tau\sigma}^{i_1,s_1}(\mathbf n)
        \cdots
        A_{\tau\sigma}^{i_N,s_N}(\mathbf n)
    \right]
    \bigotimes_{j=1}^N
    \ket{\tau_{i_j}}\ket{\sigma_{s_j}}.
\]
One verifies that the MPS tensor \( A^{i,s}_{\tau \sigma}(\mathbf{n}) \) is injective, right-normalized for all \( \mathbf{n} \in S^{2} \), and a continuous function
\( S^{2} \to \mathcal{E}(d=6,D=2) \).
Explicitly, we have:
\begin{align*}
    A^{1,1}_{\tau \sigma}(\mathbf{n}) &=
    \frac{1}{4}
    \left(
        \begin{smallmatrix}
          1+n_3 & 1+n_3 \\
          1-n_3 & 1-n_3
        \end{smallmatrix}
    \right), &
    A^{1,-1}_{\tau \sigma}(\mathbf{n}) &=
    \frac{1}{4}
    \left(
        \begin{smallmatrix}
          1+n_3 & -1-n_3 \\
          -1+n_3 & 1-n_3 \\
        \end{smallmatrix}
    \right), \\
    A^{-1,1}_{\tau \sigma}(\mathbf{n}) &=
    \frac{1}{4}
    \left(
        \begin{smallmatrix}
          1-n_3 & 1-n_3 \\
          1+n_3 & 1+n_3
        \end{smallmatrix}
    \right), &
    A^{-1,-1}_{\tau \sigma}(\mathbf{n}) &=
    \frac{1}{4}
    \left(
        \begin{smallmatrix}
          1-n_3 & -1+n_3 \\
          -1-n_3 & 1+n_3
        \end{smallmatrix}
    \right), \\
    A^{0,1}_{\tau \sigma}(\mathbf{n}) &=
    \frac{\sqrt{2}}{4}
    \left(
        \begin{smallmatrix}
          n_1+i n_2 & n_1+i n_2 \\
          -n_1-i n_2 & -n_1-i n_2
        \end{smallmatrix}
    \right), &
    A^{0,-1}_{\tau \sigma}(\mathbf{n}) &=
    \frac{\sqrt{2}}{4}
    \left(
        \begin{smallmatrix}
          n_1+i n_2 & -n_1-i n_2 \\
          n_1+i n_2 & -n_1-i n_2
        \end{smallmatrix}
    \right).
\end{align*}
Note that this MPS tensor does not descend to a function \( \RP^{2} \to \mathcal{E}(6,2) \).
Indeed, tensors at antipodal points only agree up to a gauge transformation:
\[
    A_{\tau\sigma}^{i,s}(-\mathbf n)
    = \sigma^x A_{\tau\sigma}^{i,s}(\mathbf n)\sigma^x.
\]
This fact however guarantees that this map descends to a continuous function \( \RP^{2} \to \mathcal{B}_{U} \).

\bigskip

Regarding the $\Z_2^\sigma$ symmetry action, the MPS tensors~\eqref{eq:rp2-blocked} satisfy the following push-through property:
\begin{align*}
    \sum_{s'}(\sigma^x)_{ss'} A_{\tau\sigma}^{i,s'}(\mathbf n)
    = \mathbb V_\sigma^{*} A_{\tau\sigma}^{i,s}(\mathbf n)\mathbb V_\sigma,
\end{align*}
where the projective matrix in $\PU(2)$ is given by
\begin{equation}\label{eq:rp2-V-matrix}
    \mathbb V_\sigma
    \defeq
    \bbVarg{-1, A_{\tau \sigma}(\mathbf{n})}
    = [\sigma^z]
    = \left[
        \begin{matrix}
          1&0
          \\0&-1
        \end{matrix}
    \right].
\end{equation}
for all \( \mathbf{n} \in S^{2} \).

\subsubsection{Computation of the invariant}
In this case, our invariant of \zcref{thm:universal-class} is an element of $\check{H}^1(\RP^2; \Z_2) \cong \Z_{2}$.
It turns out that to compute it, it is enough to restrict our family of states \( \RP^{2} \to \mathcal{B}_{U} \) to a nontrivial loop \( \gamma : S^{1} \to \RP^{2} \) that generates \( \pi_{1}(\RP^{2}) \cong \Z_{2} \).
Let us first prove the following
\begin{proposition}\label{prop:circle-into-X}
    Let \( X \) be a connected CW complex with \( \pi_{1}(X) \cong \Z_{n} \), and let \( \gamma : S^{1} \to X \) be a representative of a generator of \( \pi_{1}(X) \).
    Assume that the image of the pullback homomorphism
    \begin{equation*}
        \gamma^*:
        \check H^1(X;\widehat{\Z_n})
        \rightarrow
        \check H^1(S^1;\widehat{\Z_n})
    \end{equation*}
    contains a generator of \( \check H^1(S^1;\widehat{\Z_n}) \).
    Then, \( \gamma^{*} \) is an isomorphism.
\end{proposition}
\begin{proof}
    By the universal coefficient theorem~\cite[Theorem 3.2]{Hatcher} and the degree-one Hurewicz theorem~\cite[Theorem 2A.1]{Hatcher}, we have the natural isomorphisms
    \begin{equation*}
        \check H^1(X;\widehat{\bbZ_n})
        \cong H^1(X;\widehat{\bbZ_n})
        \cong \Hom(H_1(X;\mathbb Z),\widehat{\bbZ_n})
        \cong \Hom(\pi_1(X),\widehat{\bbZ_n}).
    \end{equation*}
    Here the first identification follows from the equivalence of \v{C}ech and singular cohomology for CW complexes (see, e.g.,~\cite[Chapter 6]{Spanier}).
    Since $\pi_1(X) \cong \bbZ_n$, $\widehat{\bbZ_n} \cong \bbZ_n$ and $\Hom(\bbZ_n, \bbZ_n) \cong \bbZ_n$, the above isomorphism shows that
    $\check{H}^1(X; \widehat{\bbZ_n}) \cong \bbZ_n$. We also know $\check{H}^1(S^1; \widehat{\bbZ_n}) \cong \widehat{\bbZ_n} \cong \bbZ_n$.
    Since the image of \( \gamma^{*} \) contains a generator, \( \gamma^{*} \) must clearly be surjective. Since the domain and codomain both have $n$ elements, it is also injective.
\end{proof}

In our case, a representative of such a nontrivial class is given by:
\begin{equation}\label{eq:rp2-gamma}
    \begin{aligned}
        \gamma &: [0,1] \to \RP^{2}, \\
        t &\mapsto [\cos(\pi t) : \sin(\pi t) : 0],
    \end{aligned}
\end{equation}
where we have used homogeneous coordinates, so that the points \( \pm\mathbf{n} \in S^{2} \) both project to \( [n_{1} : n_{2} : n_{3}] \).
Note that \( \gamma(0) = \gamma(1) \) so this descends to a function of \( S^{1} \).
A lift of~\eqref{eq:rp2-gamma} is given by
\( \widehat \gamma : [0,1] \to S^{2} \), \( \widehat \gamma(t) = (\cos(\pi t), \sin(\pi t), 0) \)
so we see that \( \gamma \) is not null-homotopic.\footnote{
  A based loop in \( X \) is null-homotopic if and only if its lift (for a chosen lift of the initial point) to the universal cover \( \tilde X \) is again a loop.
  In this case, the lift \( \widehat \gamma \) starts \( (1,0,0) \) and ends at \( (-1,0,0) \), so \( \gamma \) represents the nontrivial element of \( \pi_{1}(\RP^{2}) \cong \Z_{2} \).
}
The restriction of the MPS tensor~\eqref{eq:rp2-blocked} satisfies
\[
    A_{\tau\sigma}^{i,s}(\widehat \gamma(1)) = \sigma^x A_{\tau\sigma}^{i,s}(\widehat\gamma(0)) \sigma^x
\]
and thus describes a loop of states \( S^{1} \to \mathcal{B}_{U} \), for which we now compute the invariant in \( \check{H}^{1}(S^{1};\Z_{2}) \cong \Z_{2} \).
Similarly to \zcref{sec:example-sven}, it is however convenient to first perform a gauge transformation that makes \( A_{\tau \sigma}^{i,s} \circ \widehat\gamma \) into a closed loop.
Consider the following continuous path of (projective) unitary matrices:
\[
    W(t) =
    \exp\sbr{
      \frac{i\pi t}{2}(\1-\sigma^x)
    }, \qquad
    W(0) = \1, \qquad
    W(1) = \sigma^x,
\]
and define the gauge transformed MPS tensor:
\[
    \tilde A^{i,s}_{\tau\sigma}(t)
    \defeq
    W(t)^{\dagger} A_{\tau\sigma}^{i,s}(\widehat\gamma(t)) W(t).
\]
Then clearly we have \( \tilde A^{i,s}_{\tau \sigma}(1) = \tilde A^{i,s}_{\tau \sigma}(0) \).
Under this gauge transformation, the matrix~\eqref{eq:rp2-V-matrix} implementing the symmetry on the virtual indices becomes
\[
    \tilde\bbV_\sigma(t)
    = W(t)^{\dagger} \bbV_\sigma W(t) =
    \sbr{
      \begin{matrix}
        c & -i s \\
        i s & - c
      \end{matrix}
    },
\]
where we have used the definitions~\eqref{eq:sven-pump-aux-def}.
This is a continuous loop of $\mathbb PU(2)$ matrices, as $\bbV_\sigma(0) = \bbV_\sigma(1) = [\sigma^z]$.

\bigskip

The bond dimension of \( \tilde{A}_{\tau \sigma}(t) \) is fixed at 2 for all \( t \in S^{1} \), so it is enough to consider continuous local lifts of \( \tilde\bbV_{\sigma}(t) \).
Using the good of cover of \( S^{1} \) as in~\eqref{eq:circle-patches}, we can choose the following lifts:
\begin{align*}
    V_{\sigma,1}(t)
    &=
    \begin{cases}
      \begin{psmallmatrix}
        c & -is \\
        is & -c
      \end{psmallmatrix}
      & t \in [0, \frac{1}{6} +\epsilon), \\
      -\begin{psmallmatrix}
        c & -is \\
        is & -c
      \end{psmallmatrix}
      & t \in (\frac{5}{6}-\epsilon, 1],
    \end{cases} \\
    V_{\sigma,2}(t)
    &=
    \begin{psmallmatrix}
      c & -is \\
      is & -c
    \end{psmallmatrix}
    \hspace{10mm}
    t \in (\tfrac{1}{6}-\epsilon, \tfrac{1}{2}+\epsilon), \\
    V_{\sigma,3}(t)
    &=
    \begin{psmallmatrix}
      c & -is \\
      is & -c
    \end{psmallmatrix}
    \hspace{10mm}
    t \in (\tfrac{1}{2}-\epsilon, \tfrac{5}{6}+\epsilon).
\end{align*}
Note that each of these matrices squares to the identity and thus defines a linear representation of $\mathbb Z_2$.
This means that the group 2-cocycle is identically equal to 1.
Using~\eqref{eq:phi-from-V} we finally obtain \( \phi_{12}(t) = 1 \), \( \phi_{23}(t) = 1 \) and \( \phi_{31}(t) = -1 \), where \( 1 \) and \( -1 \) denote the trivial and nontrivial homomorphisms in \( \widehat{\Z_{2}} \) respectively.
We conclude that the invariant associated to the considered loop is the nontrivial element of \( \check{H}^{1}(S^{1};\widehat{\Z_{2}}) \cong \Z_{2} \).

Following \zcref{rem:pullback-parameter}, this invariant is the pullback via \( \gamma \) of the invariant associated to the original \( \RP^{2} \)-parametrized phase~\eqref{eq:rp2-blocked}.
Using now \zcref{prop:circle-into-X} we conclude that the pullback \( \gamma^{*} \) is an isomorphism and thus the desired phase invariant of~\eqref{eq:rp2-blocked} is precisely the nontrivial class in $\check{H}^1(\RP^2,\Z_2) \cong \Z_{2}$.

\subsection{\texorpdfstring{$\mathbb{Z}_2$}{ℤ₂}-Cluster State Pump}\label{sec:clsta}

The example of the Thouless pump described in \zcref{sec:example-sven} was given by a loop states contained in the trivial SPT for \( G \), as can be deduced from the fact that the loop is based at the product state (or from the fact that \( \eta \) is constantly equal to 1).
The aim of this section is to provide an example of a Thouless pump (\( X = S^{1} \)) contained in a nontrivial SPT.
Such a loop is based at the so-called \emph{cluster state}.
More in detail, the pump is made of \( (\Z_{2} \times \Z_{2}) \)-symmetric states, and it transports charge for one the two \( \Z_{2} \) factors only.

The cluster state $\ket{\Psi}$ is the ground state of the following Hamiltonian:
\begin{equation*}
    H = - \sum_{i\in\Z}\sigma^{z}_{i-1}\sigma^{x}_{i}\sigma^{z}_{i+1}.
\end{equation*}
This enjoys a \( G = \Z_{2}^{\mathrm{e}} \times \Z_{2}^{\mathrm{o}} \) symmetry generated by
\begin{equation}\label{eq:cluster-state-z2-z2-generators}
    U^{\mathrm{e}} = \prod_{i\text{ even}} \sigma^{x}_{i}, \qquad
    U^{\mathrm{o}} = \prod_{i\text{ odd}} \sigma^{x}_{i} .
\end{equation}
Consider the controlled-\( Z \) gate:
\begin{equation*}
    \mathrm{CZ}_{i,i+1} = \frac{1}{2}(1+ \sigma^z_i + \sigma^z_{i+1} - \sigma^z_{i} \sigma^z_{i+1}).
\end{equation*}
The cluster state can be obtained as
\begin{equation*}
    \ket{\Psi} = \prod_{i\text{ even}} \mathrm{CZ}_{i,i+1} \ket{\cdots + + + \cdots},
\end{equation*}
where $\ket{+}$ is the eigenstate of $\sigma^x$ with eigenvalue 1.
Notice that all operators \( \mathrm{CZ}_{i,i+1} \) commute pairwise.

The idea is as follows. To construct a pump for the \( \Z_{2}^{\mathrm{e}} \) factor\footnote{
  In both the trivial and the nontrivial SPT, Thouless pumps are classified by the group \( \Hgrp^{1}(\Z_{2}\times \Z_{2};\Unitary(1)) \cong \Z_{2} \times \Z_{2} \).
} of \( G \) in the nontrivial SPT, that is, the one classified by the generator of \( \Hgrp^{2}(\Z_{2}\times \Z_{2};\Unitary(1)) \cong \Z_{2} \), we start from a nontrivial pump for that factor in the trivial SPT and act on it with the operator \( \prod_{i} \mathrm{CZ}_{i,i+1} \) introduced above.

As in the $\mathbb{RP}^2$ example of \zcref{sec:RP2_example}, it is convenient to derive the MPS tensors by working on a finite periodic chain, here of $L$ sites.
We can write:
\begin{equation}\label{eq:cluster-state-finite-chain}
    \begin{aligned}
        \ket{\Psi} &= \sbr{
          \prod_{i=1}^{L} \mathrm{CZ}_{i,i+1}
        }
        \ket{+ \cdots + } \\
        &= \frac{1}{2^{\frac{L}{2}}} \sum_{s_1, \ldots, s_L = \pm 1} \sbr{
          \prod_{i=1}^{L} \mathrm{CZ}_{i,i+1}
        }
        \ket{s_1 \cdots s_L} \\
        &= \frac{1}{2^{\frac{L}{2}}} \sum_{s_1, \ldots, s_L = \pm 1} \sbr{
          \prod_{i=1}^{L} \frac{1}{2}(1+s_i +s_{i+1} -s_i s_{i+1})
        }
        \ket{s_1 \cdots s_L} ,
    \end{aligned}
\end{equation}
where the states $\ket{\pm} = \frac{1}{\sqrt{2}} (\ket{1} \pm \ket{-1})$ satisfy $ \sigma^{z} \ket{\pm 1} = \pm \ket{\pm 1}$.

\begin{proposition}\label{prop:cluster-state-mps}
    An MPS representation of the state \( \ket{\Psi} \) in~\eqref{eq:cluster-state-finite-chain} is given by
    \begin{equation*}
        \tilde K^s_{\alpha\beta} = \frac{1}{\sqrt{2}}\delta_{\beta, s} (-1)^{\frac{1}{4}(\alpha-1)(s-1)} ,
        \qquad
        s = 1,-1, \quad \alpha,\beta = 1,-1,
    \end{equation*}
    or alternatively:
    \begin{equation*}
        \begin{aligned}
            \tilde K^1= \frac{1}{\sqrt{2}}
            \begin{pmatrix}
              1 & 0 \\
              1 & 0
            \end{pmatrix},
            \qquad
            \tilde K^{-1}= \frac{1}{\sqrt{2}}
            \begin{pmatrix}
              0 & 1\\
              0 & -1
            \end{pmatrix}.
        \end{aligned}
    \end{equation*}
    Here, rows and columns are indexed by \( 1 \) and \( -1 \) in this order, and the tildes denote that the physical indices of MPS tensor are in the \( \sigma^{z} \)-eigenbasis.
\end{proposition}
\begin{proof}
    We can explicitly compute the following trace:
    \begin{align}\label{eq:gate-as-mps}
        &\Tr(\tilde K^{s_{1}} \cdots \tilde K^{s_L})
        = \sum_{\alpha_1 \cdots \alpha_L} \tilde K^{s_1}_{\alpha_1 \alpha_2} \cdots \tilde K^{s_L}_{\alpha_L \alpha_1} \nonumber \\
        &\qquad= \frac{1}{2^{\frac{L}{2}}} \sum_{\alpha_1 \cdots \alpha_L} (-1)^{\frac{1}{4}(\alpha_1-1)(s_1-1)} \delta_{\alpha_2, s_1}
        (-1)^{\frac{1}{4}(\alpha_2-1)(s_2-1)} \delta_{\alpha_3, s_2} \cdots \nonumber \\[-1ex]
        &\hspace{6cm} \cdots (-1)^{\frac{1}{4}(\alpha_L-1)(s_L-1)} \delta_{\alpha_1, s_L} \nonumber \\[-1ex]
        &\qquad = \frac{1}{2^{\frac{L}{2}}} \prod_{i=1}^{L} (-1)^{\frac{1}{4}(s_i-1)(s_{i+1}-1)}.
    \end{align}
    Using a simple algebraic identity, we can write:
    \begin{equation*}
        \frac{1}{2}(1+s_i +s_{i+1} -s_i s_{i+1}) = (-1)^{\frac{1}{4}(s_i-1)(s_{i+1}-1)},
    \end{equation*}
    which proves the proposition.
\end{proof}
In order to get an injective MPS tensor, we block together pairs of neighboring sites as follows:
\begin{equation*}
    \tilde K^{s_{2i}, s_{2i+1}}_{\alpha \beta}
    \defeq \sum_{\gamma}  \tilde K^{s_{2i}}_{\alpha \gamma} \tilde K^{s_{2i+1}}_{\gamma \beta}
    = \frac{1}{2} (-1)^{\frac{1}{4}(\alpha-1)(s_{2i}-1)} \delta_{\beta, s_{2i+1}} (-1)^{\frac{1}{4}(s_{2i}-1)(s_{2i+1}-1)} .
\end{equation*}
More explicitly, we have:
\begin{equation*}
    \begin{alignedat}{3}
      \tilde K^{1,1}
      &= \frac{1}{2}
        \begin{pmatrix}
          1 & 0\\
          1 & 0
        \end{pmatrix},
      &\qquad\qquad
        \tilde K^{1,-1}
      &= \frac{1}{2}
        \begin{pmatrix}
          0 & 1 \\
          0 & 1
        \end{pmatrix}, \\
      \tilde K^{-1,1}
      &= \frac{1}{2}
        \begin{pmatrix}
          1 & 0 \\
          -1 & 0
        \end{pmatrix},
      &
        \tilde K^{-1,-1}
      &= \frac{1}{2}
        \begin{pmatrix}
          0 & -1 \\
          0 & 1
        \end{pmatrix}.
    \end{alignedat}
\end{equation*}
These matrices clearly span $\mat{2}$, and are right-normalized.

\begin{proposition}
    The cluster state is invariant under the \( G = \Z_{2}^{\mathrm{e}} \times \Z_{2}^{\mathrm{o}} \) symmetry~\eqref{eq:cluster-state-z2-z2-generators}.
\end{proposition}
\begin{proof}
    Notice that \( \Z_{2}^{\mathrm{e}} \) (resp.\ \( \Z_{2}^{\mathrm{o}} \)) acts by mapping \( \ket{\pm 1} \) to \( \ket{\mp 1} \) on the even (resp.\ odd) sites.
    It is then enough to provide unitary matrices that realize the push-through property:
    \begin{align*}
        \tilde K^{s_{2i},s_{2i+1}} &\quad\overset{\Z_{2}^{\mathrm{e}}}{\longmapsto}\quad \tilde K^{-s_{2i},s_{2i+1}} = \sigma^{z} K^{s_{2i},s_{2i+1}} \sigma^{z}, \\
        \tilde K^{s_{2i},s_{2i+1}} &\quad\overset{\Z_{2}^{\mathrm{o}}}{\longmapsto}\quad \tilde K^{s_{2i},-s_{2i+1}} = \sigma^{x} K^{s_{2i},s_{2i+1}} \sigma^{x}. \qedhere
    \end{align*}
\end{proof}

Let us now construct a pump for the \( \Z_{2}^{\mathrm{e}} \) symmetry.
Since the nontrivial evolution takes place only on the even sites, we first restrict our attention to the even sublattice;
the odd sites will be reinstated at the end by taking the tensor product with their state.
A pump based at the product state \( \ket{\cdots +++ \cdots} \) can be obtained from the one in \zcref{sec:example-sven} by replacing \( \ket{0} \) with \( \ket{+} \), and both \( \ket{h} \) and \( \ket{-h} \) with \( \ket{-} \).
This replacement is natural: in that example, \( \ket{0} \) is the state that is neutral under the symmetry, while \( \ket{\pm h} \) are charged.
Here the neutral state is \( \ket{+} \), and since \( \Z_{2} \) has a single nontrivial irreducible representation, both charged states are sent to the unique charged state \( \ket{-} \).
This yields the following loop:
\begin{align*}
    A^{++}(t) &=
    \begin{cases}
      \begin{psmallmatrix}
        c & 0 \\
        0 & 0
      \end{psmallmatrix}
      & t \in [0,1/2], \\
      F_{t}\bigl(
      \begin{psmallmatrix}
        0 & 0 \\
        0 & c
      \end{psmallmatrix}
      \bigr)
      & t \in (1/2,1),
    \end{cases}
    &
    A^{+-}(t) &=
    \begin{cases}
      \begin{psmallmatrix}
        0 & 0 \\
        -is & 0
      \end{psmallmatrix}
      & t \in [0,1/2], \\
      F_{t}\bigl(
      \begin{psmallmatrix}
        0 & 0 \\
        -is & 0
      \end{psmallmatrix}
      \bigr)
      & t \in (1/2,1),
    \end{cases}
    \\
    A^{-+}(t) &=
    \begin{cases}
      \begin{psmallmatrix}
        0 & 0 \\
        c & 0
      \end{psmallmatrix}
      & t \in [0,1/2], \\
      F_{t}\bigl(
      \begin{psmallmatrix}
        0 & c \\
        0 & 0
      \end{psmallmatrix}
      \bigr)
      & t \in (1/2,1),
    \end{cases}
    &
    A^{--}(t) &=
    \begin{cases}
      \begin{psmallmatrix}
        -is & 0 \\
        0 & 0
      \end{psmallmatrix}
      & t \in [0,1/2], \\
      F_{t}\bigl(
      \begin{psmallmatrix}
        -is & 0 \\
        0 & 0
      \end{psmallmatrix}
      \bigr)
      & t \in (1/2,1),
    \end{cases}
\end{align*}
where we have used the definitions in~\eqref{eq:sven-pump-aux-def}.
The computation of the invariant is completely analogous to the one in \zcref{sec:example-sven} and yields the nontrivial element of \( \check{H}^1(S^{1};\widehat{\Z_{2}}) \cong \Z_{2} \).

We now claim that a pump in the nontrivial \( G \)-SPT can be obtained by applying the controlled-\( Z \) gate pointwise to the state above, after taking the tensor product with the state on the odd sites.
In order to compute this more easily, we are going to reexpress these MPS tensors in an orthonormal \( \sigma^{z} \)-eigenbasis.
Under a change of basis \( \ket{i}_{x} = \sum_{\tilde i} U_{i\tilde i} \ket{\tilde i}_{z} \), an MPS tensor transforms as \( \tilde A^{\tilde i} = \sum_{i} U_{i \tilde i} A^{i} \).
In this case it is
\begin{equation*}
    U_{i \tilde i} =
    \frac{1}{2}
    \begin{pmatrix}
      1 & 1 & 1 & 1 \\
      1 & -1 & 1 & -1 \\
      1 & 1 & -1 & -1 \\
      1 & -1 & -1 & 1
    \end{pmatrix}_{i \tilde i}.
\end{equation*}
We then get a closed formula for the MPS tensors in the new basis
\begin{equation*}
    \tilde A^{pq}(t) =
    \begin{cases}
      \displaystyle
      \frac{1}{2}
      \begin{pmatrix}
        e^{-i\pi t pq} & 0 \\
        p e^{-i\pi t pq} & 0
      \end{pmatrix}
      & t \in [0,1/2], \\
      \displaystyle
      \tilde A^{pq}
      = -\frac{1}{2} F_{t}
      \del{
      \begin{pmatrix}
        i pq\, s & - p\, c \\
        i q\, s & -c
      \end{pmatrix}
      }
      & t \in (1/2,1),
    \end{cases}
\end{equation*}
where \( p,q = 1,-1 \) label the eigenstates of \( \sigma^{z} \), \( \ket{\pm 1} \).
One can check that these matrices belong to \( \mathcal{E}(d=4,D=2) \).

The full state, also including the odd sites, is then (now on a \( 4N \)-sites lattice)
\begin{equation*}
    \ket{\psi(t)}
    = \sum_{\{p_{i}\},\{q_{i}\}} \Tr\sbr{
      \tilde A^{p_{1}q_{1}}(t) \cdots \tilde A^{p_{N}q_{N}}(t)
    }
    \ket{p_{1}q_{1}p_{2}q_{2} \cdots p_{N}q_{N}}_{\mathrm{e}}
    \ket{+ \cdots +}_{\mathrm{o}}.
\end{equation*}
Using \( \ket{+} = (\ket{1} + \ket{-1})/\sqrt{2} \), this can be rewritten as
\begin{equation*}
    \ket{\psi(t)} =
    \!\!\sum_{\substack{\{a_{i}\}, \{b_{i}\}\\ \{c_{i}\}, \{d_{i}\}}}\!\!
    \Tr\sbr{
      \tilde B^{a_{1}b_{1}c_{1}d_{1}}(t) \cdots \tilde B^{a_{N}b_{N}c_{N}d_{N}}(t)
    }
    \ket{a_{1}b_{1}c_{1}d_{1}}
    \cdots
    \ket{a_{N}b_{N}c_{N}d_{N}} ,
\end{equation*}
where now all sites appear in order, and are expressed in the \( \sigma^{z} \)-eigenbasis.
We have also defined \( \tilde B^{abcd} = \frac{1}{2} \tilde A^{ac} \).
The sum over \( a \) and \( c \) reproduces the pump on the even sites.
The sum over \( b \) and \( d \) reproduces the \( \ket{+} \)'s on the odd sites.
Note that blocking four sites together does not change the computation of the invariant, because \( \bbV \) and \( T \) are also unchanged.
Indeed, since the symmetry under consideration, \( \Z_{2}^{\mathrm{e}} \), only acts on the \( a \) and \( c \) indices, the same \( \bbV \) as in~\eqref{eq:sven-pump-v-matrices} realizes the symmetry action.
Moreover, the same \( T \) as in~\eqref{eq:sven-pump-t-matrices} can be used.
Indeed, this is the eigenvalue-1 eigenvector of the operator
\begin{equation*}
    \sum_{abcd} (\tilde B^{abcd})^{*} (\cdot) \tilde B^{abcd}
    = \frac{1}{4} \biggl(\sum_{bd} 1\biggr) \sum_{ac} (\tilde A^{ac})^{*} (\cdot) \tilde A^{ac}
    = \sum_{ac} (\tilde A^{ac})^{*} (\cdot) \tilde A^{ac} .
\end{equation*}
Let us now act with the gate \( \mathsf{G} = \prod_{i=1}^{4N} (CZ)_{i,i+1} \) on the state \( \ket{\psi(t)} \):
\begin{multline*}
    \mathsf{G}\ket{\psi(t)}
    = \sum_{\substack{\{a_{i}\}, \{b_{i}\}\\ \{c_{i}\}, \{d_{i}\}}}
    \Tr\sbr{
      \tilde B^{a_{1}b_{1}c_{1}d_{1}}(t) \cdots \tilde B^{a_{N}b_{N}c_{N}d_{N}}(t)
    } \\
    \prod_{i=1}^{N} \frac{1}{16}
    (1+a_{k}+b_{k}-a_{k}b_{k})
    (1+b_{k}+c_{k}-b_{k}c_{k}) \\
    \qquad\qquad\qquad (1+c_{k}+d_{k}-c_{k}d_{k})
    (1+d_{k}+a_{k+1}-d_{k}a_{k+1}) \\
    \ket{a_{1}b_{1}c_{1}d_{1}} \cdots \ket{a_{N}b_{N}c_{N}d_{N}} .
\end{multline*}
This state can be rewritten as an MPS thanks to~\eqref{eq:gate-as-mps}.
However, in order to attain translation invariance, now four sites have to be blocked into one.
More specifically, using the matrices in \zcref{prop:cluster-state-mps} we have
\begin{equation*}
    4^{N} \Tr[\tilde K^{s_{1}} \cdots \tilde K^{s_{4N}}]
    = \prod_{i=1}^{4N} \frac{1}{2}
    (1+s_{i}+s_{i+1}-s_{i}s_{i+1}) ,
\end{equation*}
and thus
\begin{multline*}
    4^{N} \Tr[\tilde K^{a_{1}b_{1}c_{1}d_{1}} \cdots \tilde K^{a_{N}b_{N}c_{N}d_{N}}] = \\
    =\prod_{i=1}^{N} \frac{1}{16}
    (1+a_{k}+b_{k}-a_{k}b_{k})
    (1+b_{k}+c_{k}-b_{k}c_{k}) \\
    (1+c_{k}+d_{k}-c_{k}d_{k})
    (1+d_{k}+a_{k+1}-d_{k}a_{k+1}) ,
\end{multline*}
where we have defined \( \tilde K^{abcd} \defeq \tilde K^{a} \tilde K^{b} \tilde K^{c} \tilde K^{d} \).
Thus
\begin{align*}
    \mathsf{G}\ket{\psi(t)} &\propto \sum_{\substack{\{a_{i}\}, \{b_{i}\}\\ \{c_{i}\}, \{d_{i}\}}}
    \Tr\sbr{
      \tilde B^{a_{1}b_{1}c_{1}d_{1}}(t) \cdots \tilde B^{a_{N}b_{N}c_{N}d_{N}}(t)
    } \\[-2ex]
    &\hspace{2.2cm}\Tr[\tilde K^{a_{1}b_{1}c_{1}d_{1}} \cdots \tilde K^{a_{N}b_{N}c_{N}d_{N}}] \
    \ket{a_{1}b_{1}c_{1}d_{1}} \cdots \ket{a_{N}b_{N}c_{N}d_{N}} \\[2ex]
    &\hspace{-5mm}= \sum_{\substack{\{a_{i}\}, \{b_{i}\}\\ \{c_{i}\}, \{d_{i}\}}}
    \Tr\sbr{
      \tilde C^{a_{1}b_{1}c_{1}d_{1}}(t) \cdots \tilde C^{a_{N}b_{N}c_{N}d_{N}}(t)
    }
    \ket{a_{1}b_{1}c_{1}d_{1}} \cdots \ket{a_{N}b_{N}c_{N}d_{N}} ,
\end{align*}
where
\begin{equation}\label{eq:c-matrix-def}
    \tilde C^{abcd}(t) \defeq \tilde B^{abcd}(t) \otimes \tilde K^{abcd} .
\end{equation}
We can interpret the \( \tilde B \) factor as being responsible for the pumping, and the \( \tilde K \) factor for the nontriviality of the \( G \)-SPT.
The explicit form of the matrices \( \tilde C \) is reported in \zcref{app:cluster-state-matrices}.
\begin{proposition}\label{prop:C-tilde-bad}
    For all \( t \in [0,1/2] \), \( \tilde C(t) \) is an element of \( \mathcal{E}(d=16,D=4,\chi=2) \) (and thus in particular of \( \mathcal{E}(d=16,D=4) \)).
    However, for \( t \in (1/2,1) \), \( \tilde C \) does not belong to \( \mathcal{E}(16,4) \), and thus our current description is not good for the computation of the invariant.
\end{proposition}
\begin{proof}
    For \( t \in [0,1/2] \), it is enough to notice that the last two columns of \( \tilde C^{abcd} \) are always zero.
    Thus, one can extract the upper-left 2-by-2 block and check by direct computation that it is an injective and right-normalized (up to a rescaling) MPS tensor.

    For \( t \in (1/2,1) \), we show the statement by contradiction.
    Assume \( \tilde C \in \mathcal{E}(16,4) \).
    Then the computation of \( Q(\tilde C) \) (see \zcref{def:L_and_Q}) should reveal its essential rank, independently of the way $\tilde C$ is presented.
    To do this, we define \( L(\tilde C) = \sum_{abcd} (\tilde C^{abcd})^\dagger \tilde C^{abcd} \) and compute its eigenvalues.
    One can verify that, for a generic $t \in (1/2,1)$, these are all nonzero.
    This means that $Q(\tilde C)=\theta(L(\tilde C)) = \1_{4}$, so the essential rank of \( \tilde C \) is 4.
    In turn, this implies that $\tilde C$ is an injective and right-normalized as an ordinary MPS tensor, since $\chi = D$ (i.e., it is an element of \( \mathcal{I}(16,4) \)).
    However, a direct computation shows that this is not true.
    In fact, these 16 matrices span an 8-dimensional subspace of the 16-dimensional space of 4-by-4 matrices.
    Hence, by contradiction $\tilde C \notin \mathcal{E}(16,4)$.
\end{proof}
The issue raised in the previous proposition can be solved by recasting the state represented by \( \tilde C \) as a \( D = 2 \) MPS tensor.
We define:
\begin{equation}\label{eq:P-Q-D-matrices}
    P =
    \begin{pmatrix}
      1 & 0 \\
      0 & 1 \\
      0 & 1 \\
      1 & 0
    \end{pmatrix},
    \quad
    Q =
    \begin{pmatrix}
      1 & 0 & 0 & 0 \\
      0 & 1 & 0 & 0
    \end{pmatrix},
    \quad
    \tilde D^{abcd}(t) = 4 Q \tilde C^{abcd}(t) P.
\end{equation}
The explicit form of the MPS tensors \( \tilde D \) is reported in \zcref{app:cluster-state-matrices}.
\begin{proposition}
    For all \( t \in S^{1} \), \( \tilde D(t) \) is an element of \( \mathcal{I}(d=16,\chi=2) \) and defines the same state as \( \tilde C(t) \).
    Moreover, the functions \( t \mapsto D^{abcd}(t) \) are continuous.
\end{proposition}
\begin{proof}
    For \( t \in [0,1/2] \), one can check that the matrices \( \tilde D^{abcd} \) coincide (up to a factor) with the upper-left 2-by-2 block of \( \tilde C^{abcd} \).
    As already discussed in the proof of \zcref{prop:C-tilde-bad}, such a block is an injective MPS tensor.
    The factor 4 in the definition of \( \tilde D \) makes sure that it is right-normalized.
    By the general discussion in \zcref{sec:mps-tensors}, such a tensor defines the same state as \( \tilde C \).

    For \( t \in (1/2,1) \), observe from the \( \tilde C \) matrices in \zcref{app:cluster-state-matrices} that, in each of them, the first row always equals the fourth and the second row equals the third.
    Using this, one can check that \( \tilde C^{abcd} = PQ \tilde C^{abcd} \).
    This implies the identity:
    \begin{align*}
        &\Tr\sbr{
          \tilde C^{a_{1}b_{1}c_{1}d_{1}} \cdots \tilde C^{a_{N}b_{N}c_{N}d_{N}}
        } = \\
        &\qquad\qquad = \Tr\sbr{
          (PQ\tilde{\mathscr{C}}^{a_{1}b_{1}c_{1}d_{1}})
          (PQ\tilde{\mathscr{C}}^{a_{2}b_{2}c_{2}d_{2}})
          \cdots
          (PQ\tilde{\mathscr{C}}^{a_{N}b_{N}c_{N}d_{N}})
        } \\
        &\qquad\qquad = \Tr\sbr{
          (Q\tilde{\mathscr{C}}^{a_{1}b_{1}c_{1}d_{1}}P)
          (Q\tilde{\mathscr{C}}^{a_{2}b_{2}c_{2}d_{2}}P)
          \cdots
          (Q\tilde{\mathscr{C}}^{a_{N}b_{N}c_{N}d_{N}}P)
        } \\
        &\qquad\qquad \propto \Tr\sbr{
          \tilde D^{a_{1}b_{1}c_{1}d_{1}} \cdots \tilde D^{a_{N}b_{N}c_{N}d_{N}}
        },
    \end{align*}
    from which it follows that \( \tilde C \) and \( \tilde D \) define the same state.

    Continuity of \( t \mapsto \tilde D^{abcd}(t) \) follows from the definition~\eqref{eq:P-Q-D-matrices} and by observing that \( \tilde D^{abcd}(t) \) is related to the original matrices \( A^{i}(t) \) only by algebraic operations (independent of \( t \)).
\end{proof}
Let \( G = \Z_{2}^{\mathrm{e}} \times \Z_{2}^{\mathrm{o}} = \{1,e,o,eo\} \), with \( e^{2} = o^{2} = 1 \).
The following proposition follows by direct computation and by noting that the \( G \) acts as
\begin{align*}
    \tilde D^{abcd} &\quad\overset{e}{\longmapsto}\quad \tilde D^{-a,b,-c,d}, \\
    \tilde D^{abcd} &\quad\overset{o}{\longmapsto}\quad \tilde D^{a,-b,c,-d}.
\end{align*}
\begin{proposition}\label{prop:cluster-state-bbv}
    The matrices that realize the symmetry action as in~\eqref{eq:g_act_K_invariance} are given by:
    \begin{align*}
        \bbvarg{e,\tilde D(t)} &=
        \begin{cases}
          [\sigma^{z}] & t \in [0,1/2], \\
          \biggl[
          \begin{matrix}
            -\cos(2\pi t) & -i\sin(2\pi t) \\
            i\sin(2\pi t) & \cos(2\pi t)
          \end{matrix}
          \biggr]
          & t \in (1/2,1),
        \end{cases} \\
        \bbvarg{o,\tilde D(t)} &= [\sigma^{x}].
    \end{align*}
\end{proposition}
Notice that, for all \( t \in S^{1} \), the matrices describing the action of the two \( \Z_{2} \) factors anticommute and square to the identity.
This means that they define a projective representation of \( \Z_{2}^{\mathrm{e}} \times \Z_{2}^{\mathrm{o}} \)
(corresponding to the nontrivial cocycle in \( \Hgrp^{2}(\Z_{2}\times \Z_{2};\Unitary(1)) \cong \Z_{2} \)),
confirming that the loop of states is contained in the nontrivial \( G \)-SPT.

\bigskip

We now proceed to compute the invariant.
Since all states in the family are described by an injective MPS tensor with bond dimension two, we will be able to find a continuous family of projective unitaries $\bbv(g, t)$ implementing the symmetry on the virtual legs of the tensors.
We will then be able to  compute the invariant from \( \bbv \) directly.

First, for the \( \Z_{2}^{\mathrm{o}} \) factor, we see from \zcref{prop:cluster-state-bbv} that \( \bbvarg{o,\tilde D(t)} \) is constant.
This means that all transition functions \( \varphi_{\alpha \beta} \) can be chosen to be trivial.
Hence, the invariant is -- as expected -- the trivial element of \( \check{H}^{1}(S^{1};\widehat{\Z_{2}^{\mathrm{o}}}) \cong \Z_{2} \).

Then, for the \( \Z_{2}^{\mathrm{e}} \) factor we can choose the following lifts (working with the open cover~\eqref{eq:circle-patches}):
\begin{align*}
    v_{1}(e,t) &=
    \begin{cases}
      \mathmakebox[10em][l]{
      \begin{psmallmatrix}
        1 & 0 \\ 0 & -1
      \end{psmallmatrix}
      }
      & t \in [0,\frac{1}{6}+\epsilon), \\
      -
      \begin{psmallmatrix}
        -\cos(2\pi t) & -i\sin(2\pi t) \\
        i\sin(2\pi t) & \cos(2\pi t)
      \end{psmallmatrix}
      & t \in (\frac{5}{6}-\epsilon,1],
    \end{cases} \\
    v_{2}(e,t) &=
    \begin{cases}
      \mathmakebox[10em][l]{
      \begin{psmallmatrix}
        1 & 0 \\ 0 & -1
      \end{psmallmatrix}
      }
      & t \in (\frac{1}{6}-\epsilon, \frac{1}{2}], \\
      \begin{psmallmatrix}
        -\cos(2\pi t) & -i\sin(2\pi t) \\
        i\sin(2\pi t) & \cos(2\pi t)
      \end{psmallmatrix}
      & t \in [\frac{1}{2},\frac{1}{2}+\epsilon),
    \end{cases} \\
    v_{3}(e,t) &=
    \begin{cases}
      \mathmakebox[10em][l]{
      \begin{psmallmatrix}
        1 & 0 \\ 0 & -1
      \end{psmallmatrix}
      }
      & t \in (\frac{1}{2}-\epsilon, \frac{1}{2}], \\
      \begin{psmallmatrix}
        -\cos(2\pi t) & -i\sin(2\pi t) \\
        i\sin(2\pi t) & \cos(2\pi t)
      \end{psmallmatrix}
      & t \in [\frac{1}{2}, \frac{5}{6}+\epsilon),
    \end{cases}
\end{align*}
whence
\( \phi_{12}(t) = 1 \), \( \phi_{23}(t) = 1 \), and \( \phi_{31}(t) = -1 \), where \( 1 \) denotes the trivial homomorphism \( \Z_{2} \to \Unitary(1) \) and \( -1 \) the nontrivial one.
We conclude that the invariant of this pump is, as expected, the nontrivial element of the group \( \check{H}^{1}(S^{1};\widehat{\Z_{2}^{\mathrm{e}}}) \cong \Z_{2} \).

\subsection{\texorpdfstring{$\mathbb{Z}_n$}{ℤₙ}-Cluster State and \texorpdfstring{$L(n,1)$}{L(n,1)}-Parametrized Phases}\label{sec:lens-pump}

In this section, we construct two explicit families of injective MPS with a fixed on-site $\mathbb Z_n$ symmetry, for odd integers $n \geq 3$,
parametrized by the three-dimensional lens space $L(n,1)$, defined below.
We start from the generalized $\mathbb Z_n$ cluster state, a representative of a nontrivial one-dimensional $\mathbb Z_n\times\mathbb Z_n$ symmetry-protected topological phase~\cite{GeraedtsMotrunich2014,HanEtAl2024}, and introduce the parameter dependence by rotating one sublattice while preserving the symmetry on the other.
Pairing adjacent sites gives a translation-invariant, injective MPS tensor.

We do not require $n$ to be odd in the beginning, so let $n \geq 2$ be any integer. Let  $\omega=e^{2\pi i/n}$. Recall that
\begin{equation*}
    L(n,1)=S^3/\Z_n,
    \qquad
    S^3 = \{ z = (z_1,z_2)\in\mathbb C^2 : \abs{z_1}^2 + \abs{z_2}^2 = 1 \},
\end{equation*}
where the generator of $\mathbb Z_n$ acts by $z \mapsto \omega z$, so that the action is free. Here we are notating $\omega(z_1, z_2) = (\omega z_1, \omega z_2)$.
Since $S^3$ is simply connected, this is the universal covering and
$\pi_1(L(n,1))\cong\mathbb Z_n$
(see~\cite[Theorem 4.41]{Hatcher}). The action defining the quotient
will be implemented by virtual gauge transformations of the MPS
which, together with the virtual representation of the physical
symmetry, give charge-pump invariants generating
$\check H^1(L(n,1),\widehat{\mathbb Z_n})\cong\mathbb Z_n$.

\subsubsection{The generalized \texorpdfstring{$\mathbb Z_n$}{ℤₙ} cluster state}\label{subsec:lens-cluster}
The reference construction below makes sense for all $n\geq2$.
Consider a periodic chain of $2N$ sites, with indices understood modulo $2N$ and one copy of $\mathbb C^n$ at each site.
Its Hilbert space and observable algebra are
\begin{equation*}
    \mathcal H_{2N}=\bigotimes_{x=1}^{2N}\mathbb C^n,
    \qquad
    \mathfrak A_{2N}=\mathcal B(\mathcal H_{2N}).
\end{equation*}
We will work on a finite chain to obtain a family of MPS tensors, which can then be used to define states on the infinite chain.
Fix a basis $\{ \ket{a} : a\in\mathbb Z_n \}$ and define the Pauli operators
\begin{equation*}%\label{eq:lens-weyl}
    X \ket{a} = \ket{a+1}, \qquad
    Z\ket{a} = \omega^a \ket{a}, \qquad
    ZX = \omega XZ.
\end{equation*}
Both have order $n$.
The basis $\ket{a}$ diagonalizes $Z$, while
\begin{equation*}
    \ket{p}_X =
    \frac{1}{\sqrt n}\sum_{a\in\mathbb Z_n}
    \omega^{-pa}\ket{a},
    \qquad X\ket{p}_X=\omega^p\ket{p}_X,
    \qquad p \in \mathbb{Z}_n
\end{equation*}
diagonalizes $X$.

\bigskip

The trivial paramagnetic state $\ket{\Omega_0} = \ket{0}_X^{\otimes 2N}$ is the unique ground state of $H_0=\sum_x(2\mathbf1-X_x-X_x^\dagger)$.
The generalized controlled-phase gate on two sites is
\begin{equation*}%\label{eq:lens-controlled-phase}
    C_{x,x+1}
    = \sum_{a,b\in\mathbb Z_n}
    \omega^{ab} \ket{a,b}\bra{a,b}_{x,x+1}.
\end{equation*}
It is unitary, diagonal in the joint $Z$ eigenbasis, and satisfies
\begin{equation*}\label{eq:lens-gate-identities}
    C_{x,x+1}X_xC_{x,x+1}^\dagger = X_xZ_{x+1},
    \qquad
    C_{x,x+1}X_{x+1}C_{x,x+1}^\dagger = Z_xX_{x+1}.
\end{equation*}
Using the commuting gates in the depth-two circuit
\begin{equation*}\label{eq:lens-cluster-circuit}
    \mathcal U_{\mathrm{cl}}
    = \prod_{j=1}^{N}C_{2j-1,2j}^\dagger C_{2j,2j+1},
    \qquad 2N+1 \equiv 1,
\end{equation*}
we obtain the generalized cluster state\footnote{
  The state $\ket{\Omega_{\mathrm{cl}}}$ is the unique ground state of
  \begin{equation*}%\label{eq:lens-cluster-hamiltonian}
      H_{\mathrm{cl}}
      =\mathcal U_{\mathrm{cl}}H_0\mathcal U_{\mathrm{cl}}^\dagger=\sum_{j=1}^{N}\Big[
      4\mathbf 1
      -\big(
      Z_{2j-2}X_{2j-1}Z_{2j}^\dagger
      +Z_{2j-1}^\dagger X_{2j}Z_{2j+1}
      +\text{h.c.}
      \big)
      \Big].
  \end{equation*}
  The Hamiltonian is uniformly gapped, being unitarily equivalent to $H_0$.
}
\begin{equation*}%\label{eq:lens-cluster-state}
    \ket{\Omega_{\mathrm{cl}}}
    \defeq
    \mathcal U_{\mathrm{cl}}\ket{\Omega_0}.
\end{equation*}
For $n=2$ the circuit reduces to the usual controlled-$Z$ preparation of the qubit cluster state discussed in \zcref{sec:clsta}.
The cluster state is invariant under the two sublattice symmetries $U_{\mathrm{odd}}=\prod_jX_{2j-1}$ and $U_{\text{phys}} \defeq \prod_j X_{2j}$.
We keep only the latter as a fixed symmetry in the parametrized construction.
\begin{equation*}\label{eq:lens-physical-symmetry}
    U_{\text{phys}} = \prod_{j=1}^{N}X_{2j}.
\end{equation*}

The reference state has an explicit MPS representation.

\begin{proposition}
    Let $K^a,\overline K^a\in \mat{n}$ be the matrices
    \begin{equation*}\label{eq:lens-mps-tensors}
        (K^a)_{rs}=\frac1{\sqrt n}\omega^{ra}\delta_{s,a},
        \qquad
        (\overline K^a)_{rs}=\frac1{\sqrt n}\omega^{-ra}\delta_{s,a}.
    \end{equation*}
    Here the bar denotes entrywise complex conjugation. Then
    \begin{equation}\label{eq:lens-cluster-mps}
        \ket{\Omega_{\mathrm{cl}}}
        =
        \sum_{\boldsymbol a,\boldsymbol b\in\mathbb Z_n^N}
        \Tr\bigl(
        K^{a_1}\overline K^{b_1}
        \cdots
        K^{a_N}\overline K^{b_N}
        \bigr)
        \ket{a_1,b_1,\ldots,a_N,b_N}.
    \end{equation}
    After blocking each pair of sites, the tensors $B^{a,b} \cong K^a\overline K^b$ span $M_n(\mathbb C)$ and satisfy the right-normalization condition.
    Consequently, the blocked MPS is translation-invariant and injective, with bond dimension $n$.
\end{proposition}

\begin{proof}
    The circuit gives the amplitudes
    \begin{equation*}\label{eq:lens-cluster-wavefunction}
        \langle a_1,b_1,\ldots,a_N,b_N | \Omega_{\mathrm{cl}} \rangle
        = \frac{1}{n^N} \ \omega^{\sum_j a_j(b_{j-1}-b_j)},
        \qquad
        b_0=b_N.
    \end{equation*}
    Since
    \begin{equation*}\label{eq:lens-blocked-tensors}
        (B^{a,b})_{rs}=\frac1n\omega^{a(r-b)}\delta_{s,b},
    \end{equation*}
    the trace in~\eqref{eq:lens-cluster-mps} reproduces these amplitudes.
    Moreover, Fourier inversion yields
    \begin{equation*}
        \sum_{a\in\mathbb Z_n} \omega^{-a(r-b)} B^{a,b}
        = \ket{r}\bra{b},
        \qquad
        r,b \in \Z_n.
    \end{equation*}
    Hence the blocked tensors span the full matrix algebra. That $B^{a,b}$ is right-normalized follows from right-normalization of $K^a$, which we leave to the reader to check.
\end{proof}

The MPS representation of the cluster state also makes the action of both $\mathbb Z_n$ symmetries particularly transparent, as expressed by the following proposition.
\begin{proposition}[Virtual implementation of the symmetries]
The action of the two sublattice symmetries on the blocked cluster-state tensors can be implemented through conjugations on their virtual indices (see \zcref{virtact}).
More precisely,
\begin{align}
    \sum_{b'\in\mathbb Z_n}
    X_{bb'}B^{a,b'}
    &=X^\dagger B^{a,b}X,
    \label{eq:cluster-even-virtual}
    \\[4pt]
    \sum_{a'\in\mathbb Z_n}
    X_{aa'}B^{a',b}
    &=Z^\dagger B^{a,b}Z.
    \label{eq:cluster-odd-virtual}
\end{align}

\begin{figure}[htbp]
    \begin{tikzpicture}[
        tensor/.style={
          draw,
          rounded corners=2pt,
          fill=blue!12,
          minimum width=1.0cm,
          minimum height=0.8cm
        },
        phys/.style={
          draw,
          fill=orange!15,
          minimum width=0.65cm,
          minimum height=0.5cm
        },
        virt/.style={
          circle,
          draw,
          fill=teal!12,
          minimum size=0.95cm,
          inner sep=1pt
        },
        wire/.style={thick}
        ]

        % =========================================
        % TOP: Physical action on even sites
        % =========================================
        \begin{scope}

            \node[tensor] (B1) at (0,0) {$B$};
            \node[phys] (P1) at (0.28,1.15) {$X$};

            \draw[wire] (-1.4,0)--(B1.west);
            \draw[wire] (B1.east)--(1.4,0);

            \draw[wire] (-0.28,0.4)--(-0.28,2);
            \draw[wire] (0.28,0.4)--(P1.south);
            \draw[wire] (P1.north)--(0.28,2);

            \node at (-0.28,2.22) {$a$};
            \node at (0.28,2.22) {$b$};

            \node at (2.1,0) {$=$};

            \node[tensor] (B2) at (4.9,0) {$B$};
            \node[virt] (XL) at (3.55,0) {$X^\dagger$};
            \node[virt] (XR) at (6.25,0) {$X$};

            \draw[wire] (2.8,0)--(XL.west);
            \draw[wire] (XL.east)--(B2.west);
            \draw[wire] (B2.east)--(XR.west);
            \draw[wire] (XR.east)--(7.0,0);

            \draw[wire] (4.62,0.4)--(4.62,2);
            \draw[wire] (5.18,0.4)--(5.18,2);

            \node at (4.62,2.22) {$a$};
            \node at (5.18,2.22) {$b$};

        \end{scope}

        % =========================================
        % BOTTOM: Physical action on odd sites
        % =========================================
        \begin{scope}[shift={(0,-3.6)}]

            \node[tensor] (B3) at (0,0) {$B$};
            \node[phys] (P2) at (-0.28,1.15) {$X$};

            \draw[wire] (-1.4,0)--(B3.west);
            \draw[wire] (B3.east)--(1.4,0);

            \draw[wire] (-0.28,0.4)--(P2.south);
            \draw[wire] (P2.north)--(-0.28,2);
            \draw[wire] (0.28,0.4)--(0.28,2);

            \node at (-0.28,2.22) {$a$};
            \node at (0.28,2.22) {$b$};

            \node at (2.1,0) {$=$};

            \node[tensor] (B4) at (4.9,0) {$B$};
            \node[virt] (ZL) at (3.55,0) {$Z^\dagger$};
            \node[virt] (ZR) at (6.25,0) {$Z$};

            \draw[wire] (2.8,0)--(ZL.west);
            \draw[wire] (ZL.east)--(B4.west);
            \draw[wire] (B4.east)--(ZR.west);
            \draw[wire] (ZR.east)--(7.0,0);

            \draw[wire] (4.62,0.4)--(4.62,2);
            \draw[wire] (5.18,0.4)--(5.18,2);

            \node at (4.62,2.22) {$a$};
            \node at (5.18,2.22) {$b$};

        \end{scope}
    \end{tikzpicture}
    \caption{\label{virtact}%
      Push-through property of the $\mathbb{Z}_n$-cluster state for the two sublattice $\mathbb{Z}_n$-symmetries.
    }
\end{figure}
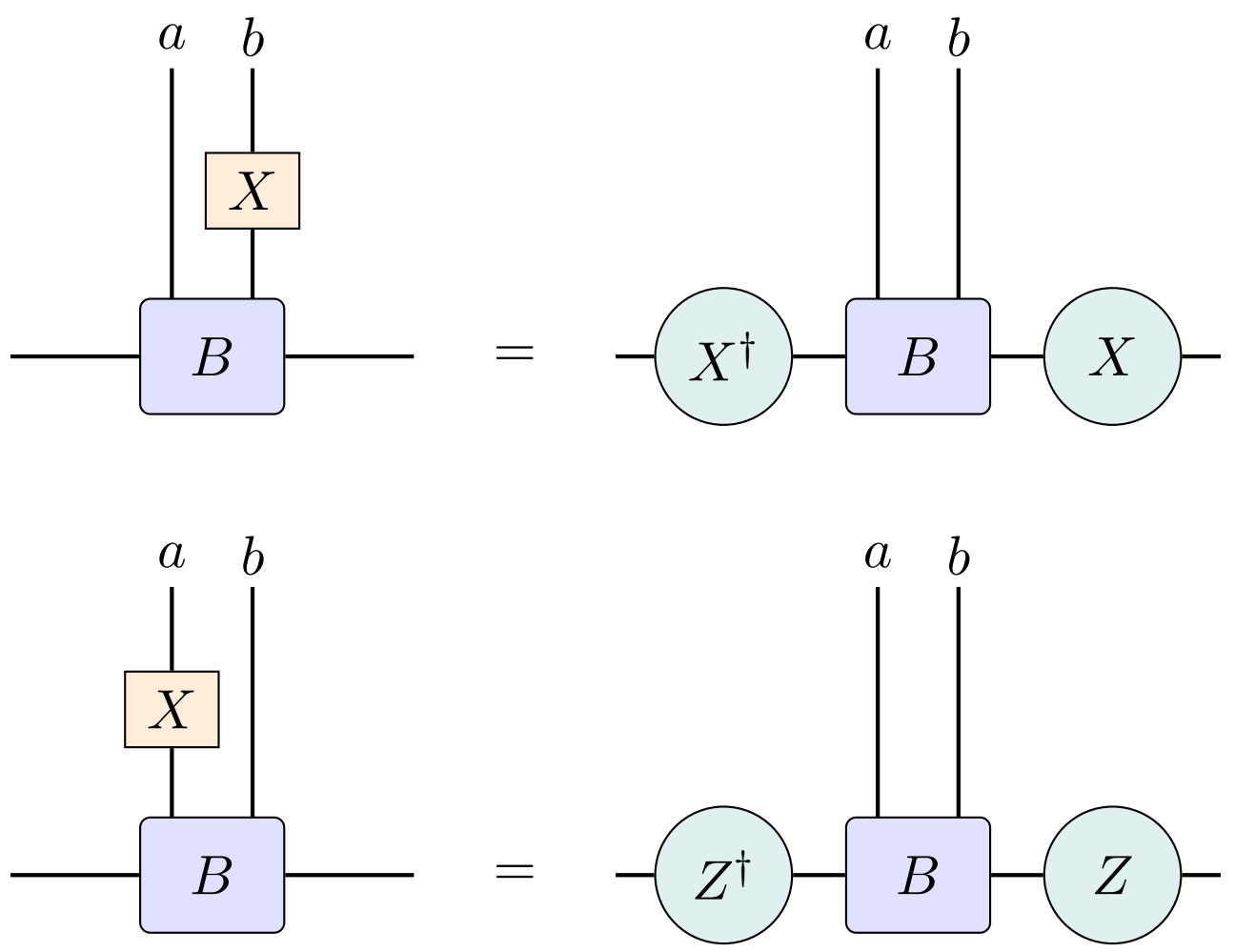

\end{proposition}

\begin{proof}
    Recall that
    \[
        (B^{a,b})_{ij}
        =\frac1n\omega^{a(i-b)}\delta_{j,b}.
    \]
    We first consider the action on the even sites.
    Since $X_{bb'}=\delta_{b,b'+1}$, we have
    \begin{align*}
        \sum_{b'\in\mathbb Z_n}
        X_{bb'}(B^{a,b'})_{ij}
        &=(B^{a,b-1})_{ij}=\frac1n
        \omega^{a(i-b+1)}\delta_{j,b-1}.
    \end{align*}
    On the other hand,
    \begin{align*}
        (X^\dagger B^{a,b}X)_{ij}
        &= \sum_{k,l \in \mathbb{Z}_n} X^{\dagger}_{ik}(B^{a,b})_{kl} X_{lj}
        = \frac{1}{n} \sum_{k, l } \delta_{i,k-1} \omega^{a(k-b)}\delta_{l,b} \delta_{l,j+1} \\
        &= \frac{1}{n} \omega^{a(i-b+1)}\delta_{j,b-1}.
    \end{align*}
    This establishes~\eqref{eq:cluster-even-virtual}.

    Similarly, for the odd sites,
    \begin{align*}
        \sum_{a'\in\mathbb Z_n}
        X_{aa'}(B^{a',b})_{ij}
        =(B^{a-1,b})_{ij} =\frac1n
        \omega^{(a-1)(i-b)}\delta_{j,b}.
    \end{align*}
    Meanwhile,
    \begin{align*}
        (Z^\dagger B^{a,b}Z)_{ij}
        = \sum_{k,l \in \mathbb{Z}_n}\omega^{-i+j} \delta_{ik} \delta_{lj}(B^{a,b})_{kl} \\
        = \frac{1}{n}
        \omega^{a(i-b)-(i-j)}\delta_{j,b}=\frac1n
        \omega^{(a-1)(i-b)}\delta_{j,b},
    \end{align*}
    This proves~\eqref{eq:cluster-odd-virtual}.
\end{proof}

\subsubsection{Families parametrized by \texorpdfstring{$L(n,1)$}{L(n,1)}}\label{subsec:lens-families}

We now restrict to odd $n \geq 3$. We identify $S^3$ with $SU(2)$ via
\begin{equation*}%\label{eq:lens-s3-su2}
    g(z)
    =
    \begin{pmatrix}
      z_1 & -\overline z_2 \\
      z_2 & \overline z_1
    \end{pmatrix},
    \qquad
    g((\omega,0)) \defeq h
    =
    \begin{pmatrix}
      \omega & 0 \\
      0 & \omega^{-1}
    \end{pmatrix}.
\end{equation*}
The relation $g(\omega z)=g(z)h$ will ensure the required equivariance.
Consider the $n$-dimensional representation
\begin{equation*}%\label{eq:lens-rep}
    \rho=\mathbf1\oplus D^{(s)} : \SU(2)\longrightarrow \Unitary(n),
    \qquad s=\frac{n-2}{2},
\end{equation*}
where $D^{(s)}$ is the spin-$s$ irreducible representation.
Writing $\ket{*}$ for the trivial summand and $\ket{m}$, $m=-s,\ldots,s$, for a weight basis, we have\footnote{
  This can be seen by the fact that $h = \exp(\frac{4 \pi i}{n} \frac{\sigma^z}{2}) = \exp(\frac{4 \pi i}{n} S^z_{s=\frac{1}{2}})$, so that $D^{(s)}(h) = \exp(\frac{4 \pi i}{n} S^z_{s})$. Therefore:
  \[
      \rho(h) \ket{m} = D^{(s)}(h) \ket{m} = e^{\frac{4 \pi i}{n} m} \ket{m} = \omega^{2m} \ket m
  \]
}
\begin{equation*}
    \rho(h)\ket{*} = \ket{*},
    \qquad
    \rho(h)\ket{m} = \omega^{2m}\ket{m}.
\end{equation*}

\begin{figure}[htbp]
    \centering
    \resizebox{\linewidth}{!}{%
      \begin{tikzpicture}[
          >=Latex,
          panel/.style={
            draw=blue!30,
            fill=blue!2,
            rounded corners=6pt,
            line width=0.8pt
          },
          target/.style={
            draw=blue!45,
            fill=blue!10,
            rounded corners=3pt,
            minimum width=1.25cm,
            minimum height=0.45cm
          },
          arrow/.style={->, thick, blue!65}
          ]

          % ====================================
          % s = 1/2, n = 3
          % ====================================
          \begin{scope}
              \draw[panel] (0,0) rectangle (6.3,-5.7);

              \node[font=\large] at (3.15,-0.45)
              {$s=\frac12,\qquad n=3$};

              \node at (1.5,-1.15) {Initial states};
              \node at (4.85,-1.15) {$X$ basis};

              \node at (1.5,-2.0) {$\ket{*}$};
              \node at (1.5,-3.2) {$\ket{+\frac 1 2}$};
              \node at (1.5,-4.4) {$\ket{-\frac 1 2}$};

              \node[target] at (4.85,-2.0) {$\ket{0}_X$};
              \node[target] at (4.85,-3.2) {$\ket{1}_X$};
              \node[target] at (4.85,-4.4) {$\ket{2}_X$};

              \draw[arrow] (2.3,-2.0)--(4.05,-2.0);
              \draw[arrow] (2.3,-3.2)--(4.05,-3.2);
              \draw[arrow] (2.3,-4.4)--(4.05,-4.4);

              \node at (3.15,-5.2) {$J \ket{m} = \ket{[2m]_3}_X$};
          \end{scope}

          % ====================================
          % s = 3/2, n = 5
          % ====================================
          \begin{scope}[shift={(6.7,0)}]

              \draw[panel] (0,0) rectangle (6.3,-5.7);

              \node[font=\large] at (3.15,-0.45)
              {$s=\frac32,\qquad n=5$};

              \node at (1.5,-1.15) {Initial states};
              \node at (4.85,-1.15) {$X$ basis};

              \node at (1.5,-1.7) {$\ket{*}$};
              \node at (1.5,-2.45) {$\ket{+\frac12}$};
              \node at (1.5,-3.2) {$\ket{-\frac32}$};
              \node at (1.5,-3.95) {$\ket{+\frac32}$};
              \node at (1.5,-4.7) {$\ket{-\frac12}$};

              \node[target] at (4.85,-1.7) {$\ket{0}_X$};
              \node[target] at (4.85,-2.45) {$\ket{1}_X$};
              \node[target] at (4.85,-3.2) {$\ket{2}_X$};
              \node[target] at (4.85,-3.95) {$\ket{3}_X$};
              \node[target] at (4.85,-4.7) {$\ket{4}_X$};

              \draw[arrow] (2.3,-1.7)--(4.05,-1.7);
              \draw[arrow] (2.3,-2.45)--(4.05,-2.45);
              \draw[arrow] (2.3,-3.2)--(4.05,-3.2);
              \draw[arrow] (2.3,-3.95)--(4.05,-3.95);
              \draw[arrow] (2.3,-4.7)--(4.05,-4.7);

              \node at (3.15,-5.2) {$J \ket{m} = \ket{[2m]_5}_X$};
          \end{scope}
      \end{tikzpicture}%
    }
    \caption{\label{fig:isometry}%
      The isometric embeddings $J\ket{*} = \ket{0}_X$ and $J\ket{m} = \ket{[2m]_n}_X$ for $s=\frac12$ ($n=3$) and $s=\frac{3}{2}$ ($n=5$).
    }
\end{figure}
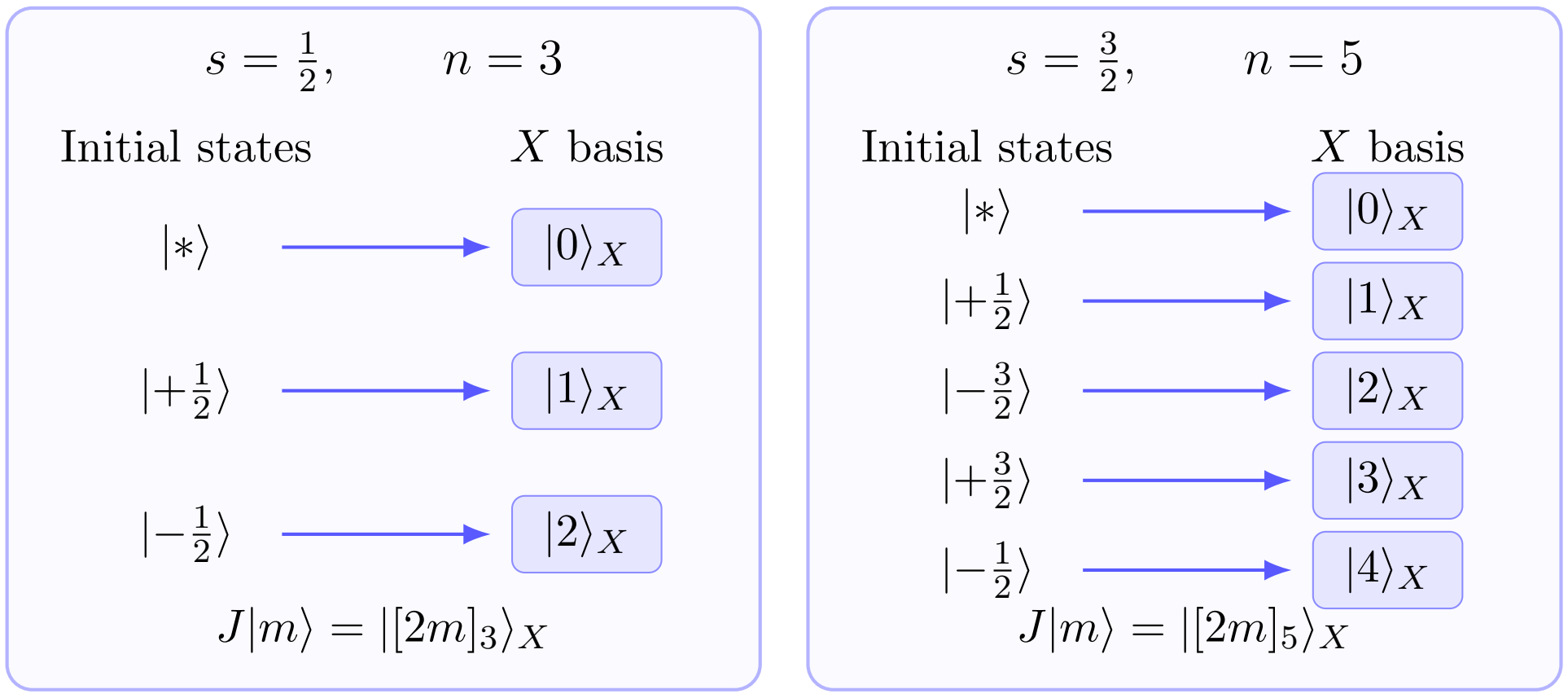

Let $J:\mathbb C\oplus\mathbb C^{n-1} \to\mathbb C^n$ be the unitary map (see \zcref{fig:isometry}) defined by:
\begin{equation*}
    J\ket{*} = \ket{0}_X \qquad
    J\ket{m} = \ket{[2m]_n}_X \quad
    (-s \leq m \leq s)
\end{equation*}
where $[2m]_n$ is the coset of $2m$ in $\bbZ_n$.

It follows that $J\rho(h)J^\dagger = X$\footnote{
  To prove this, apply $J\rho(h)$ and $X J$ to any element of the basis $\{\ket{*}, \ket{-s}, \cdots, \ket{s}\}$.
  We get:
  \begin{equation*}
      J \rho(h) \ket{*} = \ket{0}_X,
      \qquad
      J \rho(h) \ket{m} = \omega^{2m} \ket{[2m]_n}_X,
  \end{equation*}
  and
  \begin{equation*}
      X J \ket{*} = X \ket{0}_X = \ket{0}_X,
      \qquad
      XJ\ket{m} = X \ket{[2m]_n}_X = \omega^{2m} \ket{[2m]_n}_X.
  \end{equation*}
}. We are now ready to introduce the unitary operator that rotates the states.
For each $z\in S^3$, define
\begin{equation}
    V(z) \defeq J\rho(g(z))J^\dagger.
\end{equation}
Although $V(z)$ is not invariant under the $\mathbb Z_n$-action, it satisfies the covariance relation
\begin{equation*}%\label{eq:lens-V-covariance}
    V(\omega z)
    =J\rho(g(z)h)J^\dagger
    =V(z)X.
\end{equation*}
As we shall see, this covariance is sufficient to ensure that the resulting family of MPS is well-defined on the quotient $L(n,1)$.

\begin{figure}
    \centering
    \begin{tikzpicture}[
  >=Latex,
  every node/.style={font=\small},
  oddsite/.style={
    circle,draw=orange!80!black,fill=orange!14,
    minimum size=8mm,inner sep=0pt,line width=.8pt
  },
  evensite/.style={
    circle,draw=blue!70!black,fill=blue!12,
    minimum size=8mm,inner sep=0pt,line width=.8pt
  },
  orangeact/.style={-Latex,thick,draw=orange!85!black},
  blueact/.style={-Latex,thick,draw=blue!70!black},
  mybox/.style={
    draw=black!55,rounded corners=3pt,
    fill=black!4,inner sep=6pt,align=center
  }
]

% =========================================================
% (a) Physical chain
% =========================================================
\begin{scope}[shift={(0,-0.15)}]

  \draw[line width=1pt] (-.48,1.75) -- (7.48,1.75);

  % odd sites
  \foreach \j/\x in {1/0,3/2,5/4,7/6} {
    \node[oddsite] at (\x,1.75) {$\sigma_{\j}$};
    \draw[orangeact] (\x,2.80) -- (\x,2.18);
    \node[text=orange!85!black] at (\x,3.10) {$V(z)$};
  }

  % even sites
  \foreach \j/\x in {2/1,4/3,6/5,8/7} {
    \node[evensite] at (\x,1.75) {$\sigma_{\j}$};
    \draw[blueact] (\x,.70) -- (\x,1.32);
    \node[text=blue!70!black] at (\x,.48) {$X$};
  }

\end{scope}
\end{tikzpicture}
\caption{\label{VX}Construction of the family of cluster states on $L(n,1)$.}
\end{figure}
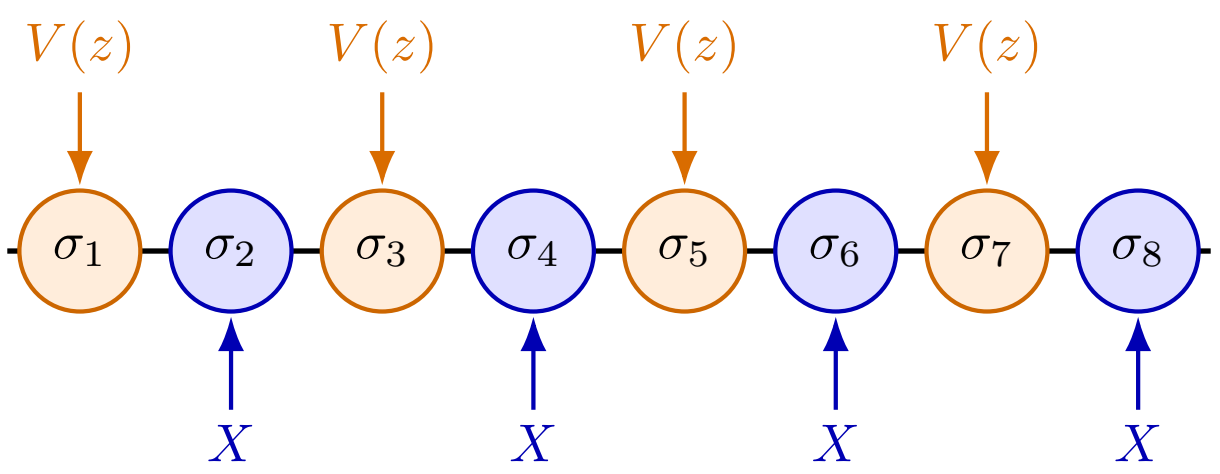

We rotate only the odd sites:
\begin{equation*}\label{eq:lens-rotated-state}
    \mathcal R_N(z)
    = \bigotimes_{j=1}^{N}V(z)_{2j-1},
    \qquad
    \ket{\Omega_N(z)}
    = \mathcal{R}_N(z) \ket{\Omega_{\mathrm{cl}}}.
\end{equation*}
Because $\mathcal{R}_N(z)$ commutes with $U_{\text{phys}}$, the family preserves the fixed on-site symmetry (see \zcref{VX}).
Moreover,
\begin{equation}\label{eq:lens-state-descent}
    \mathcal R_N(\omega z)
    = \mathcal{R}_N(z)U_{\mathrm{odd}},
    \qquad
    \ket{\Omega_N(\omega z)}
    = \ket{\Omega_N(z)},
\end{equation}
so the physical states are well defined on $L(n,1)$.
We now derive the MPS representation of the rotated cluster states.
The following proposition shows that the resulting tensors remain right-normalized, injective, and symmetric,
while the covariance of $V(z)$ is implemented by a virtual gauge transformation.

\begin{proposition}[Rotated cluster-state MPS]
    For every $z\in S^3$, the state $\ket{\Omega_N(z)}$ admits a translation-invariant MPS representation, after blocking pairs of adjacent sites, with tensors
    \begin{equation*}%\label{eq:lens-rotated-mps}
        B^{q,b}(z)
        \defeq
        \sum_{a\in\mathbb Z_n}
        V(z)_{qa}B^{a,b},
    \end{equation*}
    where $B^{a,b}=K^a\overline K^b$ are the
    cluster-state tensors introduced above. For every $z\in S^3$, the tensors are right-normalized, injective,
    and satisfy the fixed physical $\mathbb Z_n$ symmetry
    \begin{equation*}%\label{eq:lens-rotated-symmetry}
        \sum_{b'\in\mathbb Z_n}
        X_{bb'}B^{q,b'}(z)
        =
        X^\dagger B^{q,b}(z)X.
    \end{equation*}
    Moreover, they obey the covariance relation
    \begin{equation*}%\label{eq:lens-rotated-covariance}
        \boxed{
          B^{q,b}(\omega z)
          =
          Z^\dagger B^{q,b}(z)Z.
        }
    \end{equation*}
    Consequently, the rotated states define a continuous family of translation-invariant, injective, $\mathbb Z_n$-symmetric MPS parametrized by $L(n,1)$.
\end{proposition}

\begin{proof}
    We first derive the MPS representation of the rotated cluster state.
    Applying $\mathcal{R}_N(z)$ to the cluster state gives
    \begin{align*}
        \ket{\Omega_N(z)}
        &= \mathcal{R}_N(z) \ket{\Omega_{\mathrm{cl}}} \\
        &=
        \sum_{\boldsymbol a,\boldsymbol b}
        \Tr
        \bigl(B^{a_1,b_1}\cdots B^{a_N,b_N}\bigr)
        \bigotimes_{j=1}^N
        \left(
            V(z) \ket{a_j} \otimes \ket{b_j}
        \right) \\
        &=
        \sum_{\boldsymbol q,\boldsymbol b}
        \sum_{\boldsymbol a}
        \left(\prod_{j=1}^N V(z)_{q_j a_j}\right)
        \Tr\bigl(B^{a_1,b_1}\cdots B^{a_N,b_N}\bigr)
        \ket{q_1,b_1,\ldots,q_N,b_N} \\
        &=
        \sum_{\boldsymbol q,\boldsymbol b}
        \Tr
        \left[
            \prod_{j=1}^N
            \left(
                \sum_{a_j\in\mathbb Z_n}
                V(z)_{q_j a_j}B^{a_j,b_j}
            \right)
        \right]
        \ket{q_1,b_1,\ldots,q_N,b_N}.
\end{align*}
Thus the rotated state admits an MPS representation
with tensors
\begin{equation*}
    B^{q,b}(z)
    =
    \sum_{a\in\mathbb Z_n}
    V(z)_{qa}B^{a,b}.
\end{equation*}
Since $V(z)$ is unitary, the rotated tensors remain right-normalized and
span the same vector space as the original
blocked cluster-state tensors, hence the MPS is injective for every $z\in S^3$.

\bigskip

The physical symmetry follows from the corresponding
identity for the unrotated tensors (as the two transformations commute):
\begin{align*}
    \sum_{b'}X_{bb'}B^{q,b'}(z)
    &=
    \sum_a V(z)_{qa}
    \sum_{b'}X_{bb'}B^{a,b'} \nonumber\\
    &=
    \sum_a V(z)_{qa}
    X^\dagger B^{a,b}X
    \nonumber\\
    &=
    X^\dagger B^{q,b}(z)X.
\end{align*}
Finally, using $V(\omega z)=V(z)X$ and the
virtual implementation of the odd-site symmetry,
we obtain
\begin{align*}
    B^{q,b}(\omega z) &=
    \sum_a (V(\omega z))_{qa}B^{a,b}\\
    &=  \sum_a (V( z)X)_{qa}B^{a,b}\\
     &=     \sum_{a,c} V(z)_{qc} X_{ca} B^{a,b} \\
     &=
    \sum_a V(z)_{qc}
    Z^\dagger B^{a,b}Z \\
    &=
    Z^\dagger B^{q,b}(z)Z.
\end{align*}
Thus, tensors associated with $z$ and $\omega z$
differ only by a virtual gauge transformation
and therefore define the same  MPS.
Since $V(z)$ depends continuously on $z$,
the resulting family of states is well-defined
and continuous on $L(n,1)=S^3/\mathbb Z_n$.
\end{proof}

The MPS representation of a state is not unique: we are always free to redefine its tensors by a unitary conjugation, possibly depending on the parameter $z\in S^3$.
We shall exploit this gauge freedom to replace the tensors $B^{q,b}(z)$ by
\begin{equation}
    \widetilde B^{q,b}(z)
    \defeq W(z)B^{q,b}(z)W(z)^\dagger,
\end{equation}
in such a way that the new tensors are invariant under $z\mapsto\omega z$, and therefore define a continuous map
\[
    L(n,1)\longrightarrow\mathcal E(n^2,n).
\]
The price to pay is that the virtual implementation of the physical symmetry will no longer admit a globally defined representative.
To achieve this, we seek a continuous unitary map
$W:S^3\to U(n)$ satisfying
\begin{equation}\label{eq:lens-W-covariance}
    W(\omega z) = W(z)Z.
\end{equation}

\begin{proposition}[Global MPS representation]
    Let $F\in U(n)$ be the discrete Fourier transform, defined in the $Z$ basis by
    \begin{equation*}
        F_{ab}=\frac{1}{\sqrt n}\omega^{-ab}.
    \end{equation*}
    Define
    \begin{equation*}%\label{eq:lens-global-tensors}
        W(z)\defeq V(z)F.
    \end{equation*}
    Then~\eqref{eq:lens-W-covariance} holds.
    Moreover, the fixed physical $\mathbb Z_n$ symmetry is implemented virtually as
    \[
    \sum_{b'} X_{b b'} \widetilde B^{q,b'}(z) = \widetilde u(z)^\dagger \widetilde B^{q,b}(z)\widetilde u(z)
    \]
    where
    \begin{equation*}%\label{eq:lens-global-virtual-symmetry}
        \widetilde u(z)
        \defeq W(z)X W(z)^\dagger
        = V(z)Z^\dagger V(z)^\dagger.
    \end{equation*}
    This implementation satisfies
    \begin{equation*}\label{eq:lens-virtual-monodromy}
        \widetilde u(\omega z)
        =\omega \widetilde u(z).
    \end{equation*}
\end{proposition}

\begin{proof}
    The Fourier transform satisfies\footnote{
      These identities can be shown by observing that $F\ket{a} = \ket{a}_X$ and $F\ket{p}_X = \ket{-p}$.
    }
    $F^{\dagger}XF = Z$ and $F X F^{\dagger} = Z^{\dagger}$, therefore:
    \begin{equation*}
        W(\omega z) = V(\omega z) F = V(z) XF = V(z) F Z = W(z) Z
    \end{equation*}

    The fixed physical symmetry acts as
    \begin{align*}
        \sum_{b'}X_{bb'}\widetilde B^{q,b'}(z)
        &=
        W(z)
        \left(\sum_{b'}X_{bb'}B^{q,b'}(z)\right)
        W(z)^\dagger
        \nonumber\\
        &=
        W(z)X^\dagger B^{q,b}(z)XW(z)^\dagger
        \nonumber\\
        &=
        W(z)X^\dagger W(z)^{\dagger} W(z)B^{q,b}(z) W(z)^{\dagger} W(z)XW(z)^\dagger
        \nonumber\\
        &=
        \widetilde u(z)^\dagger \widetilde B^{q,b}(z)
        \widetilde u(z),
    \end{align*}
    where $\widetilde u(z)$ is as defined in the proposition statement. We observe that
    \[
        W(z) X W(z)^\dagger
        = V(z)FX F^\dagger V(z)^\dagger
        =V(z)Z^\dagger V(z)^\dagger,
    \]
    as claimed.
    Finally,
    \begin{align*}
        \widetilde u(\omega z)
        &=W(\omega z)X W(\omega z)^\dagger = W(z)Z XZ^\dagger W(z)^\dagger = \omega \widetilde u(z). \qedhere
    \end{align*}
\end{proof}

\subsubsection{Restriction to a (generating) circle and invariant}
Let us now consider the restriction of the globally defined MPS family $\widetilde B$ to the circle $\gamma : S^1 \to L(n,1)$ given by
\begin{equation*}%\label{eq:lens-generating-circle}
    \gamma(t)
    =
    \left[
        \bigl( e^{\frac{2\pi i}{n}t}, 0 \bigr)
    \right],
    \qquad t \in S^{1} = [0,1]/(0 \sim 1).
\end{equation*}
Since its lift $\widehat \gamma : [0,1] \to S^3$, $\widehat \gamma(t) = (e^{\frac{2\pi i}{n}t}, 0)$ joins $(1,0)$ to $(\omega,0)$, the loop $\gamma$ represents a generator of $\pi_1(L(n,1))\simeq\mathbb Z_n$.

Let us choose a good cover of \( S^{1} \) as in~\eqref{eq:circle-patches}.
Restricting the virtual implementation of the physical symmetry to the circle under consideration, we can define continuous local implementations of the symmetry on the three patches by setting
\begin{equation}
    u_1(t)=
    \begin{cases}
        \tilde{u}(\widehat \gamma(t)),
        &0\leq t<\frac16+\varepsilon,\\
        \omega^{-1} \tilde{u}(\widehat \gamma(t)),
        &\frac56-\varepsilon<t\leq1,
    \end{cases}
\end{equation}
and
\begin{equation}
    u_2(t) = u_3(t) = \tilde{u}(\widehat \gamma(t))
\end{equation}
on their respective domains.
Notice that $u_1$ is continuous across the
identification $0\sim1$ and on each patch they form a linear representation with 2-cocycle $\eta_i=1$. On each overlap, the two normalized virtual
implementations differ by a character
$\phi_{ij}(t)\in\widehat{\mathbb Z_n}$,
defined through
\begin{equation}
    u_j(t)^g
    =\phi_{ij}(t)(g)\,u_i(t)^g,
    \qquad g\in\mathbb Z_n.
\end{equation}
These characters are continuous and take values
in the discrete group $\widehat{\mathbb Z_n}$.
Since the pairwise intersections of our cover are connected, each $\phi_{ij}$ is constant and equal to:
\begin{equation}\label{eq:lens-circle-cocycle}
    \phi_{12}(g)=1, \qquad
    \phi_{23}(g)=1, \qquad
    \phi_{13}(g)=\omega^{g}.
\end{equation}
Thus, $\phi$ defines a Čech $1$-cocycle with coefficients in $\widehat{\mathbb Z_n}$.
Under the isomorphism $\check H^1(S^1,\widehat{\mathbb Z_n}) \simeq\widehat{\mathbb Z_n}$ derived above, its cohomology class $[\phi]$ corresponds to the character
\begin{equation}
    \phi_{12}(g)\phi_{23}(g)\phi_{31}(g) = \omega^{-g}, \qquad g\in\Z_n.
\end{equation}
Since this character generates $\widehat{\mathbb Z_n}$, the class $[\phi]$ is a generator of $\check H^1(S^1,\widehat{\mathbb Z_n})$.

According to \zcref{rem:pullback-parameter}, this means that the induced homomorphism
\begin{equation*}
    \gamma^*:
    \check H^1(L(n,1);\widehat{\Z_n})
    \rightarrow
    \check H^1(S^1;\widehat{\Z_n})
\end{equation*}
maps the parametric Thouless invariant on $L(n,1)$ to the generator $[\phi] \in \check{H}^1(S^1, \widehat{\Z_n})$.
Thanks to \zcref{prop:circle-into-X}, \( \gamma^{*} \) is an isomorphism and thus the  parametric Thouless invariant of the original family on $L(n,1)$ must also be a generator of $\check{H}^1(L(n,1); \widehat{\bbZ_n})$.

\begin{remark}[Realization of all charge-pump classes]
The construction above can be extended to realize
every element of
$\check H^1(L(n,1),\widehat{\mathbb Z_n})$. Indeed, for each $k=1,\ldots,n-1$, consider
the continuous map $f_k:S^3\to S^3$ defined by
\begin{equation}
    f_k(z_1,z_2)
    =
    \frac{
      (z_1^k,z_2^k)
    }{
      \sqrt{
        \abs{z_1}^{2k} + \abs{z_2}^{2k}
      }
    }.
\end{equation}
Since \(f_k(\omega z)=\omega^k f_k(z)\), this map descends to a continuous map
$\bar f_k:L(n,1)\to L(n,1)$.

We may therefore define a new family of
globally defined, injective, and
$\mathbb Z_n$-symmetric MPS tensors by $ \widetilde B_k
    =\widetilde B\circ\bar f_k$. Its normalized virtual symmetry implementation
is given by
$\widetilde u_k(z)=\widetilde u(f_k(z))$,
and satisfies $\widetilde u_k(\omega z)
    =\omega^{k}\widetilde u_k(z).$ Repeating the previous computation along the
generating circle, we obtain
\[
    \phi^{(k)}_{12}(g)
    \phi^{(k)}_{23}(g)
    \phi^{(k)}_{31}(g)
    =\omega^{-kg}.
\]
Consequently, the corresponding charge-pump
class is $[\phi^{(k)}]=k[\phi]$.
Together with the trivial class, realized by
a constant family, this yields every element
of $\check H^1(L(n,1);\widehat{\mathbb Z_n})$.
\end{remark}

\newpage
\appendix

\section{Measurability of Evaluation}

This appendix provides conditions under which the evaluation function $(f,x) \mapsto f(x)$ is measurable, where the input $f$ is a measurable function. This theorem was used in the proof of \zcref{prop:flatten}. For use in the proof, we recall some terminology from general topology. A collection $\cA$ of subsets of a topological space $X$ is called \emph{discrete} if every $x \in X$ has a neighborhood that intersects at most one element of $\cA$. We say $\cA$ is \emph{$\sigma$-discrete} if it can be written as a union of countably many discrete collections of subsets. A topological space is \emph{Lindel\"of} if every open cover has a countable subcover.

\begin{theorem}\label{thm:evaluation_measurable}
    Let $X$ be a topological space, let $Y$ be a bounded metric space, and let $\cL$ be a collection of Borel measurable functions $X \to Y$. Equip $\cL$ with the uniform topology. If $\cL$ is Lindel\"of with this topology, then evaluation
    \[
    \cL \times X \to Y, \quad (f, x) \mapsto f(x)
    \]
    is measurable with respect to the Borel $\sigma$-algebra on $Y$ and the $\sigma$-algebra generated by measurable rectangles on $\cL \times X$.
\end{theorem}

\begin{proof}
    Let $\cU_n = \{B_{1/n}(f): f \in \cL\}$, where $B_{1/n}(f)$ is the open ball of radius $1/n$ centered on $f$ with respect to the uniform metric $D$. Since $\cL$ is metrizable, it is paracompact Hausdorff, so there exists a $\sigma$-discrete open refinement $\cW_n$ of $\cU_n$   \cite[Thm.~5.28]{KelleyTopology}. Since $\cL$ is Lindel\"of, there exists a countable subcover $\cV_n \subset \cW_n$. Then $\cV_n$ is again $\sigma$-discrete. We may thus write $\cV_n$ as a countable union $\cV_n = \bigcup_k \cV_{n,k}$ where each $\cV_{n,k}$ is a discrete and countable collection of nonempty open sets.

    We will define a Borel measurable function $E_n:\cL \times X \to Y$ for each $n \in \bbN$ such that $\lim_{n \to \infty} E_n(f,x) = f(x)$
    for all $(f,x) \in \cL \times X$. Since the pointwise limit of a sequence of measurable functions into a metric space is measurable, this will imply that evaluation $(f,x) \mapsto f(x)$ is measurable.

    Fix $n \in \bbN$. For each $V \in \cV_n$ choose a point $f_{V} \in V$. We then define $E_n:\cL \times X \to Y$ as follows. Given $(f,x) \in \cL \times X$, let $k$ be the smallest number such that $f \in \bigcup \,\{V: V \in \cV_{n,k}\}$, and then let $E_n(f,x) = f_{V}(x)$ where $V$ is the unique element of $\cV_{n,k}$ such that $f \in V$.

    Observe that $f_V$ and $f$ are both contained in the set $V$ described above, which is contained in $B_{1/n}(f')$ for some $f' \in \cL$ since $\cV_n$ is a refinement of $\cU_n$. Thus, $D(f_V, f) < 1/(2n)$, so
    \[
    d(E_n(f,x), f(x)) \leq D(f_V, f) < \frac{1}{2n}.
    \]
    This proves that $\lim_{n \to \infty} E_n(f,x) = f(x)$ for all $(f,x) \in \cL \times X$.

    All that is left to do is show that $E_n$ is Borel measurable. Let $U \subset Y$ be open. For ease of notation, let $N_k = \bigcup\,\{V:V \in \cV_{n,k}\}$ for each $k$. Then
    \begin{align*}
    E_n^{-1}(U) &= \bigcup_k E_n^{-1}(U) \cap \left[\left( N_k \setminus \bigcup_{i < k} N_i \right) \times X \right]\\
    &= \bigcup_k \bigcup_{V \in \cV_{n,k}} E_n^{-1}(U) \cap \left[\left( V \setminus \bigcup_{i < k} N_i \right) \times X \right]
    \end{align*}
    Since there are countable number of elements of $\cV_{n,k}$, it suffices to show that each set in the union is measurable.
    Let $k \in \bbN$ and $V \in \cV_{n,k}$. If $f \in V \setminus \bigcup_{i < k} N_i$, then by definition of $E_n$ we have $E_n(f,x) = f_V(x)$ for every $x \in X$. Thus,
    \[
    E_n^{-1}(U)  \cap \left[\left( V \setminus \bigcup_{i < k} N_i \right) \times X \right] = \left( V \setminus \bigcup_{i < k} N_i \right) \times f_V^{-1}(U).
    \]
    Since $f_V$ is measurable, the right hand side is a measurable rectangle, completing the proof.
\end{proof}

\begin{corollary}\label{cor:adjoint_evaluation_measurable}
Suppose $X$ is a topological space, $Y$ is a bounded metric space, and $Z$ is a Lindel\"of space. Let $F:Z \times X \to Y$ be a function. If the adjoint function $F^\wedge :Z \to Y^X$ is continuous with respect to the uniform topology and $F^\wedge(z)$ is Borel measurable for every $z$, then $F$ is measurable with respect to the Borel $\sigma$-algebra on $Y$ and the $\sigma$-algebra generated by measurable rectangles on $Z \times X$. In particular $F$ is Borel measurable.
\end{corollary}

\begin{proof}
    Let $\cL = F^\wedge(Z)$. Since $Z$ is Lindel\"of, we know $\cL$ is Lindel\"of. By \zcref{thm:evaluation_measurable}, we know evaluation $\cL \times X \to Y$ is measurable with respect to the $\sigma$-algebra generated by measurable rectangles. We observe $F$ can be written as the composition of $F^\wedge \times \id:Z \times X \to \cL \times X$ with evaluation $\cL \times X \to Y$ and both maps are measurable with respect to the $\sigma$-algebras generated by measurable rectangles on the products.
\end{proof}

\newpage

\section{Explicit MPS Tensors for the \texorpdfstring{\( \Z_{2} \)}{ℤ₂}-Cluster State}\label{app:cluster-state-matrices}
Here we give the explicit form of the \( \tilde C \) matrices defined in~\eqref{eq:c-matrix-def}, for the Thouless pump based at the cluster state.
For \( t \in [0,1/2] \):

\begin{equation*}
    \begin{aligned}
        \tilde C^{1, 1, 1, 1} &=        \left(\begin{smallmatrix}  \frac{1}{16} e^{-i \pi  t} & 0 & 0 & 0 \\  \frac{1}{16} e^{-i \pi  t} & 0 & 0 & 0 \\  \frac{1}{16} e^{-i \pi  t} & 0 & 0 & 0 \\  \frac{1}{16} e^{-i \pi  t} & 0 & 0 & 0 \\ \end{smallmatrix}\right)     \\
        \tilde C^{1, 1, 1, -1} &=       \left(\begin{smallmatrix}  0 & \frac{1}{16} e^{-i \pi  t} & 0 & 0 \\  0 & \frac{1}{16} e^{-i \pi  t} & 0 & 0 \\  0 & \frac{1}{16} e^{-i \pi  t} & 0 & 0 \\  0 & \frac{1}{16} e^{-i \pi  t} & 0 & 0 \\ \end{smallmatrix}\right)     \\
        \tilde C^{1, 1, -1, 1} &=       \left(\begin{smallmatrix}  \frac{1}{16} e^{i \pi  t} & 0 & 0 & 0 \\  \frac{1}{16} e^{i \pi  t} & 0 & 0 & 0 \\  \frac{1}{16} e^{i \pi  t} & 0 & 0 & 0 \\  \frac{1}{16} e^{i \pi  t} & 0 & 0 & 0 \\ \end{smallmatrix}\right)         \\
        \tilde C^{1, 1, -1, -1} &=      \left(\begin{smallmatrix}  0 & -\frac{1}{16} e^{i \pi  t} & 0 & 0 \\  0 & -\frac{1}{16} e^{i \pi  t} & 0 & 0 \\  0 & -\frac{1}{16} e^{i \pi  t} & 0 & 0 \\  0 & -\frac{1}{16} e^{i \pi  t} & 0 & 0 \\ \end{smallmatrix}\right)     \\
        \tilde C^{1, -1, 1, 1} &=       \left(\begin{smallmatrix}  \frac{1}{16} e^{-i \pi  t} & 0 & 0 & 0 \\  \frac{1}{16} e^{-i \pi  t} & 0 & 0 & 0 \\  \frac{1}{16} e^{-i \pi  t} & 0 & 0 & 0 \\  \frac{1}{16} e^{-i \pi  t} & 0 & 0 & 0 \\ \end{smallmatrix}\right)     \\
        \tilde C^{1, -1, 1, -1} &=      \left(\begin{smallmatrix}  0 & \frac{1}{16} e^{-i \pi  t} & 0 & 0 \\  0 & \frac{1}{16} e^{-i \pi  t} & 0 & 0 \\  0 & \frac{1}{16} e^{-i \pi  t} & 0 & 0 \\  0 & \frac{1}{16} e^{-i \pi  t} & 0 & 0 \\ \end{smallmatrix}\right)     \\
        \tilde C^{1, -1, -1, 1} &=      \left(\begin{smallmatrix}  -\frac{1}{16} e^{i \pi  t} & 0 & 0 & 0 \\  -\frac{1}{16} e^{i \pi  t} & 0 & 0 & 0 \\  -\frac{1}{16} e^{i \pi  t} & 0 & 0 & 0 \\  -\frac{1}{16} e^{i \pi  t} & 0 & 0 & 0 \\ \end{smallmatrix}\right)     \\
        \tilde C^{1, -1, -1, -1} &=     \left(\begin{smallmatrix}  0 & \frac{1}{16} e^{i \pi  t} & 0 & 0 \\  0 & \frac{1}{16} e^{i \pi  t} & 0 & 0 \\  0 & \frac{1}{16} e^{i \pi  t} & 0 & 0 \\  0 & \frac{1}{16} e^{i \pi  t} & 0 & 0 \\ \end{smallmatrix}\right)
    \end{aligned} \qquad
    \begin{aligned}
        \tilde C^{-1, 1, 1, 1} &=       \left(\begin{smallmatrix}  \frac{1}{16} e^{i \pi  t} & 0 & 0 & 0 \\  -\frac{1}{16} e^{i \pi  t} & 0 & 0 & 0 \\  -\frac{1}{16} e^{i \pi  t} & 0 & 0 & 0 \\  \frac{1}{16} e^{i \pi  t} & 0 & 0 & 0 \\ \end{smallmatrix}\right)       \\
        \tilde C^{-1, 1, 1, -1} &=      \left(\begin{smallmatrix}  0 & \frac{1}{16} e^{i \pi  t} & 0 & 0 \\  0 & -\frac{1}{16} e^{i \pi  t} & 0 & 0 \\  0 & -\frac{1}{16} e^{i \pi  t} & 0 & 0 \\  0 & \frac{1}{16} e^{i \pi  t} & 0 & 0 \\ \end{smallmatrix}\right)       \\
        \tilde C^{-1, 1, -1, 1} &=      \left(\begin{smallmatrix}  \frac{1}{16} e^{-i \pi  t} & 0 & 0 & 0 \\  -\frac{1}{16} e^{-i \pi  t} & 0 & 0 & 0 \\  -\frac{1}{16} e^{-i \pi  t} & 0 & 0 & 0 \\  \frac{1}{16} e^{-i \pi  t} & 0 & 0 & 0 \\ \end{smallmatrix}\right)   \\
        \tilde C^{-1, 1, -1, -1} &=     \left(\begin{smallmatrix}  0 & -\frac{1}{16} e^{-i \pi  t} & 0 & 0 \\  0 & \frac{1}{16} e^{-i \pi  t} & 0 & 0 \\  0 & \frac{1}{16} e^{-i \pi  t} & 0 & 0 \\  0 & -\frac{1}{16} e^{-i \pi  t} & 0 & 0 \\ \end{smallmatrix}\right)   \\
        \tilde C^{-1, -1, 1, 1} &=      \left(\begin{smallmatrix}  -\frac{1}{16} e^{i \pi  t} & 0 & 0 & 0 \\  \frac{1}{16} e^{i \pi  t} & 0 & 0 & 0 \\  \frac{1}{16} e^{i \pi  t} & 0 & 0 & 0 \\  -\frac{1}{16} e^{i \pi  t} & 0 & 0 & 0 \\ \end{smallmatrix}\right)       \\
        \tilde C^{-1, -1, 1, -1} &=     \left(\begin{smallmatrix}  0 & -\frac{1}{16} e^{i \pi  t} & 0 & 0 \\  0 & \frac{1}{16} e^{i \pi  t} & 0 & 0 \\  0 & \frac{1}{16} e^{i \pi  t} & 0 & 0 \\  0 & -\frac{1}{16} e^{i \pi  t} & 0 & 0 \\ \end{smallmatrix}\right)       \\
        \tilde C^{-1, -1, -1, 1} &=     \left(\begin{smallmatrix}  \frac{1}{16} e^{-i \pi  t} & 0 & 0 & 0 \\  -\frac{1}{16} e^{-i \pi  t} & 0 & 0 & 0 \\  -\frac{1}{16} e^{-i \pi  t} & 0 & 0 & 0 \\  \frac{1}{16} e^{-i \pi  t} & 0 & 0 & 0 \\ \end{smallmatrix}\right)   \\
        \tilde C^{-1, -1, -1, -1} &=    \left(\begin{smallmatrix}  0 & -\frac{1}{16} e^{-i \pi  t} & 0 & 0 \\  0 & \frac{1}{16} e^{-i \pi  t} & 0 & 0 \\  0 & \frac{1}{16} e^{-i \pi  t} & 0 & 0 \\  0 & -\frac{1}{16} e^{-i \pi  t} & 0 & 0 \\ \end{smallmatrix}\right)
    \end{aligned}
\end{equation*}
\newpage
\noindent For \( t \in (1/2,1) \):
\begin{equation*}%\label{eq:c-tilde-second-part}
    \scalebox{0.8}{$
    \begin{aligned}
        \tilde C^{1, 1, 1, 1} &=      \left( \begin{smallmatrix}  -\frac{1}{16} e^{i \pi  t} & 0 & 0 & 0 \\  -\frac{1}{16} e^{i \pi  t} & 0 & 0 & 0 \\  -\frac{1}{16} e^{i \pi  t} & 0 & 0 & 0 \\  -\frac{1}{16} e^{i \pi  t} & 0 & 0 & 0 \\ \end{smallmatrix} \right) \\
        \tilde C^{1, 1, 1, -1} &=     \left( \begin{smallmatrix}  0 & -\frac{1}{16} e^{i \pi  t} & 0 & 0 \\  0 & -\frac{1}{16} e^{i \pi  t} & 0 & 0 \\  0 & -\frac{1}{16} e^{i \pi  t} & 0 & 0 \\  0 & -\frac{1}{16} e^{i \pi  t} & 0 & 0 \\ \end{smallmatrix} \right) \\
        \tilde C^{1, 1, -1, 1} &=     \left( \begin{smallmatrix}  -\frac{1}{16} e^{i \pi  t} \cos (2 \pi  t) & 0 & -\frac{1}{16} i e^{i \pi  t} \sin (2 \pi  t) & 0 \\  -\frac{1}{16} e^{i \pi  t} \cos (2 \pi  t) & 0 & -\frac{1}{16} i e^{i \pi  t} \sin (2 \pi  t) & 0 \\  -\frac{1}{16} e^{i \pi  t} \cos (2 \pi  t) & 0 & -\frac{1}{16} i e^{i \pi  t} \sin (2 \pi  t) & 0 \\  -\frac{1}{16} e^{i \pi  t} \cos (2 \pi  t) & 0 & -\frac{1}{16} i e^{i \pi  t} \sin (2 \pi  t) & 0 \\ \end{smallmatrix} \right) \\
        \tilde C^{1, 1, -1, -1} &=    \left( \begin{smallmatrix}  0 & \frac{1}{32} \left(e^{-i \pi  t}+e^{3 i \pi  t}\right) & 0 & \frac{1}{32} e^{-i \pi  t} \left(-1+e^{4 i \pi  t}\right) \\  0 & \frac{1}{32} \left(e^{-i \pi  t}+e^{3 i \pi  t}\right) & 0 & \frac{1}{32} e^{-i \pi  t} \left(-1+e^{4 i \pi  t}\right) \\  0 & \frac{1}{32} \left(e^{-i \pi  t}+e^{3 i \pi  t}\right) & 0 & \frac{1}{32} e^{-i \pi  t} \left(-1+e^{4 i \pi  t}\right) \\  0 & \frac{1}{32} \left(e^{-i \pi  t}+e^{3 i \pi  t}\right) & 0 & \frac{1}{32} e^{-i \pi  t} \left(-1+e^{4 i \pi  t}\right) \\ \end{smallmatrix} \right) \\
        \tilde C^{1, -1, 1, 1} &=     \left( \begin{smallmatrix}  -\frac{1}{16} e^{i \pi  t} & 0 & 0 & 0 \\  -\frac{1}{16} e^{i \pi  t} & 0 & 0 & 0 \\  -\frac{1}{16} e^{i \pi  t} & 0 & 0 & 0 \\  -\frac{1}{16} e^{i \pi  t} & 0 & 0 & 0 \\ \end{smallmatrix} \right) \\
        \tilde C^{1, -1, 1, -1} &=    \left( \begin{smallmatrix}  0 & -\frac{1}{16} e^{i \pi  t} & 0 & 0 \\  0 & -\frac{1}{16} e^{i \pi  t} & 0 & 0 \\  0 & -\frac{1}{16} e^{i \pi  t} & 0 & 0 \\  0 & -\frac{1}{16} e^{i \pi  t} & 0 & 0 \\ \end{smallmatrix} \right) \\
        \tilde C^{1, -1, -1, 1} &=    \left( \begin{smallmatrix}  \frac{1}{32} \left(e^{-i \pi  t}+e^{3 i \pi  t}\right) & 0 & \frac{1}{32} e^{-i \pi  t} \left(-1+e^{4 i \pi  t}\right) & 0 \\  \frac{1}{32} \left(e^{-i \pi  t}+e^{3 i \pi  t}\right) & 0 & \frac{1}{32} e^{-i \pi  t} \left(-1+e^{4 i \pi  t}\right) & 0 \\  \frac{1}{32} \left(e^{-i \pi  t}+e^{3 i \pi  t}\right) & 0 & \frac{1}{32} e^{-i \pi  t} \left(-1+e^{4 i \pi  t}\right) & 0 \\  \frac{1}{32} \left(e^{-i \pi  t}+e^{3 i \pi  t}\right) & 0 & \frac{1}{32} e^{-i \pi  t} \left(-1+e^{4 i \pi  t}\right) & 0 \\ \end{smallmatrix} \right) \\
        \tilde C^{1, -1, -1, -1} &=   \left( \begin{smallmatrix}  0 & -\frac{1}{16} e^{i \pi  t} \cos (2 \pi  t) & 0 & -\frac{1}{16} i e^{i \pi  t} \sin (2 \pi  t) \\  0 & -\frac{1}{16} e^{i \pi  t} \cos (2 \pi  t) & 0 & -\frac{1}{16} i e^{i \pi  t} \sin (2 \pi  t) \\  0 & -\frac{1}{16} e^{i \pi  t} \cos (2 \pi  t) & 0 & -\frac{1}{16} i e^{i \pi  t} \sin (2 \pi  t) \\  0 & -\frac{1}{16} e^{i \pi  t} \cos (2 \pi  t) & 0 & -\frac{1}{16} i e^{i \pi  t} \sin (2 \pi  t) \\ \end{smallmatrix} \right)
    \end{aligned} \qquad
    \begin{aligned}
        \tilde C^{-1, 1, 1, 1} &=      \left( \begin{smallmatrix}  -\frac{1}{16} e^{3 i \pi  t} & 0 & 0 & 0 \\  \frac{1}{16} e^{3 i \pi  t} & 0 & 0 & 0 \\  \frac{1}{16} e^{3 i \pi  t} & 0 & 0 & 0 \\  -\frac{1}{16} e^{3 i \pi  t} & 0 & 0 & 0 \\ \end{smallmatrix} \right) \\
        \tilde C^{-1, 1, 1, -1} &=     \left( \begin{smallmatrix}  0 & -\frac{1}{16} e^{3 i \pi  t} & 0 & 0 \\  0 & \frac{1}{16} e^{3 i \pi  t} & 0 & 0 \\  0 & \frac{1}{16} e^{3 i \pi  t} & 0 & 0 \\  0 & -\frac{1}{16} e^{3 i \pi  t} & 0 & 0 \\ \end{smallmatrix} \right) \\
        \tilde C^{-1, 1, -1, 1} &=     \left( \begin{smallmatrix}  -\frac{1}{32} e^{i \pi  t} \left(1+e^{4 i \pi  t}\right) & 0 & -\frac{1}{32} e^{i \pi  t} \left(-1+e^{4 i \pi  t}\right) & 0 \\  \frac{1}{32} \left(e^{i \pi  t}+e^{5 i \pi  t}\right) & 0 & \frac{1}{32} e^{i \pi  t} \left(-1+e^{4 i \pi  t}\right) & 0 \\  \frac{1}{32} \left(e^{i \pi  t}+e^{5 i \pi  t}\right) & 0 & \frac{1}{32} e^{i \pi  t} \left(-1+e^{4 i \pi  t}\right) & 0 \\  -\frac{1}{32} e^{i \pi  t} \left(1+e^{4 i \pi  t}\right) & 0 & -\frac{1}{32} e^{i \pi  t} \left(-1+e^{4 i \pi  t}\right) & 0 \\ \end{smallmatrix} \right) \\
        \tilde C^{-1, 1, -1, -1} &=    \left( \begin{smallmatrix}  0 & \frac{1}{32} \left(e^{i \pi  t}+e^{5 i \pi  t}\right) & 0 & \frac{1}{32} e^{i \pi  t} \left(-1+e^{4 i \pi  t}\right) \\  0 & -\frac{1}{32} e^{i \pi  t} \left(1+e^{4 i \pi  t}\right) & 0 & -\frac{1}{32} e^{i \pi  t} \left(-1+e^{4 i \pi  t}\right) \\  0 & -\frac{1}{32} e^{i \pi  t} \left(1+e^{4 i \pi  t}\right) & 0 & -\frac{1}{32} e^{i \pi  t} \left(-1+e^{4 i \pi  t}\right) \\  0 & \frac{1}{32} \left(e^{i \pi  t}+e^{5 i \pi  t}\right) & 0 & \frac{1}{32} e^{i \pi  t} \left(-1+e^{4 i \pi  t}\right) \\ \end{smallmatrix} \right) \\
        \tilde C^{-1, -1, 1, 1} &=     \left( \begin{smallmatrix}  \frac{1}{16} e^{3 i \pi  t} & 0 & 0 & 0 \\  -\frac{1}{16} e^{3 i \pi  t} & 0 & 0 & 0 \\  -\frac{1}{16} e^{3 i \pi  t} & 0 & 0 & 0 \\  \frac{1}{16} e^{3 i \pi  t} & 0 & 0 & 0 \\ \end{smallmatrix} \right) \\
        \tilde C^{-1, -1, 1, -1} &=   \left( \begin{smallmatrix}  0 & \frac{1}{16} e^{3 i \pi  t} & 0 & 0 \\  0 & -\frac{1}{16} e^{3 i \pi  t} & 0 & 0 \\  0 & -\frac{1}{16} e^{3 i \pi  t} & 0 & 0 \\  0 & \frac{1}{16} e^{3 i \pi  t} & 0 & 0 \\ \end{smallmatrix} \right) \\
        \tilde C^{-1, -1, -1, 1} &=    \left( \begin{smallmatrix}  -\frac{1}{32} e^{i \pi  t} \left(1+e^{4 i \pi  t}\right) & 0 & -\frac{1}{32} e^{i \pi  t} \left(-1+e^{4 i \pi  t}\right) & 0 \\  \frac{1}{32} \left(e^{i \pi  t}+e^{5 i \pi  t}\right) & 0 & \frac{1}{32} e^{i \pi  t} \left(-1+e^{4 i \pi  t}\right) & 0 \\  \frac{1}{32} \left(e^{i \pi  t}+e^{5 i \pi  t}\right) & 0 & \frac{1}{32} e^{i \pi  t} \left(-1+e^{4 i \pi  t}\right) & 0 \\  -\frac{1}{32} e^{i \pi  t} \left(1+e^{4 i \pi  t}\right) & 0 & -\frac{1}{32} e^{i \pi  t} \left(-1+e^{4 i \pi  t}\right) & 0 \\ \end{smallmatrix} \right) \\
        \tilde C^{-1, -1, -1, -1} &=   \left( \begin{smallmatrix}  0 & \frac{1}{32} \left(e^{i \pi  t}+e^{5 i \pi  t}\right) & 0 & \frac{1}{32} e^{i \pi  t} \left(-1+e^{4 i \pi  t}\right) \\  0 & -\frac{1}{32} e^{i \pi  t} \left(1+e^{4 i \pi  t}\right) & 0 & -\frac{1}{32} e^{i \pi  t} \left(-1+e^{4 i \pi  t}\right) \\  0 & -\frac{1}{32} e^{i \pi  t} \left(1+e^{4 i \pi  t}\right) & 0 & -\frac{1}{32} e^{i \pi  t} \left(-1+e^{4 i \pi  t}\right) \\  0 & \frac{1}{32} \left(e^{i \pi  t}+e^{5 i \pi  t}\right) & 0 & \frac{1}{32} e^{i \pi  t} \left(-1+e^{4 i \pi  t}\right) \\ \end{smallmatrix} \right)
    \end{aligned}
    $}
\end{equation*}
The matrices \( \tilde D \) defined in~\eqref{eq:c-matrix-def} are instead given by, for \( t \in [0,1/2] \):
\begin{equation*}%\label{eq:cluster-D-tilde-mats-1}
    \scalebox{0.8}{$
      \begin{aligned}
          \tilde D^{1, 1, 1, 1} &=      \left(\begin{smallmatrix} \frac{1}{4} e^{-i \pi  t} & 0 \\ \frac{1}{4} e^{-i \pi  t} & 0 \\\end{smallmatrix}\right)    \\
          \tilde D^{1, 1, 1, -1} &=     \left(\begin{smallmatrix} 0 & \frac{1}{4} e^{-i \pi  t} \\ 0 & \frac{1}{4} e^{-i \pi  t} \\\end{smallmatrix}\right) \\
          \tilde D^{1, 1, -1, 1} &=     \left(\begin{smallmatrix} \frac{1}{4} e^{i \pi  t} & 0 \\ \frac{1}{4} e^{i \pi  t} & 0 \\\end{smallmatrix}\right) \\
          \tilde D^{1, 1, -1, -1} &=    \left(\begin{smallmatrix} 0 & -\frac{1}{4} e^{i \pi  t} \\ 0 & -\frac{1}{4} e^{i \pi  t} \\\end{smallmatrix}\right) \\
          \tilde D^{1, -1, 1, 1} &=     \left(\begin{smallmatrix} \frac{1}{4} e^{-i \pi  t} & 0 \\ \frac{1}{4} e^{-i \pi  t} & 0 \\\end{smallmatrix}\right) \\
          \tilde D^{1, -1, 1, -1} &=    \left(\begin{smallmatrix} 0 & \frac{1}{4} e^{-i \pi  t} \\ 0 & \frac{1}{4} e^{-i \pi  t} \\\end{smallmatrix}\right) \\
          \tilde D^{1, -1, -1, 1} &=    \left(\begin{smallmatrix} -\frac{1}{4} e^{i \pi  t} & 0 \\ -\frac{1}{4} e^{i \pi  t} & 0 \\\end{smallmatrix}\right) \\
          \tilde D^{1, -1, -1, -1} &=   \left(\begin{smallmatrix} 0 & \frac{1}{4} e^{i \pi  t} \\ 0 & \frac{1}{4} e^{i \pi  t} \\\end{smallmatrix}\right)
      \end{aligned} \qquad
      \begin{aligned}
          \tilde D^{-1, 1, 1, 1} &=       \left(\begin{smallmatrix} \frac{1}{4} e^{i \pi  t} & 0 \\ -\frac{1}{4} e^{i \pi  t} & 0 \\\end{smallmatrix}\right)  \\
          \tilde D^{-1, 1, 1, -1} &=      \left(\begin{smallmatrix} 0 & \frac{1}{4} e^{i \pi  t} \\ 0 & -\frac{1}{4} e^{i \pi  t} \\\end{smallmatrix}\right)  \\
          \tilde D^{-1, 1, -1, 1} &=      \left(\begin{smallmatrix} \frac{1}{4} e^{-i \pi  t} & 0 \\ -\frac{1}{4} e^{-i \pi  t} & 0 \\\end{smallmatrix}\right)  \\
          \tilde D^{-1, 1, -1, -1} &=     \left(\begin{smallmatrix} 0 & -\frac{1}{4} e^{-i \pi  t} \\ 0 & \frac{1}{4} e^{-i \pi  t} \\\end{smallmatrix}\right)  \\
          \tilde D^{-1, -1, 1, 1} &=      \left(\begin{smallmatrix} -\frac{1}{4} e^{i \pi  t} & 0 \\ \frac{1}{4} e^{i \pi  t} & 0 \\\end{smallmatrix}\right)  \\
          \tilde D^{-1, -1, 1, -1} &=     \left(\begin{smallmatrix} 0 & -\frac{1}{4} e^{i \pi  t} \\ 0 & \frac{1}{4} e^{i \pi  t} \\\end{smallmatrix}\right)  \\
          \tilde D^{-1, -1, -1, 1} &=     \left(\begin{smallmatrix} \frac{1}{4} e^{-i \pi  t} & 0 \\ -\frac{1}{4} e^{-i \pi  t} & 0 \\\end{smallmatrix}\right)  \\
          \tilde D^{-1, -1, -1, -1} &=    \left(\begin{smallmatrix} 0 & -\frac{1}{4} e^{-i \pi  t} \\ 0 & \frac{1}{4} e^{-i \pi  t} \\\end{smallmatrix}\right)
      \end{aligned}
      $}
\end{equation*}
For \( t \in (1/2,1) \):
\begin{equation*}%\label{eq:cluster-D-tilde-mats-2}
    \scalebox{0.8}{$
    \begin{aligned}
        \tilde D^{1, 1, 1, 1} &=           \left(\begin{smallmatrix} -\frac{1}{4} e^{i \pi  t} & 0 \\ -\frac{1}{4} e^{i \pi  t} & 0 \\\end{smallmatrix}\right) \\
        \tilde D^{1, 1, 1, -1} &=          \left(\begin{smallmatrix} 0 & -\frac{1}{4} e^{i \pi  t} \\ 0 & -\frac{1}{4} e^{i \pi  t} \\\end{smallmatrix}\right) \\
        \tilde D^{1, 1, -1, 1} &=          \left(\begin{smallmatrix} -\frac{1}{4} e^{i \pi  t} \cos (2 \pi  t) & -\frac{1}{4} i e^{i \pi  t} \sin (2 \pi  t) \\ -\frac{1}{4} e^{i \pi  t} \cos (2 \pi  t) & -\frac{1}{4} i e^{i \pi  t} \sin (2 \pi  t) \\\end{smallmatrix}\right) \\
        \tilde D^{1, 1, -1, -1} &=         \left(\begin{smallmatrix} \frac{1}{8} e^{-i \pi  t} \left(-1+e^{4 i \pi  t}\right) & \frac{1}{8} \left(e^{-i \pi  t}+e^{3 i \pi  t}\right) \\ \frac{1}{8} e^{-i \pi  t} \left(-1+e^{4 i \pi  t}\right) & \frac{1}{8} \left(e^{-i \pi  t}+e^{3 i \pi  t}\right) \\\end{smallmatrix}\right) \\
        \tilde D^{1, -1, 1, 1} &=          \left(\begin{smallmatrix} -\frac{1}{4} e^{i \pi  t} & 0 \\ -\frac{1}{4} e^{i \pi  t} & 0 \\\end{smallmatrix}\right) \\
        \tilde D^{1, -1, 1, -1} &=         \left(\begin{smallmatrix} 0 & -\frac{1}{4} e^{i \pi  t} \\ 0 & -\frac{1}{4} e^{i \pi  t} \\\end{smallmatrix}\right) \\
        \tilde D^{1, -1, -1, 1} &=         \left(\begin{smallmatrix} \frac{1}{8} \left(e^{-i \pi  t}+e^{3 i \pi  t}\right) & \frac{1}{8} e^{-i \pi  t} \left(-1+e^{4 i \pi  t}\right) \\ \frac{1}{8} \left(e^{-i \pi  t}+e^{3 i \pi  t}\right) & \frac{1}{8} e^{-i \pi  t} \left(-1+e^{4 i \pi  t}\right) \\\end{smallmatrix}\right) \\
        \tilde D^{1, -1, -1, -1} &=        \left(\begin{smallmatrix} -\frac{1}{4} i e^{i \pi  t} \sin (2 \pi  t) & -\frac{1}{4} e^{i \pi  t} \cos (2 \pi  t) \\ -\frac{1}{4} i e^{i \pi  t} \sin (2 \pi  t) & -\frac{1}{4} e^{i \pi  t} \cos (2 \pi  t) \\\end{smallmatrix}\right)
    \end{aligned} \qquad
    \begin{aligned}
        \tilde D^{-1, 1, 1, 1} &=      \left(\begin{smallmatrix} -\frac{1}{4} e^{3 i \pi  t} & 0 \\ \frac{1}{4} e^{3 i \pi  t} & 0 \\\end{smallmatrix}\right) \\
        \tilde D^{-1, 1, 1, -1} &=     \left(\begin{smallmatrix} 0 & -\frac{1}{4} e^{3 i \pi  t} \\ 0 & \frac{1}{4} e^{3 i \pi  t} \\\end{smallmatrix}\right) \\
        \tilde D^{-1, 1, -1, 1} &=     \left(\begin{smallmatrix} -\frac{1}{8} e^{i \pi  t} \left(1+e^{4 i \pi  t}\right) & -\frac{1}{8} e^{i \pi  t} \left(-1+e^{4 i \pi  t}\right) \\ \frac{1}{8} \left(e^{i \pi  t}+e^{5 i \pi  t}\right) & \frac{1}{8} e^{i \pi  t} \left(-1+e^{4 i \pi  t}\right) \\\end{smallmatrix}\right) \\
        \tilde D^{-1, 1, -1, -1} &=    \left(\begin{smallmatrix} \frac{1}{8} e^{i \pi  t} \left(-1+e^{4 i \pi  t}\right) & \frac{1}{8} \left(e^{i \pi  t}+e^{5 i \pi  t}\right) \\ -\frac{1}{8} e^{i \pi  t} \left(-1+e^{4 i \pi  t}\right) & -\frac{1}{8} e^{i \pi  t} \left(1+e^{4 i \pi  t}\right) \\\end{smallmatrix}\right) \\
        \tilde D^{-1, -1, 1, 1} &=     \left(\begin{smallmatrix} \frac{1}{4} e^{3 i \pi  t} & 0 \\ -\frac{1}{4} e^{3 i \pi  t} & 0 \\\end{smallmatrix}\right) \\
        \tilde D^{-1, -1, 1, -1} &=    \left(\begin{smallmatrix} 0 & \frac{1}{4} e^{3 i \pi  t} \\ 0 & -\frac{1}{4} e^{3 i \pi  t} \\\end{smallmatrix}\right) \\
        \tilde D^{-1, -1, -1, 1} &=    \left(\begin{smallmatrix} -\frac{1}{8} e^{i \pi  t} \left(1+e^{4 i \pi  t}\right) & -\frac{1}{8} e^{i \pi  t} \left(-1+e^{4 i \pi  t}\right) \\ \frac{1}{8} \left(e^{i \pi  t}+e^{5 i \pi  t}\right) & \frac{1}{8} e^{i \pi  t} \left(-1+e^{4 i \pi  t}\right) \\\end{smallmatrix}\right) \\
        \tilde D^{-1, -1, -1, -1} &=   \left(\begin{smallmatrix} \frac{1}{8} e^{i \pi  t} \left(-1+e^{4 i \pi  t}\right) & \frac{1}{8} \left(e^{i \pi  t}+e^{5 i \pi  t}\right) \\ -\frac{1}{8} e^{i \pi  t} \left(-1+e^{4 i \pi  t}\right) & -\frac{1}{8} e^{i \pi  t} \left(1+e^{4 i \pi  t}\right) \\\end{smallmatrix}\right)
    \end{aligned}\qquad
    $}
\end{equation*}

\section*{Acknowledgements}

D.\ D.\ Spiegel would like to thank Agn\`es Beaudry for helpful conversations. While performing this work, D.\ D.\ Spiegel was supported by the National Science Foundation under Grant No.\ DMS 2303063 and by the Simons Foundation Award No.\ 888992 as part of the Simons Collaboration on Global Categorical Symmetries.
L.\ Goller is currently supported by the Simons Collaboration on the Probabilistic Paths to Quantum Field Theory (award SFI-MPS-PP-00012621-16).

\section*{AI Statement}

The figures in this manuscript were generated by Anthropic's Claude AI. The idea of the proof of \zcref*{thm:evaluation_measurable} and \zcref*{cor:adjoint_evaluation_measurable} is also due to Claude; the write-up is due to the authors. The central ideas of Section 5.4 are those of the authors, while the exact calculations of that section were developed by the authors with the assistance of Claude.

\section*{Data Availability Statement}

No datasets were generated or analyzed during the current study.

\section*{Conflicts of Interest}

The authors have no relevant financial or non-financial interests to disclose.

\bibliographystyle{amsalpha}
\bibliography{MPS_Symmetry-bib.bib}

\end{document}